\documentclass[12pt]{iopart}
\expandafter\let\csname equation*\endcsname\relax
\expandafter\let\csname endequation*\endcsname\relax
\usepackage{epsfig}
\usepackage{graphicx}
\usepackage{amssymb}
\usepackage{amsmath}
\usepackage{mathtools}
\usepackage{multicol}
\usepackage [english]{babel}
\usepackage{enumitem}
\usepackage{gensymb}
\usepackage{geometry}
\usepackage{cancel}
\usepackage[titletoc,title]{appendix}
\usepackage{cite}
\usepackage{epstopdf}

\usepackage{float}
\usepackage{subcaption}
\usepackage{breqn}
\usepackage{hyperref}
\hypersetup{
    colorlinks=true,
    linkcolor=blue,
    filecolor=blue,      
    urlcolor=blue,
    citecolor=blue,
}

\renewcommand\vec\mathbf

\newcommand{\dd}[1]{\!\!\! \mathrm{d}#1 \,}
\newcommand{\ee}{\mathrm{e}}
\newcommand{\ii}{\mathrm{i}}
\newcommand{\sss}{\mathrm{s}}
\newcommand{\cc}{\mathrm{cut}}
\newcommand{\pdev}[2]{\frac{\partial #1}{\partial #2}}
\newcommand{\pdevn}[3]{\frac{\partial^{#3} #1}{\partial #2^{#3}}}
\newcommand{\dev}[3]{\frac{\mathrm{d}^{#3} #1}{\mathrm{d} #2^{#3}}}
\newcommand{\gyg}[1]{\langle #1 \rangle_{\varphi}}
\newcommand{\erf}{\mathrm{erf}}
\newcommand{\delphis}{\Delta \phi_{\mathrm{DS}}}
\newcommand{\mugk}{\mu_{\mathrm{gk}}}

\begin{document}
\title{Sheath collapse at critical shallow angle due to kinetic effects}

\author{Robert J Ewart$^{1}$, Felix I Parra$^{1,2}$, Alessandro Geraldini$^{3,4}$}
\address{$^1$ Rudolf Peierls Centre for Theoretical Physics, University of Oxford, Oxford, OX1 3PU, UK}
\address{$^{2}$ CCFE - UKAEA, Culham Science Centre, Abingdon OX14 3DB, UK}
\address{$^{3}$ Swiss Plasma Center, \'Ecole Polytechnique F\'ed\'erale de Lausanne, CH-1015 Lausanne, Switzerland}
\address{$^{4}$ Institute for Research in Electronics and Applied Physics, University of Maryland, College Park, MD 20742 USA}
\ead{robert.ewart@balliol.ox.ac.uk} 

\begin{abstract}
The Debye sheath is known to vanish completely in magnetised plasmas for a sufficiently small electron gyroradius and small angle between the magnetic field and the wall. This angle depends on the current onto the wall. When the Debye sheath vanishes, there is still a potential drop between the wall and the plasma across the magnetic presheath. The magnetic field angle corresponding to the predicted sheath collapse is shown to be much smaller than previous estimates, scaling with the electron-ion mass ratio and not with the square root of the mass ratio. This is shown to be a consequence of the kinetic electron and finite ion orbit width effects, which are not captured by fluid models. The wall potential with respect to the bulk plasma at which the Debye sheath vanishes is calculated. Above this wall potential, it is possible that the Debye sheath will invert. 
\end{abstract}

\section{Introduction and orderings}
\begin{figure}
\centering 
\includegraphics[width=0.75\linewidth]{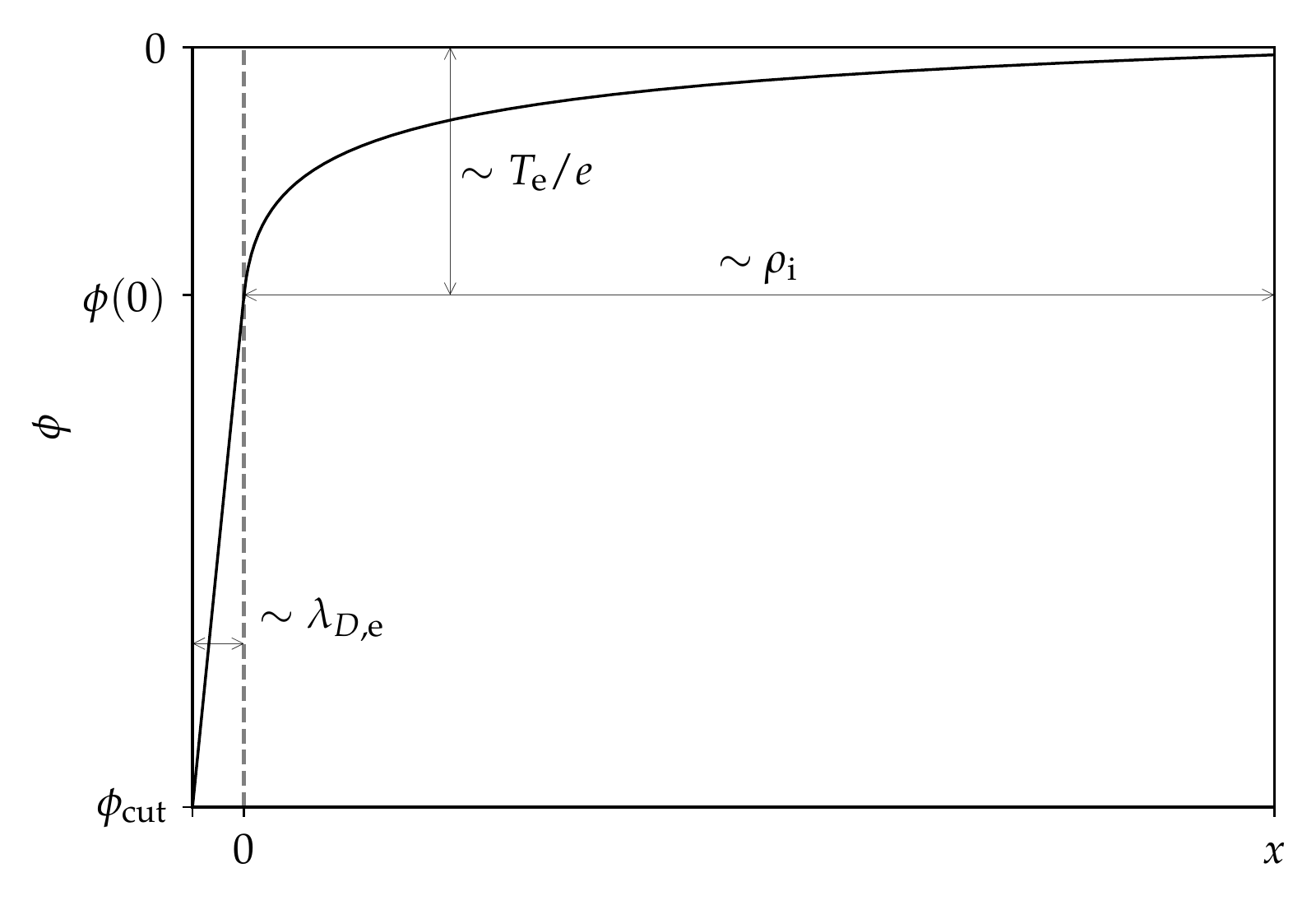}
\caption{A cartoon of the potential profile within the magnetic presheath and the Debye sheath.}
\label{fig:cartoon}
\end{figure}
At the boundary between plasmas and absorbing surfaces, non-neutral layers, called sheaths, are known to form \cite{Riemann-review, Hershkowitz-2005, Baalrud-2019-review}. The nature of these sheaths is controlled by the difference between ion and electron mobility and the physics of their interaction with the wall. Plasma wall interaction is of great importance in many devices such as spacecrafts \cite{Hastings-1995}, Hall thrusters \cite{Boeuf-2017}, plasma probes \cite{Hutchinson-2008, Hutchinson-book} and magnetic filters \cite{Anders-1995-filters}. It is of particular importance for magnetic confinement fusion, where fluxes of electrons and ions from the plasma are directed by magnetic fields onto the divertor or limiter targets, which can lead to overheating and damage \cite{Stangeby-book}. In order to reduce damage to solid targets, a common method is to incline the magnetic field at a small angle,~$\alpha$, relative to the wall, thus spreading heat flux over a larger area\cite{Loarte-2007}. 

In this paper we consider the effect of magnetisation on the plasma-wall interaction. We focus on the magnetic presheath, a quasineutral layer with a size on the order of the ion gyroradius that appears on top of the typically much thinner Debye sheath in magnetised plasmas \cite{Chodura-1982}. Within the magnetic presheath, we assume that the magnetic field is at a small angle $\alpha$ to the plane parallel to the wall of the device,
\begin{equation}
\textbf{B} = -\sin\alpha \hat{\textbf{e}}_{x} + \cos\alpha \hat{\textbf{e}}_{z}.
\end{equation}
Here we are using a set of orthonormal coordinates,~$\lbrace\hat{\textbf{e}}_{x} , \hat{\textbf{e}}_{y}, \hat{\textbf{e}}_{z} \rbrace$, such that~$\hat{\textbf{e}}_{x}$ points in the direction perpendicular to the wall and~$\hat{\textbf{e}}_{z}$ points in the direction of the component of~$\textbf{B}$ parallel to the wall. The electric field is electrostatic and points solely in the~$x$ direction, such that
\begin{equation}
\textbf{E} = -\nabla \phi = -\frac{\mathrm{d}\phi(x)}{\mathrm{d}x}\hat{\textbf{e}}_{x},
\end{equation}
where~$\phi$ is the electrostatic potential. There is assumed to be no dependence on the~$y$ or~$z$ coordinates and no time dependence. 

The first treatment of the magnetic presheath was due to Chodura \cite{Chodura-1982,Riemann-1994}. This initial study used fluid equations to describe the electrons and ions, complemented by numerical simulations with a ``particle-in-cell" kinetic model. Chodura found a solution to the potential throughout the presheath and a condition on the entrance ion distribution function, known as the Chodura condition, for the existence of this solution. The Chodura condition specifies that for cold ions, the fluid speed of ions entering the presheath along the magnetic field lines must be at least equal to the Bohm speed~$v_{B}$,
\begin{equation}
v_{B} = \sqrt{\frac{ZT_{\ee}}{m_{\ii}}},
\end{equation}
where~$Z$ is the ion charge number,~$T_{\ee}$ is the electron temperature and~$m_{\ii}$ the ion mass (we call this speed the Bohm speed to distinguish it from the sound speed~$c_{s} = \sqrt{(ZT_{\ee} + \gamma T_{\ii})/m_{\ii}}$; where the numerical factor~$\gamma$ depends on the particular ion physics). Between the magnetic presheath and the wall lies the Debye sheath. The Debye sheath has a typical thickness of several Debye lengths and, unlike the magnetic presheath, is not expected to be quasineutral. Analogous to (and predating) the Chodura condition, the Debye sheath has a `Bohm condition' \cite{Bohm-1949}, which states that ions entering the Debye sheath must be moving at least at the Bohm speed towards the wall (in contrast to along the field line).

Subsequent fluid treatments of the magnetic presheath have built upon the work of Chodura. Fluid treatments are, however, only well motivated if the ion temperature~$T_{\ii}$ is much less than the electron temperature~$T_{\ee}$ \cite{Geraldini-2017,Geraldini-2018,Geraldini-2019}. For ion temperatures comparable to the electron temperature, the strong distortion of the ion orbits means that an accurate description of the the presheath requires a kinetic treatment. The difficulty of kinetic theory and the complication of an absorbing boundary has meant that most approaches have been numerical, with further particle-in-cell simulations and Eulerian-Vlasov codes \cite{Tskhakaya-2003,Tskhakaya-2004,Tskhakaya-2017,Coulette-2014,Coulette-Manfredi-2016}, although some analytical progress has been made \cite{Daybelge-Bein-1981,Cohen-Ryutov-1998,Claassen-Gerhauser-1996}. This paper will build on the recent work of Geraldini et al \cite{Geraldini-2017,Geraldini-2018,Geraldini-2019}, whose kinetic model of the ions takes into account not only the distortion of the ion orbits, but also the contribution of ions that have recently broken free from their orbits due to the strong electric field near the Debye sheath. The electron density assumed by Geraldini et al, and a swathe of the literature, is Boltzmann distributed. This model is, in general, only valid in the limit of a very large potential drop in the Debye sheath, when most electrons are reflected. A common model to give the electrons a net flux is to truncate the entrance electron distribution in the velocity component along the magnetic field line \cite{Ingold-1972, Loizu-2011, Kawamura-2007}. The truncation velocity is determined by the wall potential because electrons considered energetic enough to reach the wall are assumed to be absorbed instead of reflected. In this paper, we will provide a drift kinetic derivation for such a model. We will treat the wall potential as a parameter, denoted~$\phi_{\mathrm{cut}}$, referred to as the `cutoff potential' or simply the `cutoff'. It is assumed that the potential and its first two derivatives are monotonic converging to zero as~$x \to\infty$ throughout this paper. Therefore, the imposed cutoff potential also sets a lower bound for the potential throughout the magnetic presheath. However, this lower bound is not usually reached within the presheath. Indeed, when the potential in the presheath does not reach the cutoff potential, this implies the remainder of the potential drop occurs in the Debye sheath (see figure \ref{fig:cartoon}). 

The width of the presheath will be set by the distance at which ion orbits begin to contact the wall, which is of the order of the ion gyroradius,~$\rho_{\ii}$. The potential drop must be sufficient to repel a considerable fraction of the electrons. Hence, we arrive at the following ordering for the potential and its first derivative,
\begin{equation}
\label{eqn:ordering1}
\phi(x) \geq \phi_{\mathrm{cut}} \sim \frac{T_{\ee}}{\ee}
\end{equation}
and
\begin{equation}
\phi' \sim \frac{T_\ee}{\ee\rho_{\ii}},
\end{equation}
where $\phi'$ denotes the derivative of~$\phi$ with respect to~$x$. One could imagine having a dial on the wall of the device that controlled the cutoff potential. Since the typical velocity of electrons is ordered as
\begin{equation}
\lvert\textbf{v}_{\ee}\rvert \sim v_{\mathrm{t,e}} = \sqrt{\frac{2T_{\ee}}{m_{\ee}}},
\end{equation}
cutoffs corresponding to energies much greater than the electron thermal energy would result in only very fast electrons reaching the wall, of which there will be exponentially few. As such, it is cutoffs at ordering (\ref{eqn:ordering1}) that we expect to significantly alter the electron distribution function. 

The typical velocity of ions is ordered as
\begin{equation}
\lvert\textbf{v}_{\ii}\rvert \sim  \sqrt{\frac{ZT_{\ee} + T_{\ii}}{m_{\ii}}}.
\end{equation}
Throughout this article we will use an~`$\ii$' subscript to refer to quantities associated with the ions and an `$\ee$' subscript for the electrons. When we wish to refer to both electrons and ions we will use the subscript~`$\sss$' for species. 

Due to the Bohm and Chodura conditions, the fluid velocity of the ions in the~$x$ direction goes from~$\alpha v_{B}$ to~$v_{B}$: the potential drop accelerates the ions towards the wall. By conservation of particle flux, the density must therefore decrease. This allowed Chodura to arrive at his prediction for the ion density at the Debye sheath entrance of \cite{Chodura-1982}
\begin{equation}
\label{eqn:density drop guess}
n_{\ii}(0) \approx \alpha n_{\ii,\infty},
\end{equation}
where~$n_{\ii,\infty}$ is the ion density at~$x \to \infty$. Since the system is quasineutral and the electron density and the potential are approximately Boltzmann related~(${n_{\ee} \approx n_{\ee,\infty}\exp(e\phi/T_{\ee})}$), the corresponding potential drop in the presheath is
\begin{equation}
\label{eqn:chodura drop guess}
\frac{e\phi(0)}{T_{\ee}} \approx \ln\alpha.
\end{equation}
Typical estimates of the total potential drop~($\phi_{\cc}$ in our notation) over the presheath and Debye sheath come from considering the case of an electrically floating wall accepting equal and opposite electron and ion currents. The typical scaling of such cutoff potentials was found to be \cite{Chodura-1982,Stangeby-book,Stangeby-2012}
\begin{equation}
\label{eqn:total drop guess}
\frac{e\phi_{\cc}}{T_{\ee}} = \frac{1}{2}\ln\left(\frac{m_{\ee}}{m_{\ii}}g\left(\frac{T_{\ii}}{T_{\ee}}\right)\right).
\end{equation}
Here~$g$ is a function of the ion to electron temperature ratio~$T_{\ii}/T_{\ee}$ which is independent of~$\alpha$. The important feature of this scaling for the total potential drop is its dependence on mass ratio and its lack of~$\alpha$ dependence. Stangeby \cite{Stangeby-2012} noted that, for sufficiently shallow magnetic field angle, the potential drop in the presheath predicted by Chodura could equal or exceed the total potential drop in the combined presheath and Debye sheath. This would imply that there is a magnetic field angle at which the Debye sheath would collapse. By direct comparison of equations (\ref{eqn:chodura drop guess}) and (\ref{eqn:total drop guess}), Stangeby predicted sheath collapse at an angle that scales as the square root of mass ratio,
\begin{equation}
\label{eqn:alpha ordering large}
\alpha = O\left(\sqrt{\frac{m_{\ee}}{m_{\ii}}} \right),
\end{equation}
and had value~$3.35\degree$ for a deuterium plasma.

The assumptions that led to this ordering for sheath collapse are, however, incorrect as they fail to account for the alteration to Chodura's fluid model due to the loss of electrons onto the wall and the resulting breakdown of the assumption of Boltzmann electrons. We will prove that, in fact, there is no Debye sheath collapse for the fluid model. Stangeby's central idea that an increasing potential drop across the presheath can cause the Debye sheath to collapse is nonetheless correct, but in fact this collapse owes its origins to finite ion orbit widths and is therefore an intrinsically kinetic effect. Furthermore, we will show that it occurs at angles which scale linearly with mass ratio, much smaller than those predicted by Stangeby.  

\begin{figure}
\centering 
\includegraphics[width=0.75\linewidth]{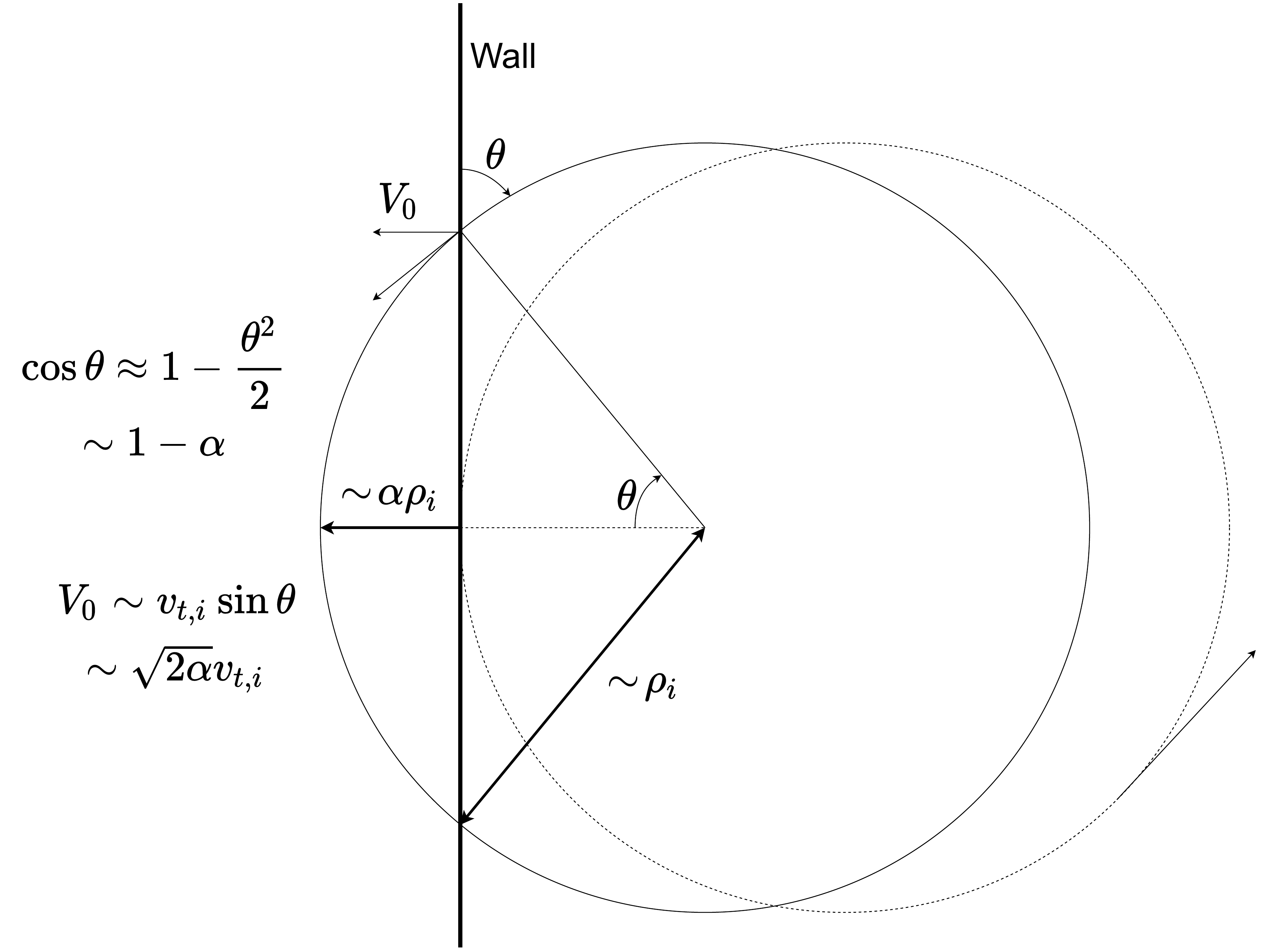}
\caption{A sketch of a circular ion orbit drifting towards the wall. The ions approaching the wall with the largest velocity perpendicular to the wall are those that just grazed the wall on the previous gyration (dashed line). In a time~$\sim 1/\Omega_{\ii}$ the ion orbit drifts a distance~$\alpha\rho_{\ii}$ towards the wall, causing these ions to make an angle of order~$\sqrt{\alpha}$ to the wall. The angles in this diagram have been exaggerated for clarity.}
\label{fig:cartoon2}
\end{figure}

The discrepancy can be understood by first noting that the approximation of Boltzmann electrons breaks down for potentials~$\phi$ near the cutoff potential~$\phi_{\cc}$. At such potentials, the electron that just grazed the wall and was reflected has a small parallel velocity
\begin{equation}
v_{\parallel,\cc} \approx \sqrt{\frac{2\Omega_{\ee}}{B}(\phi-\phi_{\cc})},
\end{equation}
and a small change in the potential therefore results in a large change in this cutoff velocity of returning electrons. This makes the derivative of electron density with respect to potential diverge as one approaches the wall potential since
\begin{equation}
\dev{n_{\ee}}{\phi}{} = \dev{}{\phi}{}\int_{-v_{\parallel,\cc}}^{\infty}f_{\ee}(v_{\parallel})\mathrm{d}v_{\parallel} \approx f_{\ee}(-v_{\parallel,\cc})\dev{v_{\parallel,\cc}}{\phi}{} \to \infty.
\end{equation}
This divergence occurs in the presheath precisely when the Debye sheath has collapsed and we will show that this implies the electric field does not require a singularity on the presheath scale to satisfy quasineutrality. As a result, equation~(\ref{eqn:density drop guess}) for the drop in density in the presheath is incorrect exactly in the case when the Debye sheath vanishes. Without the singularity, the electric force cannot overcome the magnetic force for ions with a large enough perpendicular velocity. Therefore, a fraction of ions reach the wall travelling almost tangentially to it, much like the circular orbit shown in figure~\ref{fig:cartoon2}. This gives rise to a density at the wall (or equivalently the Debye sheath entrance) that scales like~$\alpha^{1/2}$, instead of like~$\alpha$ \cite{Cohen-Ryutov-1998,Geraldini-2018}. To see this, we appeal to the geometrical picture of a circular orbit near the wall shown in figure \ref{fig:cartoon2}. Because the ions are on circular orbits, the ion reaching the wall with the largest component of velocity perpendicular to the wall is the ion that just grazed it on its previous cycle. If such an ion is moving at typical thermal velocities along the magnetic field line, in the time it takes for the ion to make a single gyration, it will have drifted closer to the wall by a distance of order $\alpha\rho_{\ii}$. The ion velocity then makes an angle~$\theta$ to the wall, shown in figure~\ref{fig:cartoon2}, which we argue scales as~$\sqrt{\alpha}$. This means the velocity onto the wall scales as~$\sqrt{\alpha}v_{\mathrm{t,i}}$. Since the ion flux at infinity scales like~$\alpha v_{B}n_{\ii,\infty}$, by particle flux conservation, the density at the wall must scale like~$\sqrt{\alpha}n_{\ii,\infty}$. Therefore, the ordering (\ref{eqn:density drop guess}) for the ion density at the Debye sheath entrance used to recover the ordering (\ref{eqn:alpha ordering large}) for the angle at which the Debye sheath vanishes is incorrect precisely when the Debye sheath vanishes, if finite ion orbit width is included. We will prove this scaling algebraically in section \ref{section:ion model}, but this simplified picture serves to gain an estimate and also shows why this could not be predicted by a fluid model.

Equipped with this estimate for the ion density, we see that the new estimate for the potential drop in the presheath is 
\begin{equation}
\label{eqn:presheath drop guess 2}
\frac{e\phi(0)}{T_{\ee}} \approx \frac{1}{2}\ln\alpha .
\end{equation}
Direct comparison between equations~(\ref{eqn:presheath drop guess 2}) and (\ref{eqn:total drop guess}) now implies that the Debye sheath will vanish at an angle $\alpha$ that scales with mass ratio rather than square root mass ratio. This motivates a different, smaller, ordering for~$\alpha$, given by
\begin{equation}
\label{eqn:alpha ordering small}
\alpha = O\left(\frac{m_{\ee}}{m_{\ii}} \right).
\end{equation}

The rest of the paper is structured as follows. In section \ref{section:Trajectories within the presheath} we will discuss the single particle motion of electrons and ions within the presheath, showing both to be quasi-periodic. We will build on this with the electron model in section \ref{section:Electron model}, detailing the simple model of the electron orbits within the presheath which we will use to find the electron density in section \ref{section:The electron density}. We give a more rigorous derivation of this electron model from drift kinetics in \ref{section:Gyrokinetic electrons} which will prove the validity of our model for the small values of~$\alpha$ required to see the collapse of the Debye sheath. For the ion density, we review the ion model developed in \cite{Geraldini-2017}\cite{Geraldini-2018} in section \ref{section:ion model}. We will use this model to derive the equation for the ion density in section \ref{section:ion density} which we will show can give rise to the~$\sqrt{\alpha}n_{\ii,\infty}$ we predict, provided the ion orbits are sufficiently undistorted. In section \ref{section:Quasineutrality} we will expand the quasineutrality equation close to and far from the wall analytically. Far from the wall, this will give rise to the modified kinetic Chodura condition, while close to the wall we will prove that, for the majority of the solutions, the presheath will satisfy a marginal kinetic Bohm condition. The truncation of the electron distribution will lead to a subsonic Bohm condition, which agrees qualitatively with the predictions of \cite{Loizu-2012}. In order to gain insight into these specific solutions, we will appeal to the limits of hot and cold ion temperatures in section \ref{section:Limits of hot and cold ions}. We will see that there exist cutoff potentials that do not admit solutions, and that this lack of solutions is tied to finite orbit effects. These effects manifest themselves most obviously in the hot ion limit. We will show they are not present in the limit of cold ions. In section~\ref{section:Numerical method} we will detail how our solutions are reached numerically and where this procedure differs from \cite{Geraldini-2018}. In section \ref{section:Results} we will present our numerical solutions to the presheath for a range of temperature ratios, angles~$\alpha$, and cutoff potentials. We will show that there are indeed ranges of cutoffs for which no solution exists. Furthermore, we will show that this lack of solutions implies that there are angles of the magnetic field for which the current from the plasma cannot be zero unless the electric field of the Debye sheath reverses. These angles are of order~$m_{\ee}/m_{\ii}$.

\section{Trajectories within the presheath and the electron model}
\label{section:Trajectories within the presheath}
In this section we will discuss the model for the electron orbits within the presheath. This will begin with a general discussion for both ions and electrons of the constants of the motion at~$\alpha=0$, and a justification of treating the electron and ion motion as periodic when~$\alpha \neq 0$. In subsection \ref{section:Electron model} we will use this insight to discuss the relevant variables for the electron model. Finally in subsection \ref{section:The electron density} we will use this model to calculate the electron density as a function of potential. We will see that this may deviate strongly from the Boltzmann response near the Debye sheath entrance.
\subsection{The $\alpha=0$ problem and quasi-periodicity}
\label{section:zeroth order problem}
The equations of motion for ions and electrons in the presheath are
\begin{equation}
\label{eqn:vxdot}
\dot{v}_{x} = -\mathrm{sgn}(Z_{\sss})\frac{\Omega_{\sss}}{B}\phi' + \mathrm{sgn}(Z_{\sss})\Omega_{\sss} v_{y}\cos\alpha ,
\end{equation}
\begin{equation}
\label{eqn:vydot}
\dot{v}_{y} = -\mathrm{sgn}(Z_{\sss})\Omega_{\sss} v_{x}\cos\alpha - \mathrm{sgn}(Z_{\sss})\Omega_{\sss} v_{z}\sin\alpha,
\end{equation}
and
\begin{equation}
\label{eqn:vzdot}
\dot{v}_{z} = \mathrm{sgn}(Z_{\sss})\Omega_{\sss} v_{y}\sin\alpha.
\end{equation}
Here~$\Omega_{\sss} = |Z_{\sss}| eB/ m_{\sss}$ is the gyrofrequency of a particular species. The factors of~$\mathrm{sgn}(Z_{\sss})$ have appeared so we may refer to~$\Omega_{\sss}$ as a positive quantity for both electrons and ions.
  
The trajectories defined by equations~(\ref{eqn:vxdot}-\ref{eqn:vzdot}), as in \cite{Geraldini-2018,Geraldini-2017}, can be solved for~$\alpha = 0$, giving the constants of the motions~$U_{\perp}$,~$U$ and~$\bar{x}$ defined by
\begin{equation}
\label{eqn:xbar}
\bar{x} = x + \mathrm{sgn}(Z_{\sss})\frac{v_{y}}{\Omega_{\sss}},
\end{equation} 
\begin{equation}
\label{eqn:Uperp}
U_{\perp} = \frac{1}{2}v_{x}^{2} + \frac{1}{2}v_{y}^{2}+ \mathrm{sgn}(Z_{\sss})\frac{\Omega_{\sss}\phi}{B},
\end{equation}
and
\begin{equation}
\label{eqn:U}
U = \frac{1}{2}v_{x}^{2} + \frac{1}{2}v_{y}^{2} + \frac{1}{2}v_{z}^{2} + \mathrm{sgn}(Z_{\sss})\frac{\Omega_{\sss} \phi}{B}.
\end{equation}
Here~$\bar{x}$ is a constant of the motion that will later be useful for characterising orbits, and~$U_{\perp}$ and~$U$ are the perpendicular and total energy, respectively. For~$\alpha \neq 0$, $\bar{x}$ and~$U_{\perp}$ are no longer constants of the motion. In this case we take equations~(\ref{eqn:xbar}) and~(\ref{eqn:Uperp}) to be the definitions of these quantities, with the understanding that they are no longer constants of the motion. Indeed, we may now use our equations of motion to determine the rates of change of~$\bar{x}$ and~$U_{\perp}$,
\begin{equation}
\label{eqn:xbar dot}
\dot{\bar{x}} = -v_{z}\alpha + O(v_{\mathrm{t,s}}\alpha^{2}),
\end{equation}
and
\begin{equation}
\label{eqn:Uperp dot estimate}
\dot{U}_{\perp} = -\mathrm{sgn}(Z_{\sss})\Omega_{\sss} v_{z}v_{y}\alpha + O(v_{\mathrm{t,s}}^{2}\alpha^{3}).
\end{equation}
Hence, while~$\bar{x}$ and~$U_{\perp}$ are no longer constants of the motion for~$\alpha \neq 0$, they are still useful as they vary slowly in~$\alpha$. 

We may use our definition (\ref{eqn:xbar}) to rewrite~$U_{\perp}$ as
\begin{dmath}
\label{eqn:Uperp with effective potential}
U_{\perp} = \frac{1}{2}v_{x}^{2} + \frac{1}{2}\Omega_{\sss}^{2}\left(\bar{x} - x \right)^2 + \mathrm{sgn}(Z_{s})\frac{\Omega_{\sss} \phi}{B} = \frac{1}{2}v_{x}^{2} + \chi(x, \bar{x}).
\end{dmath}
Here we have defined the effective potential~$\chi(x,\bar{x})$ (which will be most relevant for the ion model) as
\begin{equation}
\label{eqn:effective potential definition}
\chi(x, \bar{x}) = \frac{1}{2}\Omega_{\sss}^{2}(\bar{x}-x)^{2} + \mathrm{sgn}(Z_{\sss})\frac{\Omega_{\sss}}{B}\phi.
\end{equation}
When~$\alpha = 0$, $U_{\perp}$ and~$\bar{x}$ are constants and thus the motion is truly periodic in~$x$ for certain values of~$U_{\perp}$ and~$\bar{x}$. Equation (\ref{eqn:xbar}) implies that the range of~$x$ sampled in one period will be~$\sim \rho_{\sss} = v_{\mathrm{t,s}}/\Omega_{\sss}$ and thus the period of the motion will be~$\sim 1/\Omega_{\sss}$ by equation~(\ref{eqn:Uperp with effective potential}). When~$\alpha \neq 0$, the change in~$U_{\perp}$ over a time~$1/\Omega_{\sss}$ will be a factor of~$\alpha$ smaller than the value of~$U_{\perp}$ by equation~(\ref{eqn:Uperp dot estimate}). Hence the motion is no longer truly periodic, but we will refer to it as `quasi-periodic' since the particles are moving in a potential well that is changing slowly compared to the time taken to cross the well. 

While we have thus far used~$U_{\perp}$ to discuss the quasi-periodic motion of the electrons and ions, it is possible for this quasiperiodic motion to have another approximate conserved quantity. Such a variable is known as an adiabatic invariant. We will give an expression for the adiabatic invariant for the electrons,~$\mu$,  in the next section and adiabatic invariant for the ions,~$\mugk$, in section~\ref{section:ion density}. 

\subsection{The electron model}
\label{section:Electron model}
As we have motivated in section \ref{section:zeroth order problem}, the motion of both the electron and ion orbits is split into two timescales: one rapid oscillatory timescale and another, much longer, timescale over which the parameters of the orbits can change. With the electrons the case is further simplified by the small orbits of size~$\sim \rho_{\ee}$. These orbits, unlike the ions, sample only a small change in the electric field over one period and as such are only weakly deformed by the potential. This allows us to give an anayltical expression for another variable, the adiabatic invariant, which is approximately constant over the electron's motion. This is~$\mu$ given by
\begin{equation}
\label{eqn:mu definition main text}
\mu = \frac{1}{2\Omega_{\ee}}\left( \lvert \vec{v} - \vec{v_{E}}(x)  \rvert^{2} - \lvert \vec{v}\cdot\vec{b} \rvert^{2} \right),
\end{equation}
where~$\vec{b} = \vec{B}/|\vec{B}|$ is the normalised magnetic field and~$\vec{v_{E}}$ is the~$\vec{E}\times\vec{B}$ drift velocity given by
\begin{equation}
\label{eqn:E cross B drift main text}
\vec{v_{E}} = \frac{1}{\lvert \vec{B} \rvert}\vec{E}\times \hat{\vec{b}}.
\end{equation}
Together with~$U$, the total energy of the electron, we will use this to specify a given electrons velocity. Of course, we still need one variable that will capture the rapid variation of the electron motion. This is given by the gyrophase,~$\varphi$, defined by
\begin{equation}
\label{eqn:varphi definition main text}
\varphi = -\tan^{-1} \left( \frac{(\vec{v} - \vec{v_{E}}(x))\cdot \hat{\vec{e}}_{y}}{(\vec{v} - \vec{v_{E}}(x))\cdot \hat{\vec{e}}_{n}} \right).
\end{equation}
Here $\hat{\vec{e}}_{n} = \cos\alpha\, \hat{\vec{e}}_{x} + \sin\alpha \,\hat{\vec{e}}_{z}$ denotes the unit vector perpendicular to both the magnetic field and the $y$ direction. 

Assuming that the electron distribution function at each point~$x$ is independent of the rapidly varying gyrophase, we can cast the electron motion in terms of only the constants of the motion~$U$ and~$\mu$, $f_{\ee}(x,\vec{v})\approx f_{\ee}(U\mu)$. Therefore, in these variables the electron distribution will be constant along the trajectory of the electrons and to determine the electron density at a given position one only has to ask what values of~$U$ and~$\mu$ could still be present in this position. This will be done in the next section, where we will derive the electron density as a function of position.

In fact, the adiabatic invariant for the electrons is only approximately constant, as is the independence of the electron distribution function on gyrophase. Therefore, while the simple discussion above suffices to fully describe the electron model, it is unclear how accurate it is. Since we wish to consider the small angle ordering~(\ref{eqn:alpha ordering small}) as well as~(\ref{eqn:alpha ordering large}), it is necessary to ensure that the electron model is still valid where the electron and ion density are small in~$\alpha$. This is done in~\ref{section:Gyrokinetic electrons},~\ref{section:Gyroaverages} and~\ref{section:first order correction} using a drift kinetic derivation for the evolution of the electron distribution function. It will turn out that, despite various complications, the errors are of sufficiently small magnitude to be ignored in both ordering~(\ref{eqn:alpha ordering large}) and ordering~(\ref{eqn:alpha ordering small}).

\subsection{The electron density}
\label{section:The electron density}
The electron density at any position~$x$ is given by
\begin{equation}
n_{\ee}(x) = \iiint f_{e}(x,v_{x},v_{y},v_{z})\mathrm{d}v_{x}\mathrm{d}v_{y}\mathrm{d}v_{z},
\end{equation}
over suitable limits. The aim now is to change variables to those given in section~\ref{section:Electron model}. The Jacobian for the coordinate transformation to~${\lbrace U,\mu,\varphi \rbrace}$ is given by
\begin{equation}
\left\vert \frac{\partial(v_{x},v_{y},v_{z})}{\partial(U,\mu,\varphi)} \right\vert = \frac{\Omega_{\ee}}{|v_{\parallel}|}.
\end{equation}
So we have
\begin{dmath}
n_{\ee}(x) = \sum_{\sigma_{\parallel} = \pm 1}\int\mathrm{d}\varphi\int\Omega_{\ee}\mathrm{d}\mu\int\frac{\gyg{f_{\ee}}}{|v_{\parallel}|}\mathrm{d}U + O\left(\frac{m_{\ee}}{m_{\ii}}n_{\ee} \right),
\end{dmath}
where the sum is over electrons moving into~($\sigma_{\parallel} = 1$) and out of~($\sigma_{\parallel} = -1$) the presheath. Here, we have exploited the decomposition of~$f_{\ee}$ in equation~(\ref{eqn:Gryoaverage decomposition}) into the part independent of gyrophase,~$\gyg{f_{\ee}}$, and a correction which we find to be small in~\ref{section:Gyrokinetic electrons}. When written in the variables~$U$ and~$\mu$,~$\gyg{f_{e}}$ is independent of~$\bar{x}$ (the distinction between~$\bar{x}$ and~$x$ will turn out to be unimportant for the electrons) and as such we may use the distribution function evaluated at infinity. All that remains is to determine for which values of~$U$ and~$\mu$ a given electron will be present at position~$x$.

The effect of the potential drop within the presheath is to decelerate electrons in the direction parallel to the magnetic field. In~\ref{section:Gyroaverages}, we write the total energy~$U$ as 
\begin{equation}
U = \Omega_{\ee}\mu - \frac{\Omega_{\ee}}{B}\phi(x) + \frac{1}{2}v_{\parallel}^{2} + v_{y}v_{E} + O\left(v_{\mathrm{t,e}}^{2}\frac{m_{\ee}}{m_{\ii}} \right).
\end{equation}
Since the total energy and the adiabatic invariant are constant within the presheath, the parallel velocity must decrease to compensate for the potential drop (the~$\vec{E}\times\vec{B}$ drift will not compensate for the potential drop as it is smaller than the electron thermal speed by a factor of~$\sqrt{m_{\ee}/m_{\ii}}$). Thus, an electron that enters the presheath at a finite velocity may at some point be reflected. As such, the minimum value of the total energy for a particle at position~$x$ with adiabatic invariant~$\mu$ and gyrophase~$\varphi$ corresponds to a particle that has zero parallel velocity. Thus,
\begin{equation}
\label{eqn:U min returning}
U_{\mathrm{min}} = \Omega_{\ee}\mu - \frac{\Omega_{\ee}}{B}\phi(x) + v_{y}v_{E} + O\left(v_{\mathrm{t,e}}^{2}\frac{m_{\ee}}{m_{\ii}} \right).
\end{equation}
Note that the term~$v_{y}v_{E}$ in equation~(\ref{eqn:U min returning}) makes~$U_{\min}$ a function the gyrophase~$\varphi$. However, this correction is small (of size~$\sqrt{m_{\ee}/m_{\ii}}$), and when integrated over~$\varphi$ will be of size~$m_{\ee}/m_{\ii}$, so will turn out to be unimportant. The maximum possible parallel velocity for an electron entering the presheath is infinite since particles can enter the presheath at any speed. For electrons exiting the presheath, however, a different consideration must be made. We assume that an electron that reaches the wall of the device is absorbed onto the wall and does not return to the presheath. Then, if an electron is leaving the presheath (i.e. has negative parallel velocity), it must have been reflected before reaching the wall. There must be a maximum energy of these particles leaving the presheath that depends on the adiabatic invariant, which we denote by~$U_{\max}(\mu)$. To determine the value of~$U_{\max}(\mu)$, we must consider which particles are lost to the wall within the Debye sheath. Energy is conserved for particles entering the Debye sheath. We may parametrise~$U_{\mathrm{max}}$ by the function~$\Phi(\mu,\delphis)$ (where we have denoted the potential drop in the Debye sheath,~$\phi(0)- \phi_{\cc}$, by $\delphis$), declaring
\begin{equation}
\label{eqn:U max returning}
U_{\mathrm{max}}(\mu) = \Omega_{\ee}\mu - \frac{\Omega_{\ee}}{B}\phi_{\cc}  - \frac{\Omega_{\ee}}{B}\Phi(\mu,\delphis),
\end{equation} 
where the choice to isolate the~$\Omega_{\ee}\mu$ is arbitrary, but driven by the form of our expressions for~$U$, which in~\ref{section:Gyroaverages} we write as
\begin{equation}
\label{eqn:U as function of xbar to argue about Phi}
U = \Omega_{\ee}\mu + \frac{v_{\parallel}^{2}}{2} - \frac{\Omega_{\ee}}{B}\phi(\bar{x}),
\end{equation}
with errors small by mass ratio (note that~$\phi$ is evaluated at~$\bar{x}$ and not~$x$). In general the physics of absorption onto the wall of a plasma is complicated by the fact that the length scale of the Debye sheath, which is a few Debye lengths~$\lambda_{\mathrm{D}}$, may be of the order of the electron gyroradius, which would lead to a complex form of~$\Phi(\mu,\delphis)$, depending implicitly on the form of the electron and ion distribution functions at the Debye sheath entrance as well as on~$\mu$ and~$\delphis$.

Inserting the relevant limits on our variables, we arrive at the following expression for the electron density at position~$x$
\begin{dmath}
\label{eqn:electron density non-local}
n_{\ee}(x) = \int_{0}^{2\pi}\mathrm{d}\varphi\int_{0}^{\infty}\Omega_{\ee}\mathrm{d}\mu\left[\int_{\Omega_{\ee}\mu - \frac{\Omega_{\ee}}{B}\phi(x) + v_{y}v_{E}}^{\infty}\frac{\gyg{f_{e}}(U,\mu)\mathrm{d}U}{\sqrt{2\left(U-\Omega_{\ee}\mu + \frac{\Omega_{\ee}}{B}\phi(x) - v_{y}v_{E} \right)}} + \int_{\Omega_{\ee}\mu - \frac{\Omega_{\ee}}{B}\phi(x) + v_{y}v_{E}}^{\Omega_{\ee}\mu -\frac{\Omega_{\ee}}{B}\phi_{\cc} -  \frac{\Omega_{\ee}}{B}\Phi(\mu,\delphis)}\frac{\gyg{f_{e}}(U,\mu)\mathrm{d}U}{\sqrt{2\left(U-\Omega_{\ee}\mu + \frac{\Omega_{\ee}}{B}\phi(x) - v_{y}v_{E} \right)}}\right].
\end{dmath}
Here, the integral over~$U$ has been explicitly split into the incoming and outgoing particles, contained in the first and second terms, respectively. The factors of~$v_{y}v_{E}$ in equation (\ref{eqn:electron density non-local}) are due to the fact that we aim to evaluate the potential at a given position~$x$ rather than at~$\bar{x}$. However, these factors are small by~$\sqrt{m_{\ee}/m_{\ii}}$ and the integral can be Taylor expanded in them. After expanding, the lowest order correction due to~$v_{y}v_{E}$ vanishes when integrating over~$\varphi$ leaving a higher order correction of size~$m_{\ee}/m_{\ii}$, which we neglect. We therefore arrive at an expression for the electron density that depends only on the input distribution function, the local potential, and the cutoff potential
\begin{dmath}
\label{eqn:electron density local}
n_{\ee}(x) = 2\pi\int_{0}^{\infty}\Omega_{\ee}\mathrm{d}\mu\left[\int_{\Omega_{\ee}\mu - \frac{\Omega_{\ee}}{B}\phi(x)}^{\infty}\frac{\gyg{f_{e}}(U,\mu)\mathrm{d}U}{\sqrt{2\left(U-\Omega_{\ee}\mu + \frac{\Omega_{\ee}}{B}\phi(x)\right)}} + \int_{\Omega_{\ee}\mu - \frac{\Omega_{\ee}}{B}\phi(x)}^{\Omega_{\ee}\mu - \frac{\Omega_{\ee}}{B}\phi_{\cc} -\frac{\Omega_{\ee}}{B}\Phi(\mu,\delphis)}\frac{\gyg{f_{e}}(U,\mu)\mathrm{d}U}{\sqrt{2\left(U-\Omega_{\ee}\mu + \frac{\Omega_{\ee}}{B}\phi(x) \right)}}\right].
\end{dmath}
Important features of this density can be seen easily by making the change of variables from~$U$ to~$\bar{v}_{\parallel}$ defined by 
\begin{equation}
\label{eqn:v parallel bar definition}
\bar{v}_{\parallel} = \sqrt{2\left(U - \Omega_{\ee}\mu + \frac{\Omega_{\ee}}{B}\phi(x)\right)}.
\end{equation} 
This changes equation~(\ref{eqn:electron density local}) to 
\begin{dmath}
\label{eqn:electron density local with v parallel bar}
n_{\ee}(x) =  2\pi\int_{0}^{\infty}\Omega_{\ee}\mathrm{d}\mu\left[\int_{0}^{\infty}\gyg{f_{e}}\left(\Omega_{\ee}\mu + \frac{\bar{v}_{\parallel}^2}{2} - \frac{\Omega_{\ee}}{B}\phi(x),\mu \right)\mathrm{d}\bar{v}_{\parallel}+ \int_{0}^{\sqrt{2\frac{\Omega_{\ee}}{B}\left(\phi(x)- \phi_{\cc} - \Phi(\mu,\delphis) \right)}}\gyg{f_{e}}\left(\Omega_{\ee}\mu + \frac{\bar{v}_{\parallel}^2}{2} - \frac{\Omega_{\ee}}{B}\phi(x),\mu\right)\mathrm{d}\bar{v}_{\parallel}\right].
\end{dmath}
\begin{figure}[htbp]
    \begin{subfigure}[t]{0.5\textwidth}
    \includegraphics[width = 1\linewidth]{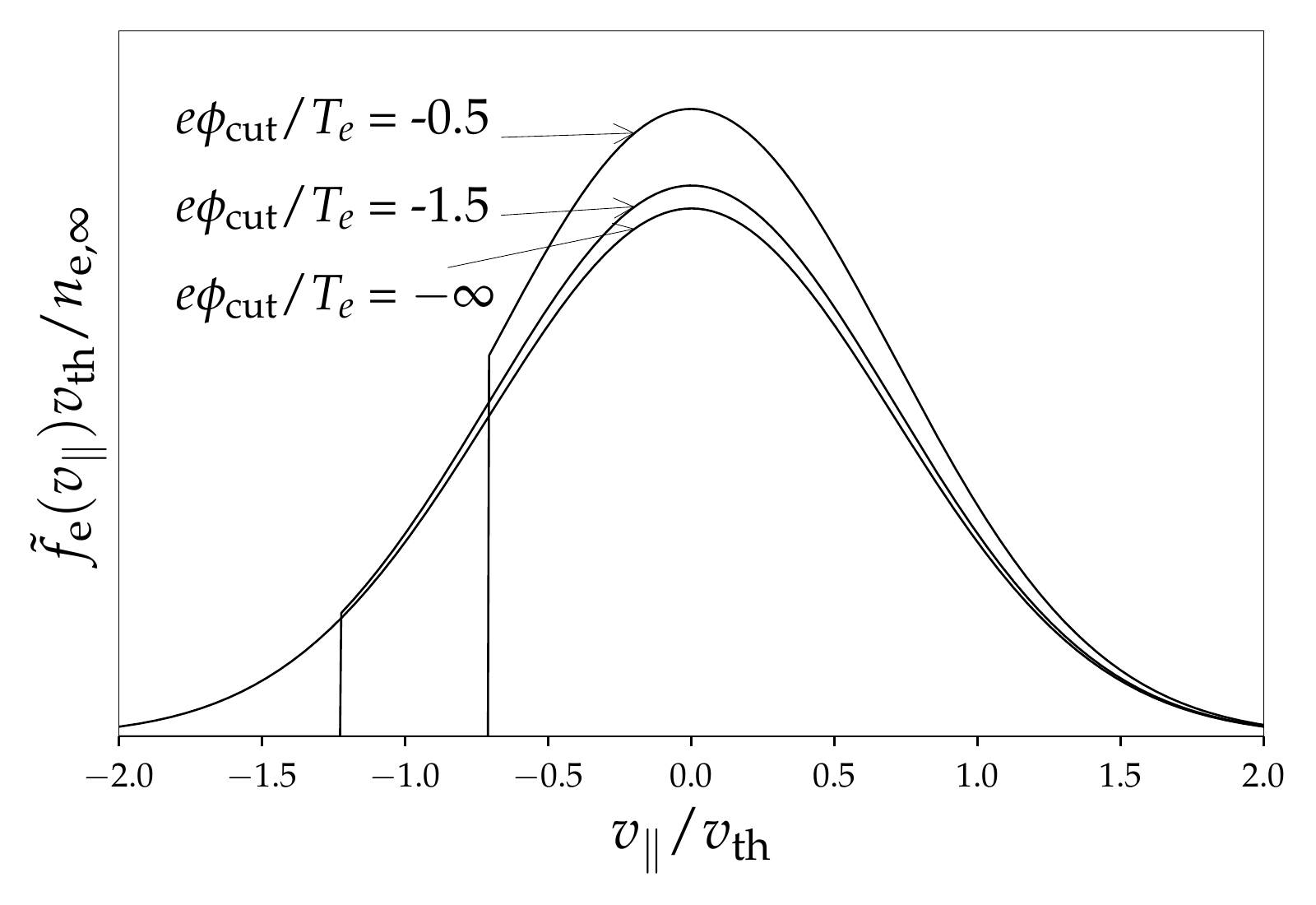}
    \caption{Example marginalised electron distribution function in $v_{\parallel}$}
    \end{subfigure}
    \begin{subfigure}[t]{0.5\textwidth}
    \includegraphics[width = 1\linewidth]{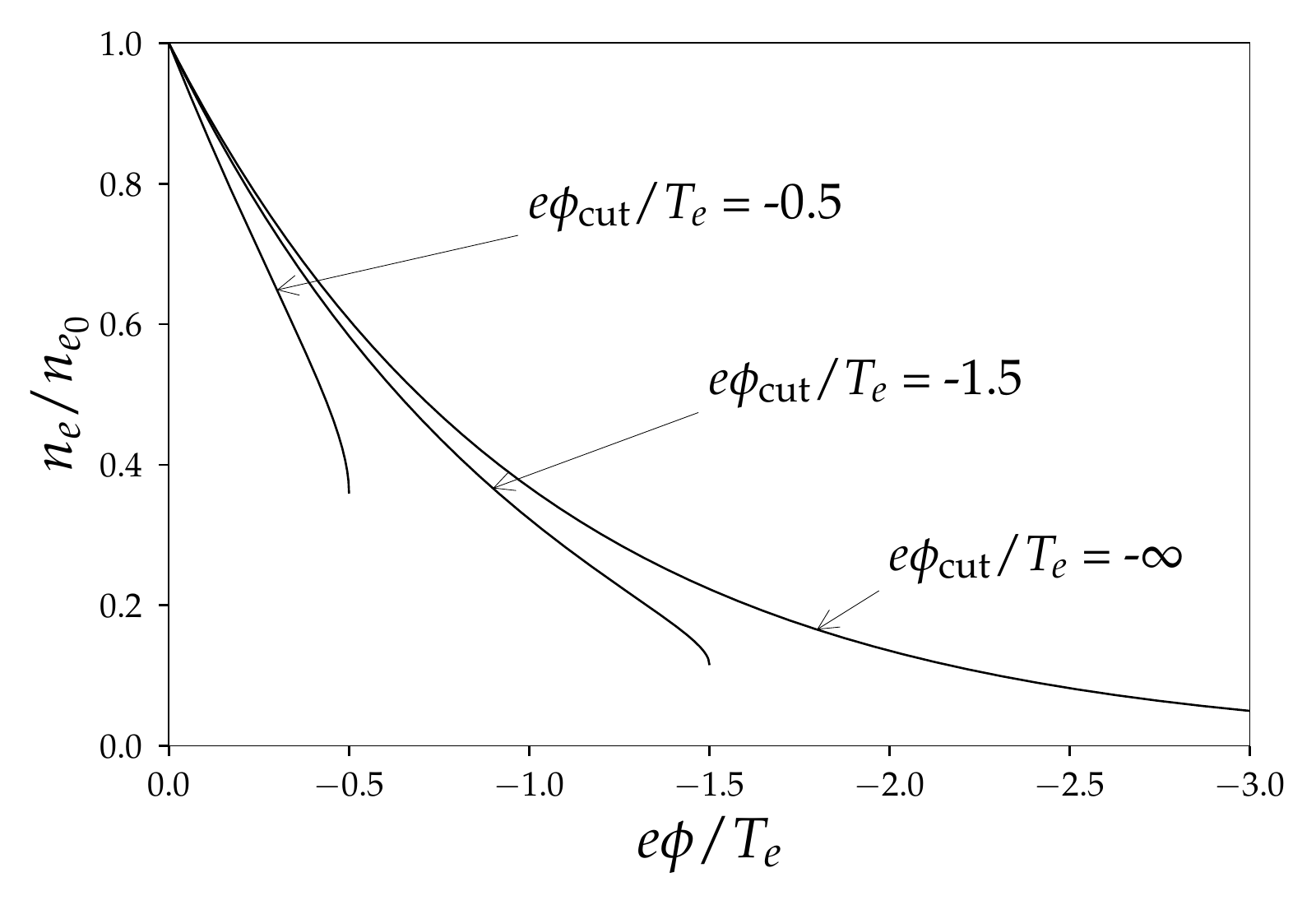}
    \caption{Electron density as a function of $\phi$}
    \end{subfigure}
    \caption{Here we see an example entrance electron distribution (the truncated Maxwellian) that has a discontinuity. Note that as the~$\phi_{\cc}$ is taken to infinity the distribution tends to a Maxwellian and the density profile to a Boltzmann distribution, as equation~(\ref{eqn:electron density local}) would imply.}
    \label{fig:density}
\end{figure}

In the limit~$\rho_{\ee}/\lambda_{\mathrm{D}} \ll 1$, we expect that the electron orbit width can be neglected. Neglecting the electron orbit width is equivalent to assuming that the electrons are absorbed when they reach a value of~$\bar{x}$ corresponding to the wall, such that~$\phi(\bar{x}) = \phi_{\cc}$. Appealing to equations~(\ref{eqn:U max returning}) and~(\ref{eqn:U as function of xbar to argue about Phi}), this corresponds to the assumption of~$\Phi(\mu,\delphis) = 0$. This justifies our convention on~$\Phi(\mu,\delphis)$ in equation~(\ref{eqn:U max returning}) as it isolates the potential required to reach the wall in this limit. When the electron gyroradius is comparable to the Debye length, the form of~$\Phi(\mu,\delphis)$ is unknown. As such, for the remainder of this paper we will only consider the limit where~$\rho_{\ee}/\lambda_{\mathrm{D}} \ll 1$, and~$\Phi = 0$. In this particular limit, it will prove very important to consider the derivative of the electron density with respect to the potential at the origin. This is given by
\begin{dmath}
\label{eqn:first derivative o density}
\left.\dev{n_{\ee}}{\phi}{}\right|_{x=0} = \frac{2\pi \Omega_{\ee}}{B} \int_{0}^{\infty}\Omega_{\ee}\mathrm{d}\mu\left[-\int_{0}^{\infty}\pdev{\gyg{f_{\ee}}}{U}\left( \Omega_{\ee}\mu +\frac{\bar{v}_{\parallel}^{2}}{2} - \frac{\Omega_{\ee}}{B}\phi(0),\mu\right)\mathrm{d}\bar{v}_{\parallel} -\int_{0}^{\sqrt{2\frac{\Omega_{\ee}}{B}(\phi(0)-\phi_{\cc})}}\pdev{\gyg{f_{\ee}}}{U}\left( \Omega_{\ee}\mu +\frac{\bar{v}_{\parallel}^{2}}{2} - \frac{\Omega_{\ee}}{B}\phi(0),\mu\right)\mathrm{d}\bar{v}_{\parallel} + \frac{1}{\sqrt{2\frac{\Omega_{\ee}}{B}(\phi(0)-\phi_{\cc})}}\gyg{f_{\ee}}\left(\Omega_{\ee}\mu - \frac{\Omega_{\ee}}{B}\phi_{\cc},\mu \right) \right].
\end{dmath}
It follows that the derivative of the electron density with respect to potential at~$x=0$ will diverge as~$\phi(0) \to \phi_{\cc}$. This can be seen in figure~\ref{fig:density}, where we plot~$n_{\ee}$ as a function of~$\phi$ with a truncated Maxwellian input electron distribution function. In section~\ref{section:Quasineutrality} we will prove this infinite derivative in~$n_{\ee}$ to be an important distinction from the Maxwell-Boltzmann response.

\section{The ion model and density}
\subsection{The ion model}
\label{section:ion model}
In this section we will review the ion model presented by \cite{Geraldini-2017,Geraldini-2018}. Like the electrons, the aim is to change to convenient variables to describe the ion orbits. However, unlike the electrons, these orbits will be strongly distorted by the electric field and will be classified into two groups: `closed orbits' where the ion will not contact the wall within the next period of the motion and `open orbits' where, at some point within the next period of the motion, the ion on this orbit will contact the wall. Both must be included to give a correct expression for the density of ions near the wall.

\subsubsection{Closed orbits.}
\begin{figure}
    \begin{subfigure}{0.5\textwidth}
    \includegraphics[width = 1\linewidth]{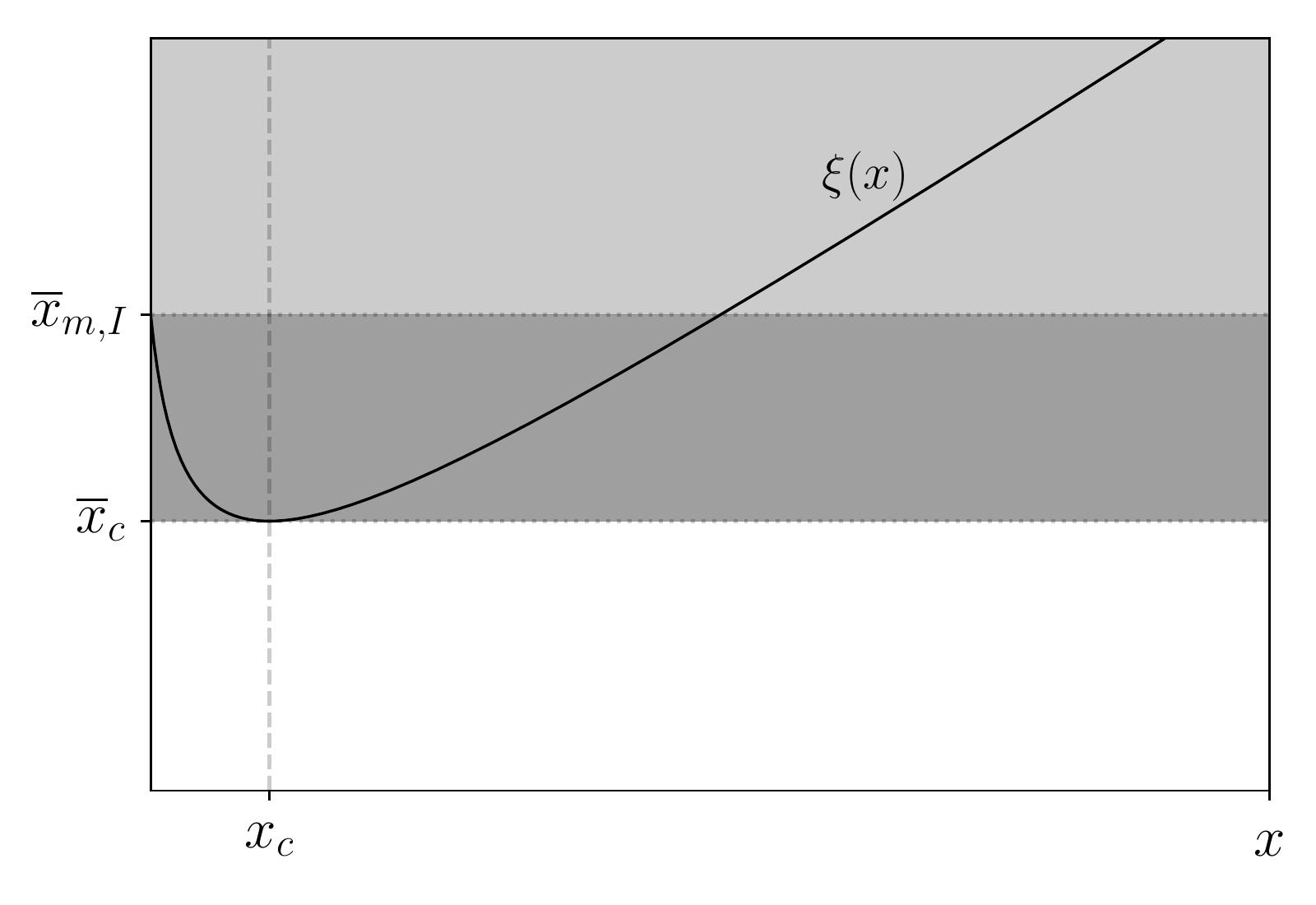}
    \caption{$\xi(x)$ with finite $\phi'(0)$}
    \end{subfigure}
    \begin{subfigure}{0.5\textwidth}
    \includegraphics[width = 1\linewidth]{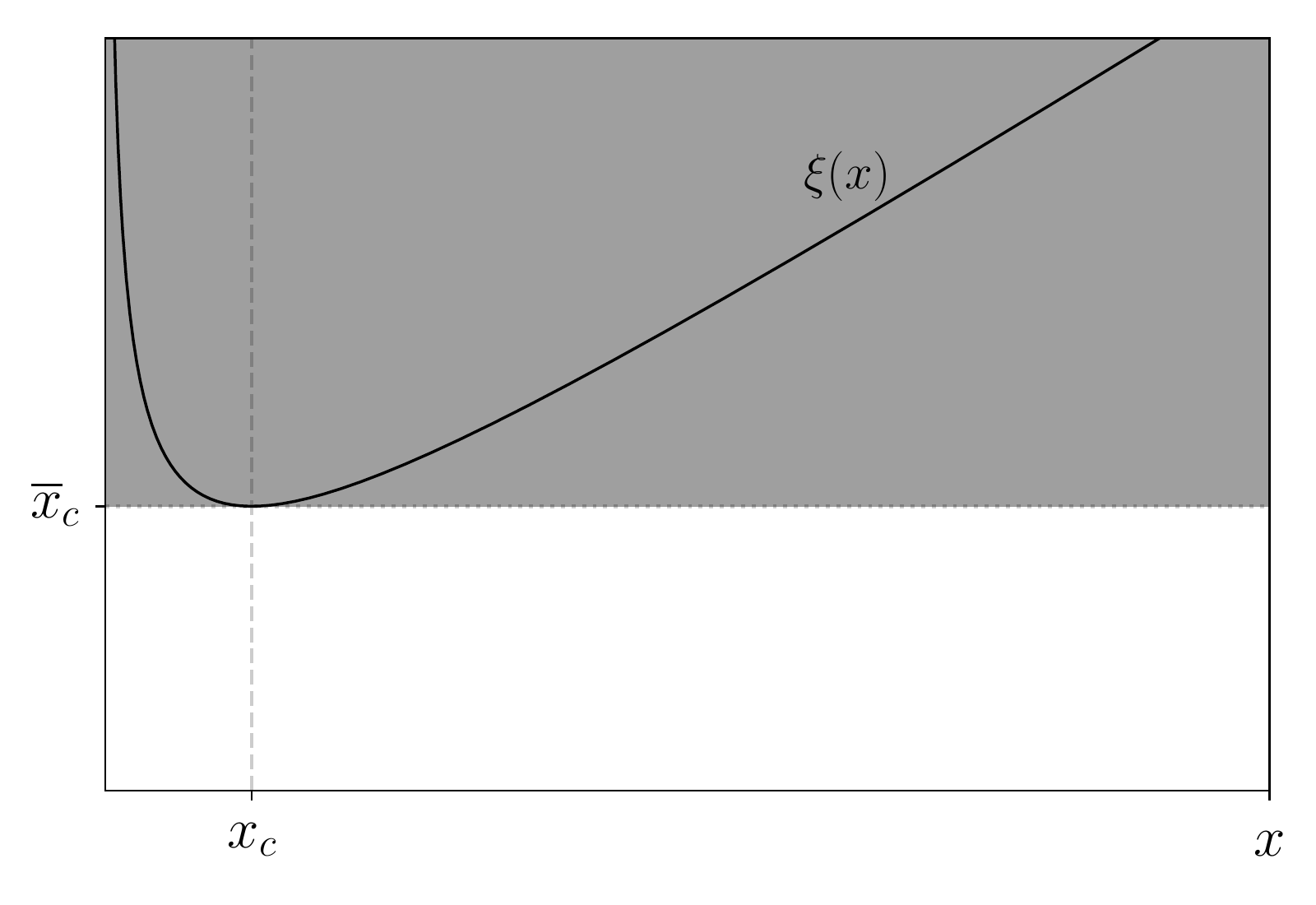}
    \caption{$\xi(x)$ with infinite $\phi'(0)$}
    \end{subfigure}
    \caption{Curves~$\xi(x)$, defined in equation~(\ref{eqn:xi definition}), for two possible potentials in the presheath: one with finite gradient at~$x=0$, the other with infinite gradient. For a given value of~$\bar{x}$, the effective potential~$\chi(x,\bar{x})$ will have a stationary point wherever~$\xi(x)=\bar{x}$. The effective potential will have negative gradient wherever~${\bar{x}>\xi(x)}$, and positive gradient wherever~$\bar{x}<\xi(x)$. Thus, the effective potential gains a well for~$\bar{x}>\bar{x}_{c}$. For~$\bar{x}_{\mathrm{m,I}}>\bar{x}>\bar{x}_{\mathrm{c}}$ (shaded dark grey), the effective potential will then be a well with two extrema, creating a type 2 orbit. For~$\bar{x}>\bar{x}_{\mathrm{m,I}}$ (shaded light grey), the potential has only one extremum; a minimum, creating a type 1 orbit. Note that, in the right hand plot, the gradient of the potential is infinite at~$x=0$. Hence, there are always two solutions to equation (\ref{eqn:conditionformin}) meaning all~$\bar{x}$ give type 2 orbits. One possible~$\xi(x)$ curve not shown is one where~$\xi(x)$ is an increasing function of~$x$, implying~$x_{c}=0$. Such a~$\xi(x)$ would give rise to only type 1 orbits.}
	\label{fig:xi curves}
\end{figure}

Equation~(\ref{eqn:effective potential definition}) defines the effective potential in which an ion with a given~$\bar{x}$ and~$U_{\perp}$ oscillates. We refer to an orbit as being closed if in the future it will have a bounce point, which is a point at which the~$x$ directed velocity of the particle changes sign. Since we have defined closed orbits in relation to bounce points, and bounce points are characterised by~$\chi(x,\bar{x}) = U_{\perp}$, it is relevant to consider where the extrema of the effective potentials are. The locations of the extrema are given by the equation
\begin{equation}
\label{eqn:derivative of effective potential}
\pdev{\chi}{x}(x,\bar{x}) = \Omega_{\ii}^{2}(x-\bar{x}) + \frac{\Omega_{\ii}}{B}\phi'(x) = 0. 
\end{equation}
Thus,
\begin{equation}
\label{eqn:conditionformin}
\bar{x}-x = \frac{\phi'(x)}{\Omega_{\ii}B}.
\end{equation}
Since we consider potentials with positive and monotonic decreasing, convex up, derivative (see figure~\ref{fig:cartoon}), we expect equation~(\ref{eqn:conditionformin}) to have zero, one or two solutions. This can be shown by defining the function~$\xi(x)$,
\begin{equation}
\label{eqn:xi definition}
\xi(x) = x + \frac{\phi'(x)}{\Omega_{\ii}B}.
\end{equation}
Whenever~$\xi(x) = \bar{x}$, the effective potential will have a stationary point. Furthermore, the effective potential will have positive gradient when~$\xi(x)>\bar{x}$  and negative gradient when~$\xi(x)<\bar{x}$. From figure~\ref{fig:xi curves}, it is evident that, at low~$\bar{x}$, there will be no solutions to equation~(\ref{eqn:conditionformin}) and the gradient of the effective potential will be positive for all~$x$. The first~$\bar{x}$ for which equation~(\ref{eqn:conditionformin}) has a solution is~$\bar{x}_{\mathrm{c}}$, defined by 
\begin{equation}
\bar{x}_{\mathrm{c}} = \min_{x\in [0,\infty]}\left(x + \frac{\phi'(x)}{\Omega_{\ii}B} \right) = x_{\mathrm{c}} + \frac{\phi'(x_{\mathrm{c}})}{\Omega_{\ii}B}.
\end{equation}
For~$\bar{x}>\bar{x}_{c}$ it is possible that equation~(\ref{eqn:conditionformin}) will have one or two solutions. Having two solutions implies the effective potential will have two stationary points (one maximum and one minimum), creating a potential well in which an ion could oscillate, provided its energy did not exceed that of the effective potential maximum. This is not the only type of potential well in which ions could oscillate, as all that is required for a potential well is a local minimum. Appealing once more to figure \ref{fig:xi curves}, we see that equation (\ref{eqn:conditionformin}) will only have one solution when~$\bar{x}$ exceeds~$\bar{x}_{\mathrm{m,I}}$, defined by
\begin{equation}
\label{eqn:xbarmI}
\bar{x}_{\mathrm{m,I}} = \frac{\phi'(0)}{\Omega_{\ii}B}.
\end{equation}
The parameter region~$\bar{x}>\bar{x}_{\mathrm{m,I}}$ corresponds to the effective potential decreasing initially from~$x=0$ before reaching a minima and increasing. Thus, this also defines a potential well in which the ion can oscillate. 

\begin{figure}
    \begin{subfigure}{0.5\textwidth}
    \includegraphics[width = 1\linewidth]{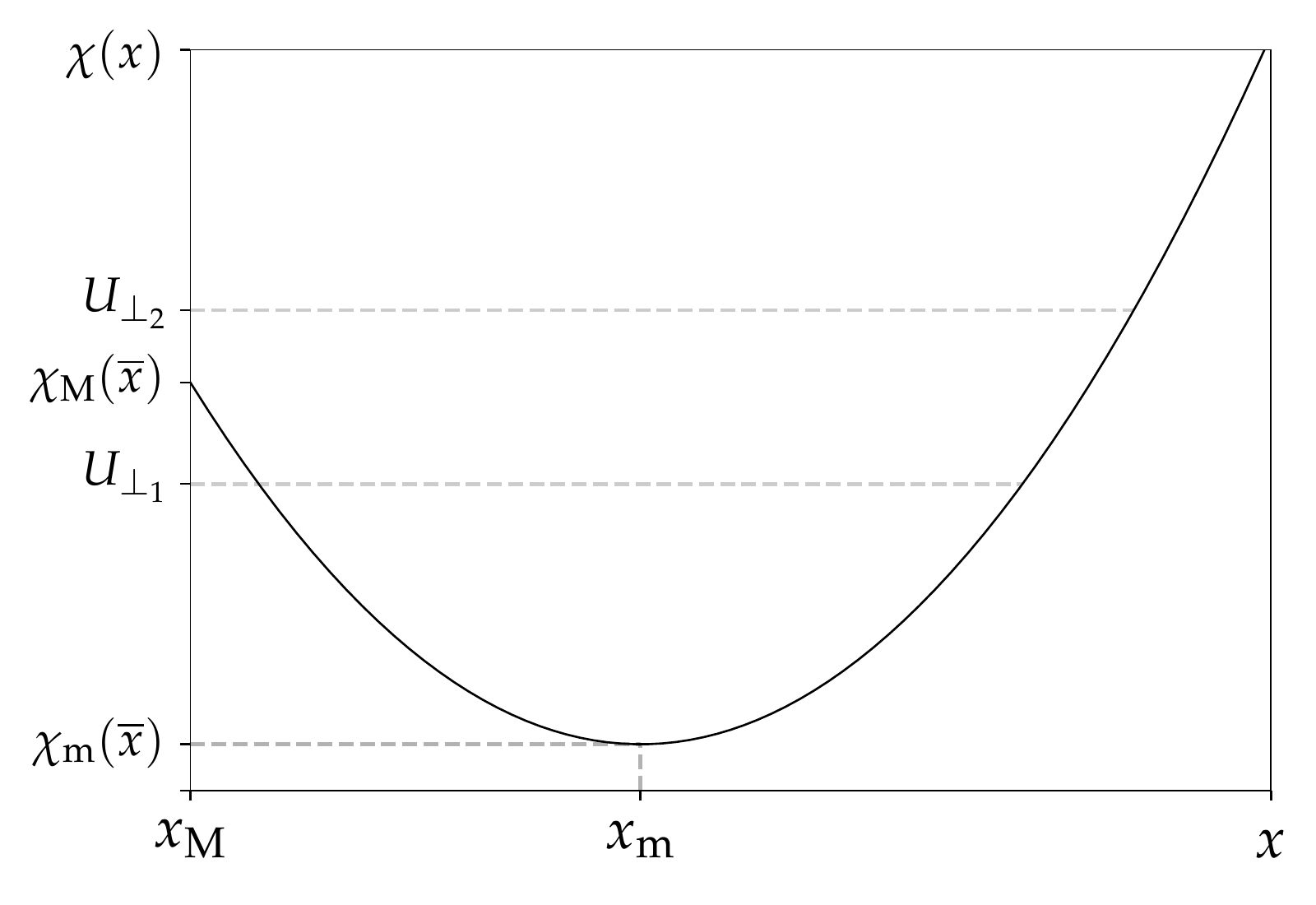}
    \caption{type 1 orbit}
    \end{subfigure}
    \begin{subfigure}{0.5\textwidth}
    \includegraphics[width = 1\linewidth]{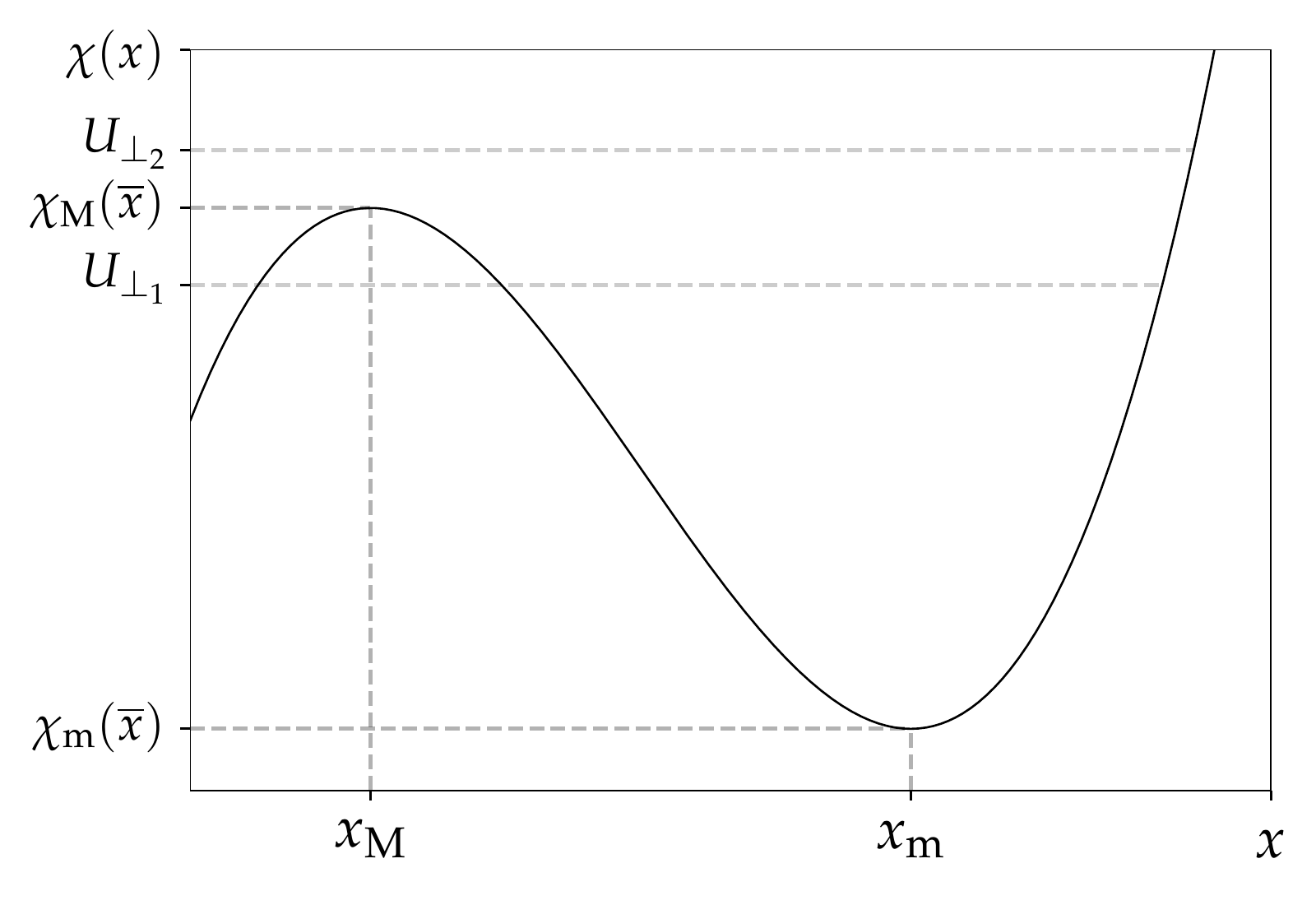}
    \caption{type 2 orbit}
    \end{subfigure}
    \caption{Examples of type~1 and type~2 orbits. Type~1 orbits have~$x_{M}=0$ and type~2 orbits have~$x_{M} \geq 0$. Orbits with perpendicular energy~$U_{\perp_{1}}$ will be closed orbits in both cases while orbits at~$U_{\perp_{2}}$ will by open orbits. Note that the typical size of~$U_{\perp_{2}}$ has been exaggerated for this sketch as open orbits will only have perpendicular energies marginally higher than~$\chi_{\mathrm{M}}(\bar{x})$.}
	\label{fig:orbittypes}
\end{figure}

We now introduce the nomenclature used in \cite{Cohen-Ryutov-1998 ,Geraldini-2018,Geraldini-2017} to classify orbits. An orbit is referred to as being of `type 1' if it has a single local minimum, and it is referred to as being of `type 2' if it has a maximum and a minimum. Effective potentials corresponding to these orbits are shown in figure~\ref{fig:orbittypes} and we see from the above discussion that~$\bar{x}>\bar{x}_{\mathrm{m,I}}$ gives rise to type 1 orbits while~$\bar{x}_{\mathrm{m,I}}>\bar{x}>\bar{x}_{\mathrm{c}}$ gives rise to type 2 orbits. Note that the derivative of the potential at~$x=0$ need not be finite and in such a case all orbits will be of type 2. 

An ion associated with any of these effective potentials will be in a closed orbit provided the perpendicular energy,~$U_{\perp}$, is not sufficient to allow the ion to reach the wall. In other words, the perpendicular energy must be less than the effective potential maximum,~$\chi_{\mathrm{M}}(\bar{x})$, defined by
\begin{equation}
\label{eqn:ChiM definition}
\chi_{\mathrm{M}}(\bar{x}) = \max_{x \in [0,x_{\mathrm{c}}]}(\chi(x,\bar{x})) = \chi(x_{\mathrm{M}}(\bar{x}),\bar{x}).
\end{equation}
Here we have also defined~$x_{\mathrm{M}}(\bar{x})$, the~$x$ coordinate of the effective potential maximum for a given~$\bar{x}$. Note that for type 1 orbits,~$\chi_{\mathrm{M}}(\bar{x}) = \chi(0,\bar{x})$ and~$x_{\mathrm{M}}(\bar{x}) = 0$.

In section \ref{section:zeroth order problem}, we showed that~$\bar{x}$ slowly decreases in time. As such, for a given ion,~$\chi_{\mathrm{M}}(\bar{x})$ will also decrease slowly, eventually falling below~$U_{\perp}$ (which is also decreasing, but more slowly). At this point, the ion will be free to travel towards~$x=0$ and is no longer considered to be on a closed orbit. To account for this effect, it is necessary not only to consider closed orbits, but also `open orbits'

\subsubsection{Open orbits.}
While closed orbits dominate throughout most of the presheath, in the area close to~$x=0$, there are much fewer closed orbits. Clearly we expect more orbits to become unbound as the ions drift slowly towards the wall. This motivates the fact that the presheath cannot be understood using only closed orbits and instead we need to include open orbits. We define an open orbit to be an ion trajectory which at a past time had bounce points and has no bounce points at future times. These orbits are of type 1 or type 2 depending on the type of closed orbit they last `escaped' from.
In reference \cite{Geraldini-2018}, it was shown that the perpendicular energies of these open orbits was strongly concentrated around the effective potential maximum. This implies that the~$v_{x}$ distribution of the ions with a particular value of $\bar{x}$ near the wall is strongly concentrated around~$V_{0}$, defined by
\begin{equation}
V_{0} = -\sqrt{\chi_{\mathrm{M}}(\bar{x}) - \chi(0,\bar{x})}.
\end{equation} 
Physically, this makes sense as the orbits in consideration are those that have only recently become unbound and therefore their perpendicular energy will be only a small amount above~$\chi_{_{\mathrm{M}}}(\bar{x})$ because~$U_{\perp}$ does not increase appreciably over a single orbit, by equation~(\ref{eqn:Uperp dot estimate}). Moreover, Geraldini et al \cite{Geraldini-2018} found that the velocity spread of ions in the~$x$ direction in open orbits is
\begin{equation}
\label{eqn:delvx}
\Delta v_{x} = \sqrt{2\left(\Delta_{\mathrm{M}} + \chi_{\mathrm{M}}(\bar{x}) - \chi(x,\bar{x}) \right)} - \sqrt{2\left(\chi_{\mathrm{M}}(\bar{x}) - \chi(x,\bar{x}) \right)},
\end{equation}
where the function~$\Delta_{\mathrm{M}}(\bar{x},U)$ is defined by
\begin{equation}
\label{eqn:delta M for delta vx}
\Delta_{\mathrm{M}}(\bar{x},U) = 2\alpha\Omega_{\ii}^{2}\sqrt{2(U-\chi_{\mathrm{M}}(\bar{x}))}\int_{x_{b}}^{x_{t}}\frac{s-x_{b}}{\sqrt{2(\chi_{\mathrm{M}}(\bar{x})-\chi(s,\bar{x}))}}\mathrm{d}s,
\end{equation}
and~$x_t$ and~$x_b$ are the top and bottom bounce points: the locations where the velocity in the~$x$ direction reverses in the quasiperiodic motion. Therefore~$x_b = x_{\mathrm{M}}$ and~$x_{t}$ solves~$\chi(x_{t},\bar{x}) = \chi_{\mathrm{M}}(\bar{x})$ with~$x_{t}>x_{b}$. Although~$\Delta_{\mathrm{M}}$ is, for most cases, not analytically tractable, its scaling will be sufficient for our purposes. We first examine the integral in equation~(\ref{eqn:delta M for delta vx}). The denominator of the integral vanishes at both end points of the integral but it is otherwise finite. Dealing with the lower bounce point first, we see that the integrand is always finite since~$\chi_{\mathrm{M}}(\bar{x}) - \chi(s,x)$ is at worst quadratic in~$s-x_{b}$ as~$s\to x_{b}$ (this occurs for type 2 orbits, see figure~\ref{fig:orbittypes}) and so the denominator~$\sqrt{\chi_{\mathrm{M}}(\bar{x}) - \chi(s,x)}$ is at worst linear in~$s-x_{b}$, which is cancelled by the numerator. At the upper bounce point,~$\chi_{\mathrm{M}}(\bar{x}) - \chi(s,x)$ vanishes linearly as~$s\to x_{t}$ for both type 1 and 2 orbits. Therefore, as the integrand contains an integrable singularity (which is present in regular orbits), the integral in equation~(\ref{eqn:delta M for delta vx}) will scale as~$v_{\mathrm{t,i}}/\Omega_{\ii}^{2}$ regardless of the type of orbit, and we arrive at the overall scaling of~$\Delta_{M} = O\left(\alpha v_{\mathrm{t,i}}^{2}\right)$. As a result of the size of~$\Delta_{\mathrm{M}}$, when the ion is near the effective potential maximum, such that~$\chi_{\mathrm{M}}(\bar{x}) - \chi(x,\bar{x}) \sim \Delta_{\mathrm{M}}$,~$\Delta v_{x} = O\left(\sqrt{\alpha}v_{\mathrm{t,i}}\right)$ (by equation~(\ref{eqn:delvx})). However, away from the effective potential maximum, we find~$\Delta_{\mathrm{M}} \ll \chi_{\mathrm{M}}(\bar{x}) - \chi(x,\bar{x})$ giving~$\Delta v_{x} = O\left(\alpha v_{\mathrm{t,i}}\right)$.

\subsection{The ion density}
\label{section:ion density}
As with the electrons, to compute the ion density, we need the integral
\begin{equation}
\label{eqn:ion density integral}
n_{\ii}(x) = \iiint f_{\ii}(x,v_{x},v_{y},v_{z})\mathrm{d}v_{x}\mathrm{d}v_{y}\mathrm{d}v_{z},
\end{equation}
over the regions that will not fictitiously include ions that ought to have encountered the wall in their prior trajectory. We have found these regions in the variables~$\bar{x}$,~$U_{\perp}$ and~$U$ in section~\ref{section:ion model} and so just need to change coordinates. The Jacobian of the transformation~$\lbrace v_{x},v_{y},v_{z}\rbrace \to \lbrace \bar{x}, U_{\perp},U\rbrace$ can be found from equations~(\ref{eqn:xbar}),~(\ref{eqn:Uperp}) and~(\ref{eqn:U}) and is
\begin{equation}
\pdev{(v_{x},v_{y},v_{z})}{(\bar{x},U_{\perp},U)} = \frac{\Omega_{\ii}}{v_{x}v_{z}} = \Omega_{\ii}\frac{1}{\sqrt{2\left(U_{\perp} - \chi(x,\bar{x}) \right)}}\frac{1}{\sqrt{2\left(U - U_{\perp} \right)}}.
\end{equation}
The ion distribution function is most convenient when written in the conserved variables of the ions which are the energy~$U$ and the adiabatic invariant for the ions,~$\mugk$, given in \cite{Geraldini-2017} as
\begin{equation}
\mugk(\bar{x},U_{\perp}) = \frac{1}{\pi}\int_{x_{b}}^{x_{t}}\sqrt{2(U_{\perp} - \chi(x,\bar{x}))}\mathrm{d}x.
\end{equation}
Here~$x_{b}$ and~$x_{t}$ are the top and bottom bounce points respectively, as defined in section~\ref{section:ion model}. In line with~\cite{Geraldini-2018}, we may now denote the closed orbit ion distribution function as a function of the conserved variables~$\lbrace\mugk,U\rbrace$ to be~$F_{\mathrm{cl}}(\mugk,U)$. Hence the density integral~(\ref{eqn:ion density integral}) is transformed to
\begin{equation}
\label{eqn:general ion density integral}
n_{\ii}(x) = \sum_{\sigma_{\parallel}= \pm 1}\int \Omega_{\ii}\mathrm{d}\bar{x} \int \frac{2\mathrm{d}U_{\perp}}{\sqrt{2\left(U_{\perp} - \chi(x,\bar{x}) \right)}} \int \frac{F_{\mathrm{cl}}(\mugk(\bar{x},U_{\perp}),U,\sigma_{\parallel})}{\sqrt{2\left(U - U_{\perp} \right)}}\mathrm{d}U.
\end{equation}
The sum is over the particles with positive and negative~$z$ directed velocities. A similar sum exists over the positive and negative~$x$ directed velocities but, since the distribution function is gyrophase independent at lowest order, this simply becomes a factor of 2. 

All that remains is to find the appropriate limits on the coordinates to ensure that only true closed orbits are being integrated over. Consider an ion at position~$x$. For this ion to be in a closed orbit, two conditions must be met: first, there must be a point in the ions effective potential at~$s < x$ such that~$\chi(s,\bar{x}) > \chi(x,\bar{x})$. This implies
\begin{equation}
\frac{1}{2}\Omega_{\ii}^{2}(x-\bar{x})^2 + \frac{\Omega_{\ii} \phi(x)}{B} < \frac{1}{2}\Omega_{\ii}^{2}(s-\bar{x})^{2} + \frac{\Omega_{\ii} \phi(s)}{B},
\end{equation}
which rearranges to give
\begin{equation}
\bar{x} > \frac{1}{2}(x+s) + \frac{\phi(x)-\phi(s)}{\Omega_{\ii} B (x-s)}.
\end{equation}
The minimum possible value of~$\bar{x}$ for which this condition could be satisfied is~$\bar{x}_{\mathrm{m}}(x)$, defined by
\begin{equation}
\label{eqn:xbar m defintion}
\bar{x}_{\mathrm{m}}(x) = \min_{s \in [0,x)}\left( \frac{1}{2}(x+s) + \frac{\phi(x)-\phi(s)}{\Omega_{\ii} B (x-s)} \right),
\end{equation}
giving the condition~$\bar{x}> \bar{x}_{\mathrm{m}}(x)$. Having chosen a possible~$\bar{x}$, the second condition for a closed orbit is that the particle must not have sufficient perpendicular energy to pass the maximum of the effective potential. This amounts to an upper bound on~$U_{\perp}$ of~$\chi_{_{\mathrm{M}}}(\bar{x})$. The only condition on the parallel energy is that the velocity in the~$z$ direction must be real valued. The resulting ion density integral is given by 
\begin{equation}
\label{eqn:closedorbitint}
n_{\ii,\mathrm{cl}}(x) = \int_{\bar{x}_{m}(x)}^{\infty} \Omega_{\sss}\mathrm{d}\bar{x} \int_{\chi(x,\bar{x})}^{\chi_{\mathrm{M}}(\bar{x})} \frac{2\mathrm{d}U_{\perp}}{\sqrt{2\left(U_{\perp} - \chi(x,\bar{x}) \right)}} \int_{U_{\perp}}^{\infty} \frac{F_{\mathrm{cl}}(\mugk(\bar{x},U_{\perp}),U)}{\sqrt{2\left(U - U_{\perp} \right)}}\mathrm{d}U.
\end{equation} 
Here we have also removed the sum over parallel directions due to that fact that no ions are reflected back from the sheath into the presheath.

As well as closed orbits, we must also consider open orbits. Just as with closed orbits, we make the same change of variables from~$\lbrace v_{x},v_{y},v_{z}\rbrace$ to~$\lbrace \bar{x}, U_{\perp},U\rbrace$. However, we must impose different limits on the allowed values of these coordinates at a given~$x$. To do so, we note that an ion on an open orbit will hit the wall on a timescale of order~${1}/{\Omega_{\ii}}$. Thus we may exploit the smallness of equations~(\ref{eqn:xbar dot}) and~(\ref{eqn:Uperp dot estimate}) to argue that the value of~$\bar{x}$ and~$U_{\perp}$ that this ion had  a time of order~$1/\Omega_{\ii}$ before will not change appreciably. An orbit becomes open when~$U_{\perp}$ grows minutely larger than the value of the effective potential maximum~$\chi_{M}(\bar{x})$ that had stopped the ion from reaching the wall. This small increase will however leave the upper bounce point~$x_{t}$ virtually unchanged. This means that, as an ion transitions into an open orbit, it can be found anywhere between~$x=0$ and the upper bounce point of the closed orbit it previously occupied. Therefore, ions in open orbits at positions~$x>x_{c}$ cannot have come from orbits with~$\bar{x} < \bar{x}_{\mathrm{m}}(x)$ for the same reason closed orbits at this position could not come from this~$\bar{x}$: the perpendicular energy required to reach~$x$ from that~$\bar{x}$ would imply the orbit had been open for much longer than one period of the motion and thus no ions could exist on such an orbit. If~$x<x_{c}$, however, all open orbits are free to move towards~$x=0$ (by definition) and hence an ion with~$x<x_{c}$ could have come from any valid closed orbit:~$\bar{x}>\bar{x}_{c}$. We denote the minimum~$\bar{x}$ for an open orbit at position~$x$ to be~$\bar{x}_{\mathrm{m,o}}$ given by
\begin{equation}
\label{eqn:xbar m,o definition}
\bar{x}_{\mathrm{m,o}} = \begin{cases}
\bar{x}_{\mathrm{c}} \text{  if } x \leq x_{\mathrm{c}}, \\
\bar{x}_{\mathrm{m}}(x) \text{  if } x > x_{\mathrm{c}}.
\end{cases}
\end{equation}
Once we have established the value of~$\bar{x}$ that an ion in an open orbit previously had, we can determine the range of possible~$v_{x}$. As argued above, it is only a minute change in~$U_{\perp}$ that causes an orbit to become open so~$v_{x}$ will be tightly concentrated around a single value. This is made precise by~(\ref{eqn:delvx}). Thus,
\begin{equation}
\label{eqn:openorbitint}
n_{\ii,\mathrm{op}}(x) = \int_{\bar{x}_{\mathrm{m,o}}}^{\infty}\Omega_{\ii}\mathrm{d}\bar{x}\int_{\chi_{\mathrm{M}}(\bar{x})}^{\infty}\frac{F_{\mathrm{cl}}(\mugk(\bar{x},\chi_{\mathrm{M}}(\bar{x})),U)}{\sqrt{2\left(U- \chi_{\mathrm{M}}(\bar{x})\right)}}\Delta v_{x}\,\mathrm{d}U.
\end{equation}
\begin{figure}
    \begin{subfigure}{0.5\textwidth}
    \includegraphics[width = 1\linewidth]{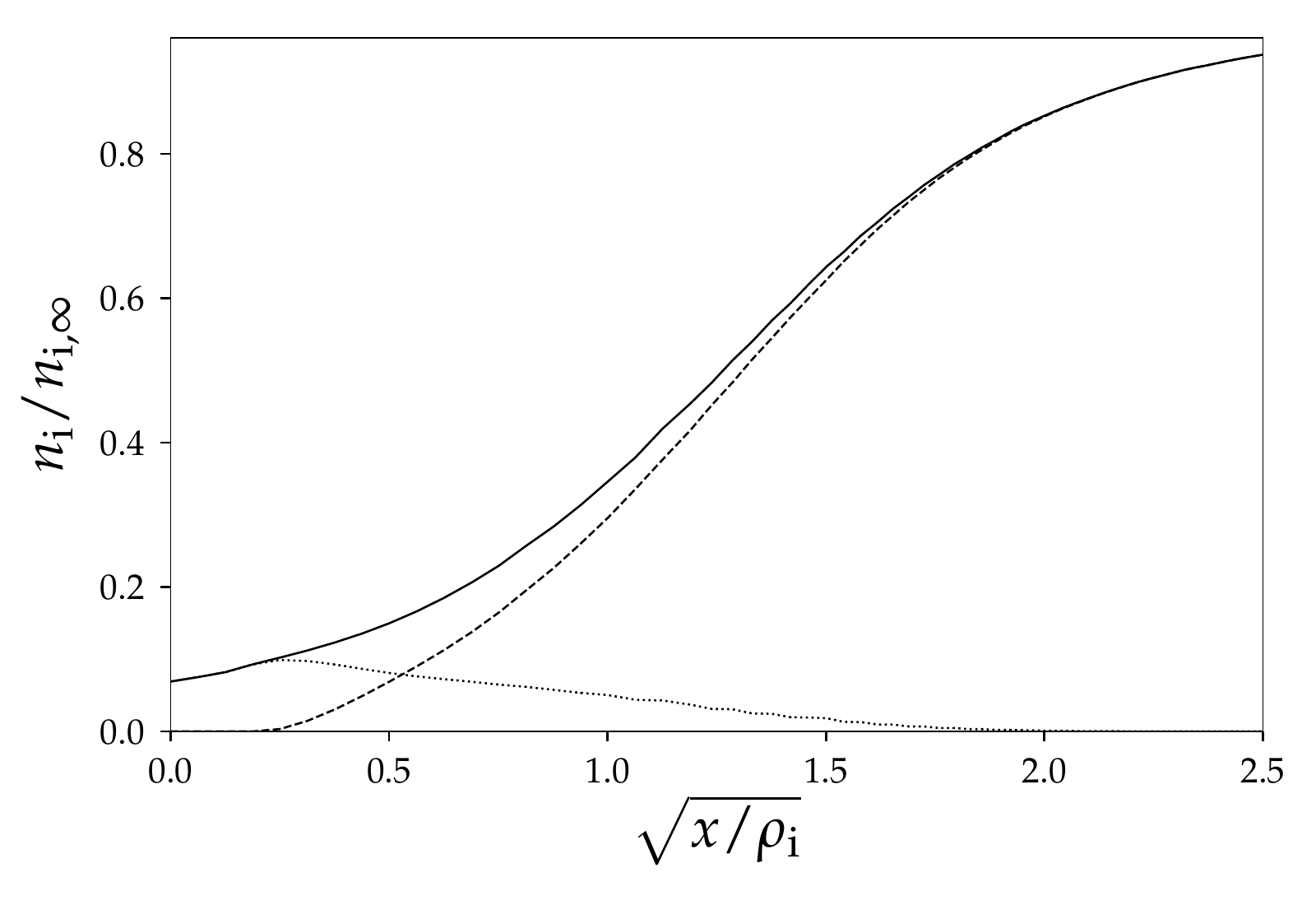}
    \caption{$\phi_{\cc} \to -\infty$}
    \end{subfigure}
    \begin{subfigure}{0.5\textwidth}
    \includegraphics[width = 1\linewidth]{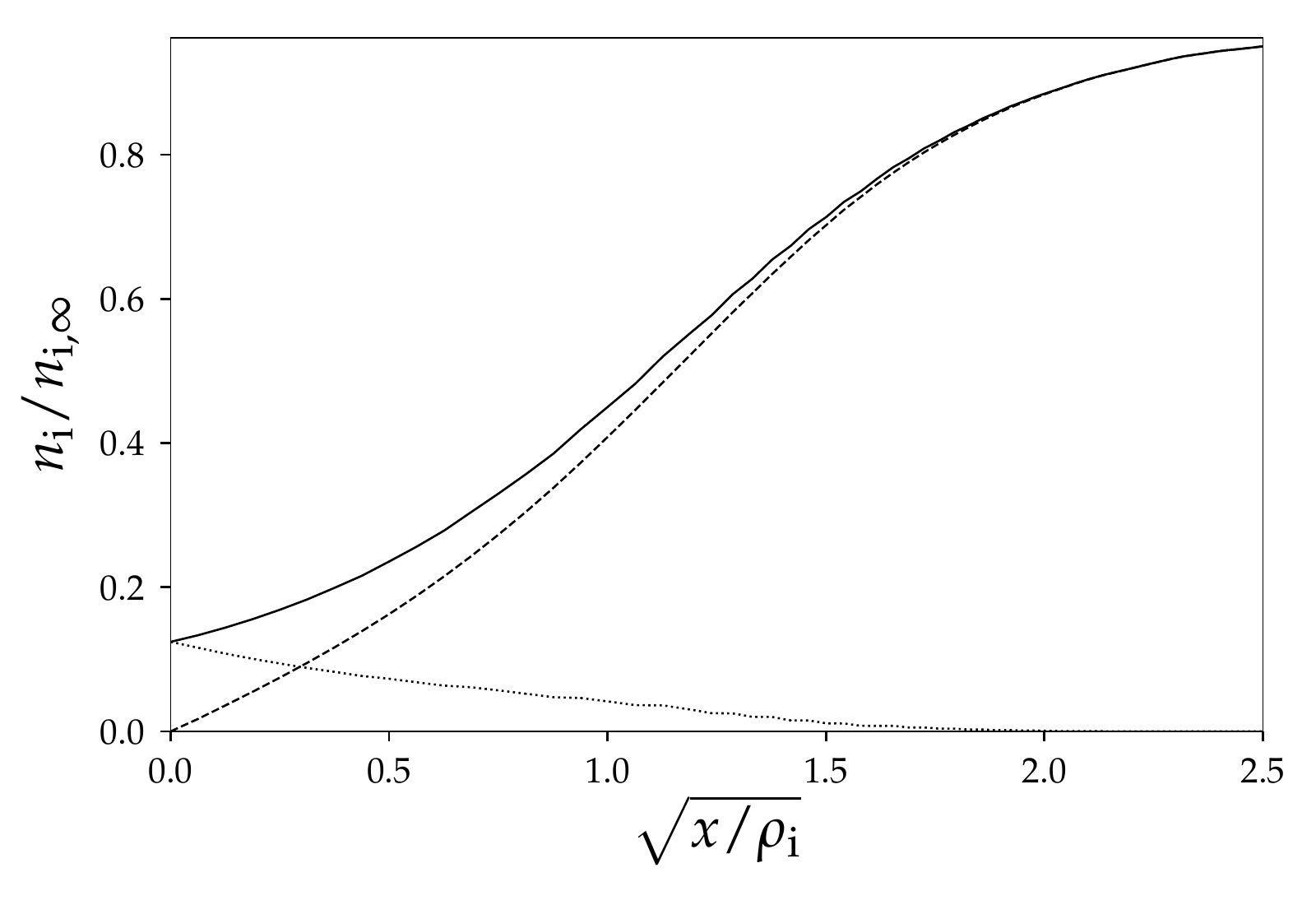}
    \caption{$\phi_{\cc} = \phi(0)$}
    \end{subfigure}
    \caption{Ion density (solid line) composed of closed (dashed) and open (dotted) orbit densities for two solutions with different cutoff potentials. For these simulations,~${\alpha = 0.05}$ radians, and we use the boundary conditions specified in section~\ref{section:Numerical method} with~$T_{\ee} = T_{\ii}$}
    \label{fig:ion densities}
\end{figure}While the presence of~$\Delta v_{x}$ ensures that the open orbit density is everywhere small, it is still the dominant contribution to the ion density near~$x=0$, where the closed orbit ion density tends to zero. We can see from figure~\ref{fig:ion densities} that this is indeed true and that in all cases it would be a mistake to neglect the contribution of the open orbits to the ion density. 

As discussed in the introduction, the scaling of~$n_{\ii,\mathrm{op}}(x)$ with~$\alpha$ near~$x=0$ is of special importance due to its relation to the total potential drop in the presheath. By definition, type 1 orbits have their effective potential maximum at~${x=0}$, while the locations of the maxima of type 2 effective potentials occupy a continuous range of~$x$. Recalling the discussion at the end of section~\ref{section:ion model}, we see that, away from~${x=0}$, the vast majority of ions are not near their effective potential maxima, and so~${\Delta v_{x} = O\left(\alpha v_{\mathrm{t,i}} \right)}$ which therefore implies~${n_{\ii,\mathrm{op}}(x) = O(\alpha n_{\ii,\infty})}$. As we approach~${x=0}$, however, if there is a significant number of type 1 orbits present, these orbits will be near their effective potential maxima, and as such will have~$\Delta v_{x} = O(\sqrt{\alpha}v_{\mathrm{t,i}})$. This means that if type 1 orbits are present,~${n_{\ii,\mathrm{op}}(0)= n_{\ii}(0) = O(\sqrt{\alpha}n_{\ii,\infty})}$. On the other hand, if only type 2 orbits are present,~${n_{\ii}(0) = O(\alpha n_{\ii,\infty})}$. This is the algebraic counterpart to the argument based on circular orbits in figure~\ref{fig:cartoon2} as orbits that are not significantly deformed by the potential are naturally of type 1. 

By conservation of particle flux, we know that the product of the ion density and fluid velocity is constant. Therefore, when only type 2 orbits are present, the fluid velocity towards the wall is much larger than when type 1 and type 2 orbits are present. In section~\ref{section:expansion at zero} we will see that this is an early sign of how the Bohm condition will emerge from the theory, and how the presence of type 1 orbits can have a startling impact on the presheath. 

\section{Quasineutrality}
\label{section:Quasineutrality}
The solutions to the collisionless magnetic presheath are expected to be quasineutral and thus satisfy the equation
\begin{equation}
\label{eqn:Quasi}
n_{\ee}(x) = Zn_{\ii}(x).
\end{equation}
The problem of solving the magnetic presheath amounts to finding a potential throughout the presheath such that equation (\ref{eqn:Quasi}) is satisfied. In this section we will study the tractable asymptotic limits of this. These limits are at large and small~$x$ where the Chodura and Bohm conditions are expected to manifest, respectively. Compared to section 5 of \cite{Geraldini-2018} only the electron physics has been changed and therefore the expansions of the ion density will be identical. We will find in section~\ref{section:expansion at infty} that the new electron physics creates only a small modification to the Chodura condition in the large~$x$ expansion. However, in section~\ref{section:expansion at zero}, we will take more care in deriving the modification to the Bohm condition, which, despite looking only superficially different, we will see can be qualitatively different depending on~$\phi(0)$ and~$\phi_{\cc}$. Indeed, we will show in section \ref{section:criticalcutoff} how the breakdown of the Bohm condition for~$\phi(0) = \phi_{\cc}$ (a collapsed Debye sheath) arises naturally as the self-consistent culmination of the Bohm condition's weakening as~$\phi(0)$ approaches~$\phi_{\cc}$.

\subsection{Expansion at $x \to \infty$}
\label{section:expansion at infty}
At large~$x$, ion and electron orbits are far from the wall. This means that the value of~$x$ itself becomes less relevant and the more relevant quantity is the relation between the potential and the density of the plasma. In \cite{Geraldini-2018} this relation was found in the asymptotic limit of a small slowly varying field, defined by the two equations
\begin{equation}
\frac{e\phi}{T_{\ee}} \sim \epsilon^{2} \ll 1,
\end{equation}
and
\begin{equation}
\frac{\rho_{\ii}\phi'}{\phi} \sim \epsilon.
\end{equation}
Equipped with the small parameters~$\phi$ and~$\phi'$ one can then expand the ion and electron density. The calculation can therefore be ported across with the only difference being the more exact electron density expression
\begin{equation}
n_{\ee}(\phi) = n_{\ee,\infty} + \left.\dev{n_{\ee}}{\phi}{}\right|_{\phi=0}\phi + \left.\dev{n_{\ee}}{\phi}{2}\right|_{\phi = 0}\phi^{2}
\end{equation}
as the extension to non-Boltzmann electrons. The central result of the expansion is then the ``kinetic chodura condition"
\begin{equation}
\label{eqn:Chodura condition}
Z\iiint\frac{f_{\ii,\infty}(\vec{v})}{v_{z}^{2}}\mathrm{d}{v_{x}}\mathrm{d}{v_{y}}\mathrm{d}{v_{z}} \leq  \frac{1}{v_{B}^{2}}\frac{T_{\ee}}{e}\left.\dev{n_{\ee}}{\phi}{}\right|_{\phi=0},
\end{equation}
which is a condition for there to be non-oscillatory solutions of the potential at~$x \to \infty$. This alters the result of \cite{Geraldini-2018} to include the effect of kinetic electrons at the presheath entrance. Despite this being an inequality, it is expected that in most cases the plasma beyond the magnetic presheath self-consistently marginally satisfies the kinetic Chodura condition just as unmagnetised plasma beyond a Debye sheath usually marginally satisfies a kinetic Bohm condition \cite{Riemann-review}. We will later impose this condition numerically in section \ref{section:Numerical method}.
\subsection[]{Expansion around x \(\to\) 0}
\label{section:expansion at zero}
While at infinity we were safe to assume that the potential had a Taylor expansion in~$x$, care must be taken as we approach the Debye sheath entrance. Indeed, it was shown in \cite{Geraldini-2018} that an infinite gradient at the origin is always present in a magnetic presheath with Boltzmann electrons. 

With this in mind, in this section we will first expand quasineutrality in the parameter~$\delta\phi$ defined by
\begin{equation}
\delta\phi = \phi(x) - \phi(0).
\end{equation}
Thus we will not make any assumptions about the nature of~$\delta\phi$ save for the fact that it is finite (equivalent to the statement that the potential is finite at the origin). We conduct this expansion for two cases: first, in the usual case of~$\phi(0) \neq \phi_{\cc}$, and secondly, in the special case of~$\phi(0) = \phi_{\cc}$.
\subsubsection[]{$\phi(0) \neq \phi_{\cc}$.}
\label{secsec:normalcutoff}

From equation~(\ref{eqn:electron density local with v parallel bar}), we see that the electron density will have a Taylor expansion in~$\delta\phi$, giving
\begin{dmath}
\label{eqn:netaylor}
n_{\ee}(\phi) - n_{\ee}(\phi(0)) = \dev{n_{\ee}}{\phi}{}\delta\phi + \frac{1}{2}\dev{n_{\ee}}{\phi}{2}\delta\phi^{2} + O\left(n_{\ee}(0)\left(\frac{\ee}{T_{\ee}}\delta\phi\right)^{3}\right),
\end{dmath}
where the derivatives of density with respect to potential are evaluated at the origin. The ion density at the origin has contributions from both open and closed orbits. In \cite{Geraldini-2018}, it was shown that the leading order term of~$n_{\ii,\mathrm{cl}}$ is due to closed type 1 orbits and gives a contribution of
\begin{equation}
n_{\ii,\mathrm{cl,I}}(x) = 2\int_{\bar{x}_{\mathrm{m,I}}}^{\infty}\sqrt{2\delta\chi}\Omega_{\ii} \mathrm{d}\bar{x}\int_{\chi_{\mathrm{M}}(\bar{x})}^{\infty}\frac{F_{\mathrm{cl}}(\mugk(\bar{x},\chi_{\mathrm{M}}(\bar{x})),U)}{\sqrt{2\left(U - \chi_{\mathrm{M}}(\bar{x}) \right)}} \mathrm{d}U,
\end{equation}
where~$\delta\chi$ is defined as
\begin{equation}
\delta \chi = \chi(0,\bar{x}) - \chi(x,\bar{x}) = \Omega_{\ii}^{2}x\bar{x} - \frac{\Omega_{\ii}}{B}\delta\phi + O\left(\left(\frac{x}{\rho_{\ii}}\right)^{2}v_{\mathrm{t,i}}^{2} \right).
\end{equation}
Thus, all that remains to consider is the contribution from open orbits. To this end, we rewrite equation~(\ref{eqn:openorbitint}) as
\begin{dmath}
\label{eqn:openorbitint2}
n_{\ii,\mathrm{op}}(x) - n_{\ii,\mathrm{op}}(0) = \int_{\bar{x}_{\mathrm{c}}}^{\infty}\Omega_{\ii}\mathrm{d}\bar{x}\int_{\chi_{\mathrm{M}}(\bar{x})}^{\infty}\frac{F_{\mathrm{cl}}(\mugk,U)}{\sqrt{2\left(U - \chi_{\mathrm{M}}(\bar{x}) \right)}}\left[\Delta v_{x} - \Delta v_{x_{0}} \right]\mathrm{d}U - \int_{\bar{x}_{\mathrm{c}}}^{\bar{x}_{\mathrm{m,o}}}\Omega_{\ii}\mathrm{d}\bar{x}\int_{\chi_{\mathrm{M}}(\bar{x})}^{\infty}\frac{F_{\mathrm{cl}}(\mugk,U)}{\sqrt{2\left(U - \chi_{\mathrm{M}}(\bar{x}) \right)}}\Delta v_{x}\mathrm{d}U,
\end{dmath}
where we have defined~$\Delta v_{x_{0}}$ as
\begin{equation}
\Delta v_{x_{0}} = \sqrt{2\left(\Delta_{\mathrm{M}} + \chi_{\mathrm{M}}(\bar{x}) - \chi(0,\bar{x}) \right)} - \sqrt{2\left(\chi_{\mathrm{M}}(\bar{x}) - \chi(0,\bar{x})  \right)}.
\end{equation}
If type 2 orbits are present, then~$x_{\mathrm{c}}>0$. Hence, sufficiently close to~$x=0$, we will have~$x<x_{\mathrm{c}}$ and the second integral in equation~(\ref{eqn:openorbitint2}) will be zero by equation~(\ref{eqn:xbar m,o definition}). If no type 2 orbits are present, then the second integral scales like~$\bar{x}_{\mathrm{m}}(x)$, defined in equation~(\ref{eqn:xbar m defintion}), which in turn scales like~$x$ or~$\delta\phi$ and is small compared to~$\sqrt{\delta\phi}$. Hence, only the first term in equation~(\ref{eqn:openorbitint2}) can give a contribution comparable to that of~$n_{\ii,\mathrm{cl}}$. The size of the first term in equation~(\ref{eqn:openorbitint2}) is set by~$\Delta v_{x} - \Delta v_{x_{0}}$ which is given by
\begin{dmath}
\label{eqn:DeltaminusDelta}
\Delta v_{x} - \Delta v_{x_{0}} = \sqrt{2\left(\Delta_{\mathrm{M}} + \chi_{\mathrm{M}}(\bar{x}) - \chi(x,\bar{x}) \right)} - \sqrt{2\left(\chi_{\mathrm{M}}(\bar{x})  - \chi(x,\bar{x}) \right)} - \sqrt{2\left(\Delta_{\mathrm{M}} + \chi_{\mathrm{M}}(\bar{x}) - \chi(0,\bar{x}) \right)} + \sqrt{2\left( \chi_{\mathrm{M}}(\bar{x}) - \chi(0,\bar{x}) \right)}.
\end{dmath}
Depending on~$\bar{x}$, $\Delta v_{x} - \Delta v_{x_{0}}$ will be evaluated for a type 1 orbit or a type 2 orbit. For a type 1 orbit,~$\chi_{M}(\bar{x}) = \chi(0,\bar{x})$ so the last term in equation~(\ref{eqn:DeltaminusDelta}) is zero and the first and third terms differ by a term of order~$\delta \chi$, while the second term, of order~$\delta\chi^{1/2}$, would be the largest. If instead we are on a type 2 orbit, since~$\chi_{\mathrm{M}}(\bar{x})\geq \chi(0,\bar{x})$, we may choose~$x$ sufficiently small that the above terms cancel leaving a piece of size at most~$\delta \chi$ (first term cancels with third, second cancels with fourth). As such, to lowest order, we need only consider the part of the integral for which we are on a type 1 orbit, corresponding to a lower limit on~$\bar{x}$ of~$\bar{x}_{\mathrm{m,I}}$. 

After this, the lowest order term in the quasineutrality equation is
\begin{equation}
\label{eqn:Infgradatorigin}
0 = \int_{\bar{x}_{\mathrm{m,I}}}^{\infty}\sqrt{2\delta\chi}\Omega_{\ii}\mathrm{d}\bar{x}\int_{\chi_{\mathrm{M}}(\bar{x})}^{\infty}\frac{F_{\mathrm{cl}}(\mugk,U)}{\sqrt{2\left(U-\chi_{\mathrm{M}}(\bar{x})\right)}}\mathrm{d}U.
\end{equation}
Given that the integrand is always positive, this equation is only satisfied if~$\bar{x}_{\mathrm{m,I}} \to \infty$, which by equation~(\ref{eqn:xbarmI}) implies the gradient of the potential is divergent at the origin.

\begin{figure}
	\begin{subfigure}{0.5\textwidth}
	\includegraphics[width=1\linewidth]{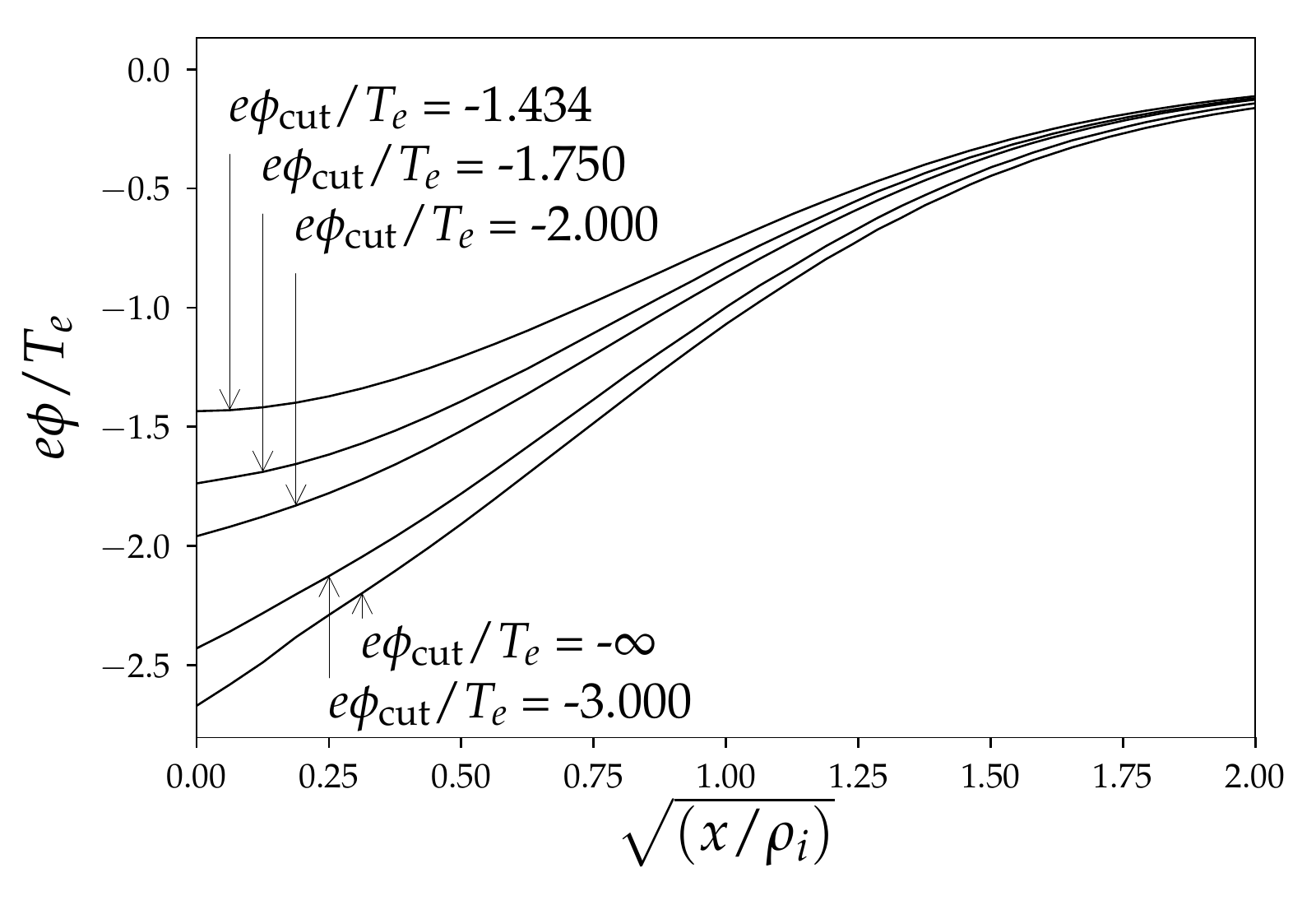}
	\caption{Potentials}
	\label{fig:cutoff variations potential}
	\end{subfigure}
	\begin{subfigure}{0.5\textwidth}
	\includegraphics[width=1\linewidth]{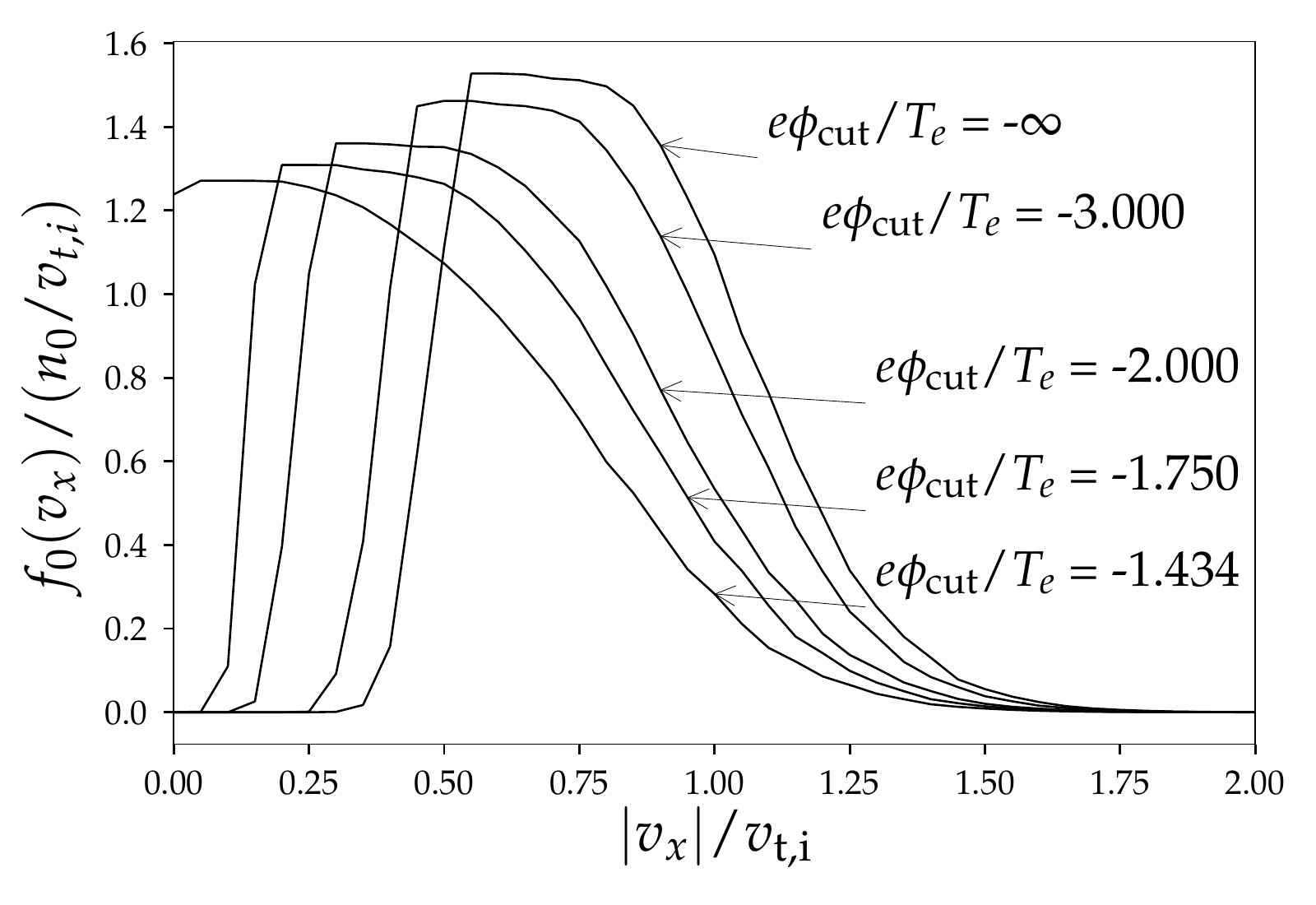}
	\caption{marginalised distribution function in $v_{x}$}
	\label{fig:cutoff variation pdf}
	\end{subfigure}
	\caption{Solutions to the electrostatic potential and the corresponding marginalised ion distribution function at the entrance to the Debye sheath for a range of cutoff potentials. Here~$\alpha = 0.05$ radians and we have used the boundary conditions given in section~\ref{section:Numerical method} with~$T_{\ee}=T_{\ii}$ }
	\label{fig:cutoff_variations}
\end{figure}

This divergence offers a considerable simplification to the quasineutrality equation: with an infinite gradient at the origin, all effective potentials are of type 2 and so all ion trajectories will be open a finite distance before~$x=0$. By examining equation~(\ref{eqn:xbar m defintion}), we see that~$\bar{x}_{\mathrm{m}}(x)$ (the minimum allowed~$\bar{x}$ for a closed orbit to be present) diverges at small~$x$ (as we now expect). From this, we see that~$\bar{x}$ must be increased greatly to allow for closed orbits closer to the origin. However, as~$\bar{x}$ is increased, the value of $U_{\perp}$ necessary to reach the origin also increases. Thus, at sufficiently small~$x$, only exponentially few trajectories will be present that are still in quasi-closed orbits. Since the majority of ions are therefore far from their effective potential maximum, we may expand equation~(\ref{eqn:DeltaminusDelta}) in the limit of~$\chi(x,\bar{x})-\chi(0,\bar{x}) \ll \chi_{\mathrm{M}}(\bar{x}) - \chi(0,\bar{x})$, giving
\begin{equation}
\Delta v_{x} - \Delta v_{x_{0}} = -\Delta\left[\frac{1}{v_{x_{0}}}\right]\delta\chi + \frac{1}{2}\Delta\left[\frac{1}{v_{x_{0}}^{3}} \right]\delta \chi^2 + O\left(\frac{\delta\chi^{3}}{v_{\mathrm{t,i}}^{5}}\right),
\end{equation}
with
\begin{equation}
\Delta\left[\frac{1}{v_{x_{0}}^n} \right] = \frac{1}{\left(2\left(\chi_{\mathrm{M}}(\bar{x}) - \chi(0,\bar{x}) \right)\right)^{\frac{n}{2}}} - \frac{1}{\left(2\left(\Delta_{\mathrm{M}}+ \chi_{\mathrm{M}}(\bar{x}) - \chi(0,\bar{x})   \right) \right)^{\frac{n}{2}}}.
\end{equation}
Collecting what are now the lowest order terms in the quasineutrality equation gives
\begin{dmath}
\dev{n_{\ee}}{\phi}{}\delta\phi = -Z\Omega_{\ii} x\int_{\bar{x}_{\mathrm{c}}}^{\infty}\Omega_{\ii}^{2}\bar{x}\mathrm{d}\bar{x}\int_{\chi_{\mathrm{M}}(\bar{x})}^{\infty}\frac{F_{\mathrm{cl}}(\mugk,U)}{\sqrt{2\left(U-\chi_{\mathrm{M}}(\bar{x})\right)}}\Delta\left[\frac{1}{v_{x_{0}}} \right]\mathrm{d}U + Z\frac{\Omega_{\ii}}{B}\delta\phi\int_{\bar{x}_{\mathrm{c}}}^{\infty}\Omega_{\ii}\mathrm{d}\bar{x}\int_{\chi_{\mathrm{M}}(\bar{x})}^{\infty}\frac{F_{\mathrm{cl}}(\mugk,U)}{\sqrt{2\left(U-\chi_{\mathrm{M}}(\bar{x})\right)}}\Delta\left[\frac{1}{v_{x_{0}}} \right]\mathrm{d}U.
\end{dmath}
Here the~$\mathrm{d}n_{\ee}/\mathrm{d}\phi$ is evaluated at the entrance to the Debye sheath. This could be solved in the form~$\delta\phi = qx$, but this solution would break our assumption that the potential has infinite gradient at the origin. Instead, the terms of order~$\delta\phi$ must cancel. This cancellation gives the \textit{marginal kinetic Bohm condition}, modified compared to that derived in \cite{Geraldini-2018}
\begin{equation}
\label{eqn:kinetic bohm condition}
\left.\dev{n_{\ee}}{\hat{\phi}}{}\right|_{x=0} = Zv_{\mathrm{B}}^{2}\int_{\bar{x}_{\mathrm{c}}}^{\infty}\Omega_{\ii}\mathrm{d}\bar{x}\int_{\chi_{\mathrm{M}}(\bar{x})}^{\infty}\frac{F_{\mathrm{cl}}(\mugk,U)}{\sqrt{2\left(U - \chi_{\mathrm{M}}(\bar{x}) \right)}}\Delta\left[\frac{1}{v_{x_{0}}} \right]\mathrm{d}U.
\end{equation}
This recovers \cite{Geraldini-2018}'s kinetic Bohm condition in the limit of~\(\phi_{\cc} \to -\infty\). Furthermore, we may note that this condition can be recast in the following form
\begin{equation}
\label{eqn:kinetic bohm condition alternative form}
\left.\dev{n_{\ee}}{\hat{\phi}}{}\right|_{x=0} = Zv_{\mathrm{B}}^{2}\iiint \frac{f_{i,0}(\vec{v})}{v_{x}^2} \mathrm{d}\vec{v},
\end{equation}
by using the following identity
\begin{dmath}
\label{eqn:identity I need}
Zv_{\mathrm{B}}^{2}\int_{\bar{x}_{\mathrm{c}}}^{\infty}\Omega_{\ii}\mathrm{d}\bar{x}\int_{\chi_{\mathrm{M}}(\bar{x})}^{\infty}\frac{F_{\mathrm{cl}}(\mugk,U)}{\sqrt{2\left(U - \chi_{\mathrm{M}}(\bar{x}) \right)}}\Delta\left[\frac{1}{v_{x_{0}}} \right]\mathrm{d}U = Zv_{\mathrm{B}}^{2}\int_{\bar{x}_{\mathrm{c}}}^{\infty}\Omega_{\ii}\mathrm{d}\bar{x}\int_{\chi_{\mathrm{M}}(\bar{x})}^{\infty}\frac{F_{\mathrm{cl}}(\mugk,U)}{\sqrt{2\left(U - \chi_{\mathrm{M}}(\bar{x}) \right)}}\mathrm{d}U\int_{-\infty}^{0}\frac{1}{v_{x}^2}\hat{\Pi}\left(v_{x},V_{0} - \Delta v_{x_{0}},V_{0} \right)\mathrm{d}v_{x} = Zv_{\mathrm{B}}^{2}\iiint\frac{f_{i,0}(\vec{v})}{v_{x}^{2}} \mathrm{d}\vec{v}.
\end{dmath}
Here, the top hat function~$\hat{\Pi}$ is defined by
\begin{equation}
\hat{\Pi}(v_{x},v_{1},v_{2}) = \begin{cases}
1 \text{  if } v_{1}<v_{x}<v_{2}, \\
0 \text{  otherwise}.
\end{cases}	
\end{equation}
Equation~(\ref{eqn:kinetic bohm condition alternative form}) is the marginal (equality) form of equation~(79) of \cite{Riemann-review}. The inequality form of equation~(\ref{eqn:kinetic bohm condition alternative form}) was originally derived (with adiabatic electrons) in reference~\cite{Harrison-Thompson-1959} as a generalisation of the Bohm condition to kinetic ions. As such, what we call the `kinetic Bohm condition' is often referred to as the `generalised Bohm condition'. We note that the kinetic Chodura and Bohm conditions (equations~(\ref{eqn:Chodura condition}) and~(\ref{eqn:kinetic bohm condition alternative form})) are both similar in form, as the kinetic Chodura condition demands that ions enter the presheath on average above the Bohm speed while the kinetic Bohm condition says that the presheath must accelerate these ions in the~$x$ direction into the Debye sheath. In the context of the magnetic presheath however, the kinetic Chodura condition arises as a constraint on the boundary conditions (in this case the entrance ion distribution function) while the marginal kinetic Bohm condition is an analytical property of the solution to the magnetic presheath. We further note that, while the derivative of density with respect to potential plays a small role in the kinetic Chodura condition, it can have a large impact on the marginal kinetic Bohm condition since, as discussed after equation~(\ref{eqn:first derivative o density}), the derivative of the electron density with respect to potential at~$x=0$ will diverge as~$\phi(0)\to \phi_{\cc}$. This means that as~$\phi(0)\to \phi_{\cc}$ the marginal kinetic Bohm condition becomes weaker, in the sense that the divergence of~$\mathrm{d}n_{\ee}/\mathrm{d}\hat{\phi}$ makes the ion distribution move into lower regions of~$v_{x}$ so the ions are being accelerated less strongly in agreement with figure~\ref{fig:cutoff variation pdf}.

Despite giving important information about the distribution function at~$x=0$, the marginal kinetic Bohm condition does not tell us about the form of the potential. To find the true dependence of~$\delta\phi$ on~$x$, we must expand to a further order, keeping only terms that scale like~$\delta\phi^{2}$ or $x$. Thus,
\begin{dmath}
\frac{1}{2}\dev{n_{\ee}}{\phi}{2}\delta\phi^{2} = -Z\Omega_{\ii} x\int_{\bar{x}_{\mathrm{c}}}^{\infty}\Omega_{\ii}^{2}\bar{x}\mathrm{d}\bar{x}\int_{\chi_{\mathrm{M}}(\bar{x})}^{\infty}\frac{F_{\mathrm{cl}}(\mugk,U)}{\sqrt{2\left(U-\chi_{\mathrm{M}}(\bar{x})\right)}}\Delta\left[\frac{1}{v_{x_{0}}} \right]\mathrm{d}U + Z\frac{1}{2}\left(\frac{\Omega_{\ii}\delta\phi}{B} \right)^{2}\int_{\bar{x}_{\mathrm{c}}}^{\infty}\Omega_{\ii}\mathrm{d}\bar{x}\int_{\chi_{\mathrm{M}}(\bar{x})}^{\infty}\frac{F_{\mathrm{cl}}(\mugk,U)}{\sqrt{2\left(U-\chi_{\mathrm{M}}(\bar{x})\right)}}\Delta\left[\frac{1}{v_{x_{0}}^{3}} \right]\mathrm{d}U,
\end{dmath}
and this is solved by
\begin{equation}
\delta\hat{\phi} = px^{\frac{1}{2}},
\end{equation}
with~$p$ defined as
\begin{equation}
\label{eqn:psqrtfactor}
p = \left[\frac{Z\int_{\bar{x}_{\mathrm{c}}}^{\infty}\Omega_{\ii}^{3}\bar{x}\mathrm{d}\bar{x}\int_{\chi_{\mathrm{M}}(\bar{x})}^{\infty}\frac{F_{\mathrm{cl}}(\mugk,U)}{\sqrt{2\left(U-\chi_{\mathrm{M}}(\bar{x})\right)}}\Delta\left[\frac{1}{v_{x_{0}}} \right]\mathrm{d}U}{\frac{1}{2}Zv_{\mathrm{B}}^4\int_{\bar{x}_{\mathrm{c}}}^{\infty}\Omega_{\ii}\mathrm{d}\bar{x}\int_{\chi_{\mathrm{M}}(\bar{x})}^{\infty}\frac{F_{\mathrm{cl}}(\mugk,U)}{\sqrt{2\left(U-\chi_{\mathrm{M}}(\bar{x})\right)}}\Delta\left[\frac{1}{v_{x_{0}}^{3}} \right]\mathrm{d}U - \frac{1}{2}\dev{n_{\ee}}{\hat{\phi}}{2}}\right]^{\frac{1}{2}}.
\end{equation}
Together with the modified kinetic Bohm condition given in equation~(\ref{eqn:kinetic bohm condition}), we may exhibit some important features of the factor~$p$. First, in \ref{section:Vanishing Divergence of Electric field} we find a sufficient condition for~$p$ to be finite. This condition is
\begin{equation}
\label{eqn:sufficient condition for thing}
\frac{3}{n_{\ee}}\left(\dev{n_{\ee}}{\phi}{}\right)^{2} > \dev{n_{\ee}}{\phi}{2},
\end{equation} 
where the derivatives are evaluated at~$\phi(0)$. We further prove that the entrance electron distribution function we choose in section \ref{section:Numerical method} satisfies this condition, and hence that we do not need to further expand the quasineutrality condition to compute the asymptotic behaviour of the potential at~$x=0$. We can also show that $p$ tends to zero as~$\phi(0)\to \phi_{\cc}$. To see this, we differentiate equation~(\ref{eqn:first derivative o density}) with respect to~$\phi$ giving
\begin{dmath}
\label{eqn:second derivative o density}
\left.\dev{n_{\ee}}{\hat{\phi}}{2}\right|_{x=0} = \frac{2\pi T_{\ee}^{2}}{m_{\ee}^{2}} \int_{0}^{\infty}\Omega_{\ee}\mathrm{d}\mu\left[\int_{0}^{\infty}\pdevn{\gyg{f_{\ee}}}{U}{2}\left( \Omega_{\ee}\mu +\frac{\bar{v}_{\parallel}^{2}}{2} - \frac{\Omega_{\ee}}{B}\phi(0),\mu\right)\mathrm{d}\bar{v}_{\parallel} + \int_{0}^{\sqrt{2\frac{\Omega_{\ee}}{B}(\phi(0)-\phi_{\cc})}}\pdevn{\gyg{f_{\ee}}}{U}{2}\left( \Omega_{\ee}\mu +\frac{\bar{v}_{\parallel}^{2}}{2} - \frac{\Omega_{\ee}}{B}\phi(0),\mu\right)\mathrm{d}\bar{v}_{\parallel} - \frac{1}{\sqrt{2\frac{\Omega_{\ee}}{B}(\phi(0)-\phi_{\cc})}}\pdev{\gyg{f_{\ee}}}{U}\left(\Omega_{\ee}\mu - \frac{\Omega_{\ee}}{B}\phi_{\cc},\mu \right) - \frac{1}{\left(2\frac{\Omega_{\ee}}{B}(\phi(0)-\phi_{\cc})\right)^{3/2}}\gyg{f_{\ee}}\left(\Omega_{\ee}\mu - \frac{\Omega_{\ee}}{B}\phi_{\cc},\mu \right) \right].
\end{dmath}
Provided that the distribution function~$\gyg{f_{\ee}}(\Omega_{\ee}\mu - \Omega_{\ee}\phi_{\cc}/B,\mu)$ is non-zero for some range of~$\mu$, it is guaranteed that the first derivative of~$n_{\ee}$ with respect to~$\phi$ (given in equation (\ref{eqn:first derivative o density})) will diverge to~$+\infty$ as~$\phi(0)\to\phi_{\cc}$ while the second derivative will diverge to~$-\infty$. This implies that the denominator of~$p$ diverges like~$1/\left(\phi(0) - \phi_{\cc} \right)^{3/2}$. While the numerator in the bracket in~$p$ will also diverge, it will diverge with the same divergence as the term in the marginal kinetic Bohm condition~(\ref{eqn:kinetic bohm condition}), since the additional presence of~$\bar{x}$ does not change the form of the divergence. Thus the numerator diverges only as~$1/\left(\phi(0) - \phi_{\cc} \right)^{1/2}$. It follows that~$p$ will tend to zero as~$\phi(0)$ tends to~$\phi_{\cc}$. Therefore, the divergence of the electric field vanishes as the magnitude of the potential drop across the Debye sheath decreases. This is in agreement with the numerical results shown in figure~\ref{fig:cutoff variations potential} which shows the weakening of the electric field as the magnitude of the potential drop in the Debye sheath is decreased. While the vanishing divergence of the electic field is consistent with the absence of the Debye sheath implied by~$\phi(0) = \phi_{\cc}$, it indicates the electron density must be expanded with more care in this regime. This is done in the following section.

\subsubsection[]{$\phi(0) =  \phi_{\cc}$.}
\label{section:criticalcutoff}
From equation~(\ref{eqn:electron density local with v parallel bar}), we see that when~$\phi(0)= \phi_{\cc}$, it will no longer be possible to Taylor expand the electron density in $\delta\phi$ and instead we must expand in~$\sqrt{\delta\phi}$,
\begin{equation}
n_{\ee}(x) = n_{\ee}(0) + \dev{n_{\ee}(x)}{\sqrt{\delta\phi}}{}\sqrt{\delta\phi} + O\left(n_{\ee}(0)\frac{e\delta\phi}{T_{\ee}}\right).
\end{equation} 
Since the ion physics has remained unchanged, the only alteration to equation~(\ref{eqn:Infgradatorigin}) is that now there is another term of size~$\sqrt{\delta\phi}$ on the left hand side, giving the square root order quasineutrality equation 
\begin{equation}
\label{eqn:Finite gradient at orgin}
\sqrt{\frac{T_{\ee}}{e}}\dev{n_{\ee}(x)}{\sqrt{\delta\phi}}{} \sqrt{\frac{e\delta\phi}{T_{\ee}}} = Z\int_{\bar{x}_{\mathrm{m,I}}}^{\infty}\sqrt{2\delta\chi}\Omega_{\ii}\mathrm{d}\bar{x}\int_{\chi_{\mathrm{M}}(\bar{x})}^{\infty}\frac{F_{\mathrm{cl}}(\mugk,U)}{\sqrt{2\left(U-\chi_{\mathrm{M}}(\bar{x})\right)}}\mathrm{d}U.
\end{equation}
Having assumed the potential to be finite at the origin, this equation is solved by a finite gradient in potential at~$x=0$. Therefore, we are motivated to (for sufficiently small~$x$) rewrite~$\delta \phi$ as~$\phi'(0)x$. This makes equation~(\ref{eqn:Finite gradient at orgin}) an implicit equation for~$\bar{x}_{\mathrm{m,I}}$, related to the gradient at the origin through equation (\ref{eqn:xbarmI}),
\begin{equation}
\dev{n_{\ee}}{\sqrt{\delta\hat{\phi}}}{}  = v_{B}Z\int_{\bar{x}_{\mathrm{m,I}}}^{\infty}\sqrt{2\left(\frac{\bar{x}}{\bar{x}_{\mathrm{m,I}}} - 1 \right)}\Omega_{\ii}\mathrm{d}\bar{x}\int_{\chi_{\mathrm{M}}(\bar{x})}^{\infty}\frac{F_{\mathrm{cl}}(\mugk,U)}{\sqrt{2\left(U-\chi_{\mathrm{M}}(\bar{x})\right)}}\mathrm{d}U.
\end{equation}
Thus, for~$\phi_{\cc} = \phi(0)$, the electric field at the origin is finite and there exist type~1 closed orbits. The transition from a divergent electric field to a finite one can be seen clearly in figure~\ref{fig:cutoff variations potential}. 
\section{The limits of hot and cold ion temperature}
\label{section:Limits of hot and cold ions}
In section~\ref{section:expansion at zero} we considered the distinction between the cases~${\phi(0) = \phi_{\cc}}$ and~${\phi(0) > \phi_{\cc}}$. Given that the cutoff potential is a parameter of the problem, it is natural to ask whether solutions exist for all~$\phi_{\cc}$ and whether they have~$\phi(0) = \phi_{\cc}$ or $\phi(0) > \phi_{\cc}$. We can answer these questions by appealing to the hot ion theory developed in \cite{Cohen-Ryutov-1998} and \cite{Geraldini-2019}. The hot ion theory models the limit~$T_{\ee} \ll T_{\ii}$ in which the potential cannot substantially distort the ion orbits. In this section we will first show that, for hot ions, there exists a `critical cutoff': a cutoff potential above which quasineutral solutions that adhere to our monotonicity constraints on the elecrostatic potential do not exist. At this critical cutoff, we find~$\phi(0) = \phi_{\cc}$, while for all solutions below this cutoff, we will have~$\phi(0) > \phi_{\cc}$. 

The existence of critical cutoffs is not unique to hot ions and it is possible to show that they apply to all finite ion temperatures. For zero ion temperature, we prove that no critical cutoffs exist due to the lack of finite ion orbit widths. This has an important implication: fluid models of the magnetic presheath fail at very shallow magnetic field angles
\subsection{The hot ion limit}
\label{sec:hotions}
As the ion temperature is increased, the distortion of ion orbits due to the potential becomes weaker. In the limit of a very large ion temperature, the effect of this is that the ion motion at lowest order becomes independent of the potential. This was investigated in \cite{Geraldini-2019} with the input ion distribution function
\begin{equation}
\label{eqn:Hot ion distribution function}
F_{\mathrm{cl}}(U,v_{z}) = 2n_{\ii,\infty}\left(\frac{m_{\ii}}{2\pi T_{\ii}} \right)^{\frac{3}{2}}\exp\left(-\frac{m_{\ii}U}{T_{\ii}}\right)\Theta(v_{z}),
\end{equation}
where~$\Theta$ is the Heaviside function,
\begin{equation}
\Theta(s) = \begin{cases}
1 \text{  for } s \geq 0 ,\\
0 \text{  for } s < 0.
\end{cases}
\end{equation}
For an electric field negligible for the ions, it is purely the finite orbit sizes that result in the depletion of ions in the presheath. It was shown in \cite{Geraldini-2019} that the density at~$x=0$ is given by
\begin{equation}
\label{eqn:Hot ion density at wall}
n_{\ii,\mathrm{op}}(0) = n_{\ii,\infty}\frac{\Gamma^{2}(3/4)}{\pi}\sqrt{\alpha},
\end{equation}
where the gamma function~$\Gamma(x)$ is defined by~$\Gamma(x) = \int_{0}^{\infty}t^{x-1}e^{-t}\mathrm{d}t$. Note that equation~(\ref{eqn:Hot ion density at wall}) is predictably only a function of~$\alpha$. This is a manifestation of the fact that the ions are unaltered by the potential in the presheath. 

Since the ion density is fixed regardless of potential, and the presheath is quasineutral, the potential will take the required value to make the electron density equal to the ion density. The electron density at~$x=0$ is 
\begin{dmath}
\label{eqn:Electron density at origin (local)}
n_{\ee}(0) =  2\pi\int_{0}^{\infty}\Omega_{\ee}\mathrm{d}\mu\left[\int_{0}^{\infty}\gyg{f_{e}}\left(\Omega_{\ee}\mu + \frac{\bar{v}_{\parallel}^2}{2} - \frac{\Omega_{\ee}}{B}\phi(0),\mu \right)\mathrm{d}\bar{v}_{\parallel}+ \int_{0}^{\sqrt{2\frac{\Omega_{\ee}}{B}\left(\phi(0)-\phi_{\cc}\right)}}\gyg{f_{e}}\left(\Omega_{\ee}\mu + \frac{\bar{v}_{\parallel}^2}{2} - \frac{\Omega_{\ee}}{B}\phi(0),\mu\right)\mathrm{d}\bar{v}_{\parallel}\right].
\end{dmath}
If the input electron distribution function is a decreasing function of energy, then the lower bound of equation (\ref{eqn:Electron density at origin (local)}) is reached when~$\phi(0)=\phi_{\cc}$ (see figure~\ref{fig:density}). We denote the value of~$n_{\ee}(0)$ when~$\phi(0) = \phi_{\cc}$ by~$n_{\ee,\mathrm{min}}$,
\begin{dmath}
\label{eqn:Minimum electron density in presheath}
n_{\ee,\mathrm{min}}(\phi_{\cc}) =  2\pi\int_{0}^{\infty}\Omega_{\ee}\mathrm{d}\mu\int_{0}^{\infty}\gyg{f_{e}}\left(\Omega_{\ee}\mu + \frac{\bar{v}_{\parallel}^2}{2} - \frac{\Omega_{\ee}}{B}\phi_{\cc},\mu \right)\mathrm{d}\bar{v}_{\parallel}.
\end{dmath}

If~$n_{\mathrm{e,min}}(\phi_{\cc}) < n_{\ii,\mathrm{op}}(0)$, then there is a quasineutral solution with~$\phi(0)> \phi_\cc$ such that~$n_{\ee}(0) = n_{\ii,\mathrm{op}}(0)$. If~$n_{\mathrm{e,min}}(\phi_{\cc}) = n_{\ii,\mathrm{op}}(0)$, this implies that the quasineutral solution has~$\phi(0)= \phi_{\cc}$. For such critical solutions, the Debye sheath has collapsed as the wall potential is reached at~$x=0$ on the magnetic presheath scale. Noting that equation~(\ref{eqn:Minimum electron density in presheath}) increases to~$n_{\ee,\infty}$ as $\phi_{\cc}$ approaches zero, there is a `critical cutoff'~$\phi_{\mathrm{crit}}$ for which~$n_{\mathrm{e,min}}(\phi_{\mathrm{crit}}) = n_{\ii,\mathrm{op}}(0)$. For~$\phi_{\cc} > \phi_{\mathrm{crit}}$, the electron density is constrained, sufficiently close to the wall, to be greater than the ion density and thus no quasineutral solutions can exist.
Heuristically, the reason for the existence of the critical cutoffs is as follows. The electron density within the presheath is controlled entirely by the potential. A greater potential drop in the presheath results in more electrons being repelled and the density of electrons at~$x=0$ lowering. Raising the cutoff potential~$\phi_{\cc}$ limits the potential drop in the presheath and thus limits the minimum electron density as fewer electrons can be repelled. By raising the cutoff, the minimum electron density can be made arbitrarily close to the electron density far from the wall. The ion density, however, is not only reduced by the presence of an electric field but also by the finite width of the ion orbits. This finite width means that the density will reduce as we let~$x \to 0$ as there is a smaller chance an ion could be found close to the wall without having previously touched the absorbing wall. Thus, the cutoff can be raised to a point where the minimum electron density is larger than the fixed ion density at~$x=0$ and, as a result, there can be no quasineutral solution within our assumption of a monotonic increasing potential profile. The critical cutoffs can be recovered numerically for finite~$T_{\ee}/T_{\ii}$ (see section \ref{section:Numerical method} for the numerical method).
\begin{figure}
	\begin{subfigure}{0.5\textwidth}
    \includegraphics[width = 1\linewidth]{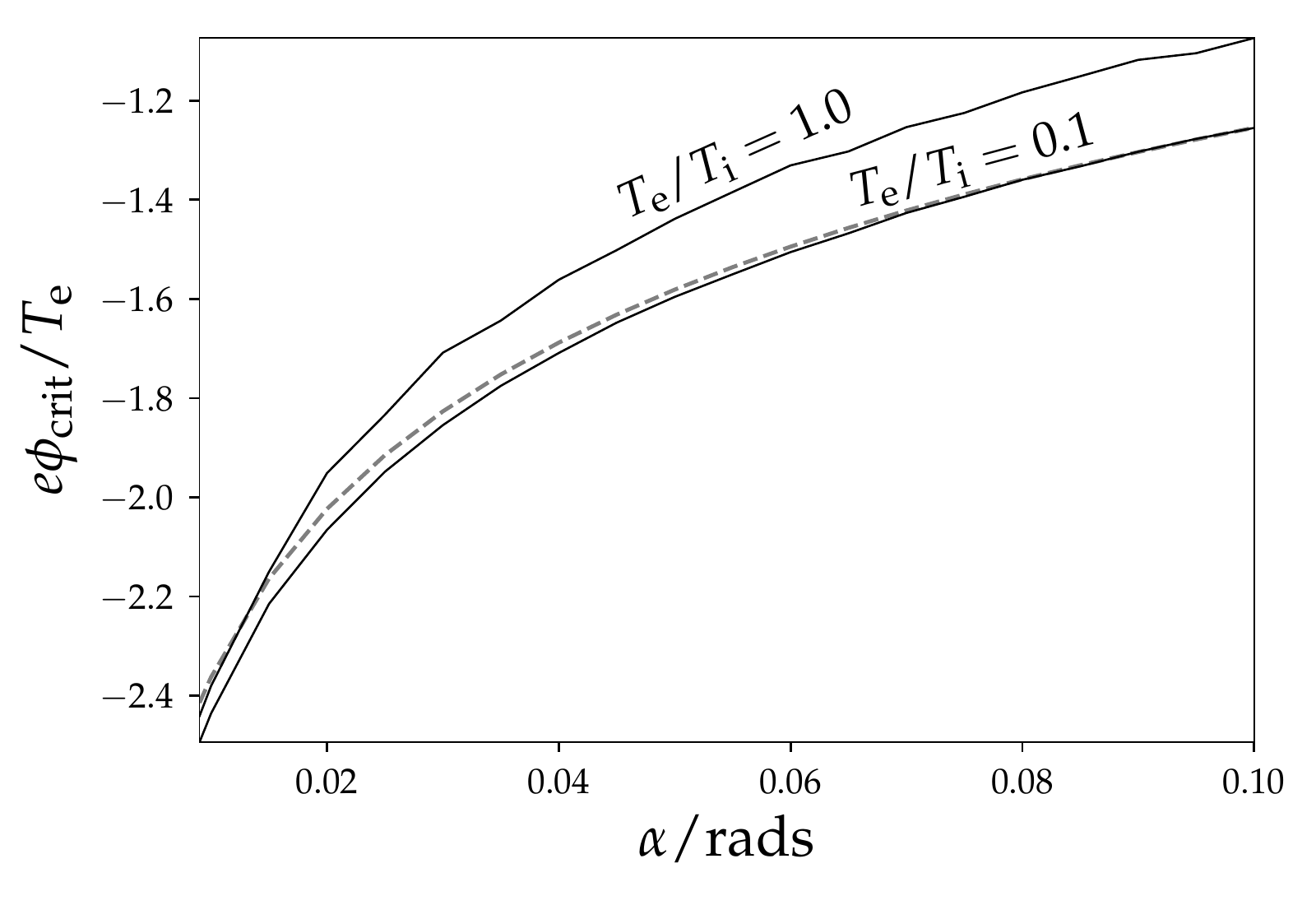}
    \caption{Critical cutoffs}
    \label{fig:comparison_vs_alpha_b}
    \end{subfigure}
    \begin{subfigure}{0.5\textwidth}
    \includegraphics[width = 1\linewidth]{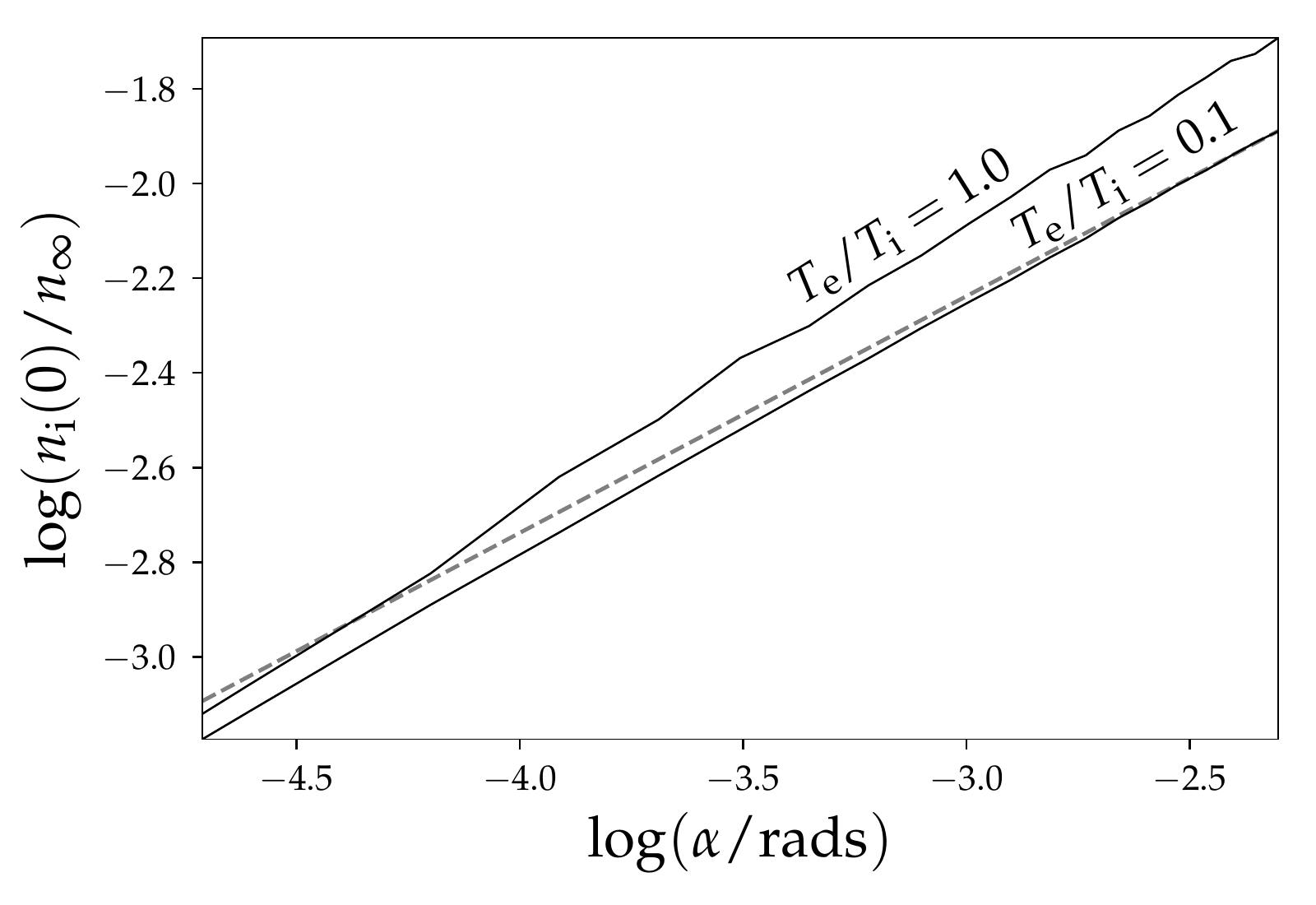}
    \caption{Density at critical cutoff}
    \label{fig:comparison_vs_alpha_a}
    \end{subfigure}
    \caption{Here we see the critical cutoffs as a function of~$\alpha$ for two values of the temperature ratio with the boundary conditions specified in section~\ref{section:Numerical method}. The dashed line represents the theoretical value of these quantities in the limit of hot ions.}
    \label{fig:comparison_vs_alpha}
\end{figure} Figures~\ref{fig:comparison_vs_alpha_b} and~\ref{fig:comparison_vs_alpha_a} show the critical cutoffs as a function of~$\alpha$ for two electron to ion temperature ratios. Even in the case of equal electron and ion temperatures, the solutions lie close to the theoretical hot ion limit of equation~(\ref{eqn:Hot ion density at wall}). This is a manifestation of the reduced distortion of the ion orbits near the wall due to the weaker electric field present in the critical solutions as these solutions are not longer forced to satisfy the marginal kinetic Bohm condition. In section~\ref{section:cold ion limit} we show that this effect is indeed only present when modelling finite orbit widths, as modifying Chodura's fluid model does not give rise to any critical cutoffs.

\subsection{The cold ion limit}
\label{section:cold ion limit}
In the limit of~$T_{\ii}/T_{\ee} \to 0$, the ions can be considered mono-energetic upon entering the magnetic presheath. As such, the ions will follow a single trajectory. This means that all ions move at the same velocity~$\mathbf{u}$, which is a function of the position~$x$. This allows us to rewrite equations~(\ref{eqn:vxdot}),~(\ref{eqn:vydot}) and~(\ref{eqn:vzdot}) for the ions as equations for the fluid velocities~$u_{x}$,~$u_{y}$ and~$u_{z}$ as a function of position
\begin{equation}
\label{eqn:uxdot}
u_{x}\dev{u_{x}}{x}{} = -\frac{\Omega_{\ii}}{B}\phi' + \Omega_{\ii} u_{y}\cos\alpha,
\end{equation}
\begin{equation}
\label{eqn:uydot}
u_{x}\dev{u_{y}}{x}{} = -\Omega_{\ii} u_{x}\cos\alpha - \Omega_{\ii} u_{z}\sin\alpha,
\end{equation}
and
\begin{equation}
\label{eqn:uzdot}
u_{x}\dev{u_{z}}{x}{} =\Omega_{\ii} u_{y}\sin\alpha.
\end{equation}
As well as this, we make use of the fluid continuity equation
\begin{equation}
\label{eqn:Flow}
\frac{\mathrm{d}}{\mathrm{d}x}(n_{\ii}u_{x}) = \frac{1}{Z}\frac{\mathrm{d}}{\mathrm{d}x}(n_{\ee}u_{x}) = 0.
\end{equation}
Here, we have invoked quasineutrality in order to relate the flow equation to our much simpler equations involving the electron density. We can then integrate equation~(\ref{eqn:Flow}) to give
\begin{equation}
\label{eqn:flow as function of x}
    u_{x}(x) = \frac{n_{\ee,\infty}u_{x,\infty}}{n_{\ee}(\phi(x))},
\end{equation}
where we have further specified that the electron density is a function of only~$\phi$ locally, as in equation (\ref{eqn:electron density local with v parallel bar}). Using equation~(\ref{eqn:uxdot}), we find an expression for the~$y$ directed velocity in terms of the potential and its first derivative,
\begin{equation}
\label{eqn:uy in terms of potential}
u_{y}(x) = \frac{1}{\Omega_{\ii}\cos\alpha}\left(-\frac{n_{\ee,\infty}^{2}u_{x,\infty}^2}{n_{\ee}^3}\dev{n_{\ee}}{\hat{\phi}}{} + \frac{ZT_{\ee}}{m_{\ii}} \right)\hat{\phi}'.
\end{equation}
This then allows us to integrate equation~(\ref{eqn:uzdot}),
\begin{dmath}
u_{z}(x)- u_{z,\infty} = \tan\alpha\left[u_{x}(x)-u_{x,\infty} + \frac{ZT_{\ee}}{m_{\ii}}\int_{0}^{\hat{\phi}(x)}\frac{1}{u_{x}(\hat{\phi})}\,\mathrm{d}\hat{\phi} \right] = \tan\alpha\left[u_{x}(x)-u_{x,\infty} + \frac{ZT_{\ee}}{n_{\ee,\infty}u_{x,\infty}m_{\ii}}\int_{0}^{\hat{\phi}(x)}n_{\ee}(\hat{\phi}) \,\mathrm{d}\hat{\phi} \right].
\end{dmath}
We note that equation~(\ref{eqn:uy in terms of potential}) is a function of the first derivative of~$\phi$. Together with the energy conservation equation~(\ref{eqn:U}), this allows us to write a first order ODE in~$\hat{\phi}$,
\begin{equation}
\hat{\phi}' = \frac{\Omega_{\ii}\cos\alpha}{v_{B}^{2}-\frac{n_{\ee,\infty}^{2}u_{x,\infty}^2}{n_{\ee}^{3}(\phi)}\dev{n_{\ee}}{\hat{\phi}}{}}\sqrt{u_{\infty}^{2}- u_{x}^{2}(\hat\phi(x)) - u_{z}^{2}(\hat\phi(x)) - 2v_{B}^{2}\hat{\phi}},
\end{equation}
A salient property of this ODE is that the potential gradient need not be finite. Indeed, motivated by the discussion of the gradient in the general case, we see that the infinite gradient should mark the transition from the quasineutral presheath into the Debye sheath. This occurs when
\begin{equation}
\label{eqn:potentialdrop}
v_{B}^2 = u_{x,\infty}^2\frac{n_{\ee,\infty}^{2}}{n_{\ee}^3(\phi)}\dev{n_{\ee}}{\hat{\phi}}{} \simeq \alpha^{2}u_{z,\infty}^2\frac{n_{\ee,\infty}^{2}}{n_{\ee}^3(\phi)}\dev{n_{\ee}}{\hat{\phi}}{},
\end{equation}
which can be recast, using equation~(\ref{eqn:flow as function of x}),  as a Bohm-like condition, consistent with equation~(\ref{eqn:kinetic bohm condition alternative form}) on the ions entering the Debye sheath,
\begin{equation}
\label{eqn:potential drop in BohmChodura form}
\frac{n_{\ee}(0)}{u_{x}(0)^2} = \frac{1}{v_{B}^2}\dev{n_{\ee}}{\hat{\phi}}{}.
\end{equation}
In the case of Maxwellian electrons, equation~(\ref{eqn:potentialdrop}) recovers the result of Chodura \cite{Chodura-1982} for the potential drop across the magnetic presheath. Note that equation~(\ref{eqn:potentialdrop}) has a solution for any~$\alpha$ and~$\phi_{\cc}$. For smaller~$\alpha$ the density at the origin will reduce itself so as to solve the equation. Thus, lower~$\alpha$ results in a more negative potential at the Debye sheath entrance as in \cite{Geraldini-2018,Chodura-1982}. In our electron model, there is a minimum density at the origin since the potential at the origin cannot be reduced below the cutoff potential. Yet, equation (\ref{eqn:potentialdrop}) still has solutions for all~$\alpha$ because the derivative of the density with respect to potential can be made arbitrarily large by making~$\phi(0)$ arbitrarily close to~$\phi_{\cc}$, as was shown explicitly in equation~(\ref{eqn:first derivative o density}). Thus, unlike with the finite temperature ions, there is no critical cutoff. 

Solutions exist for every~$\phi_{\cc}$ and~$\alpha$, implying that the fluid ion model will be unable to predict the collapse of the Debye sheath. This is in contradiction to the results of \cite{Stangeby-2012} where purely Boltzmann electrons were assumed. 
\begin{figure}
\centering
	\includegraphics[width=0.75\linewidth]{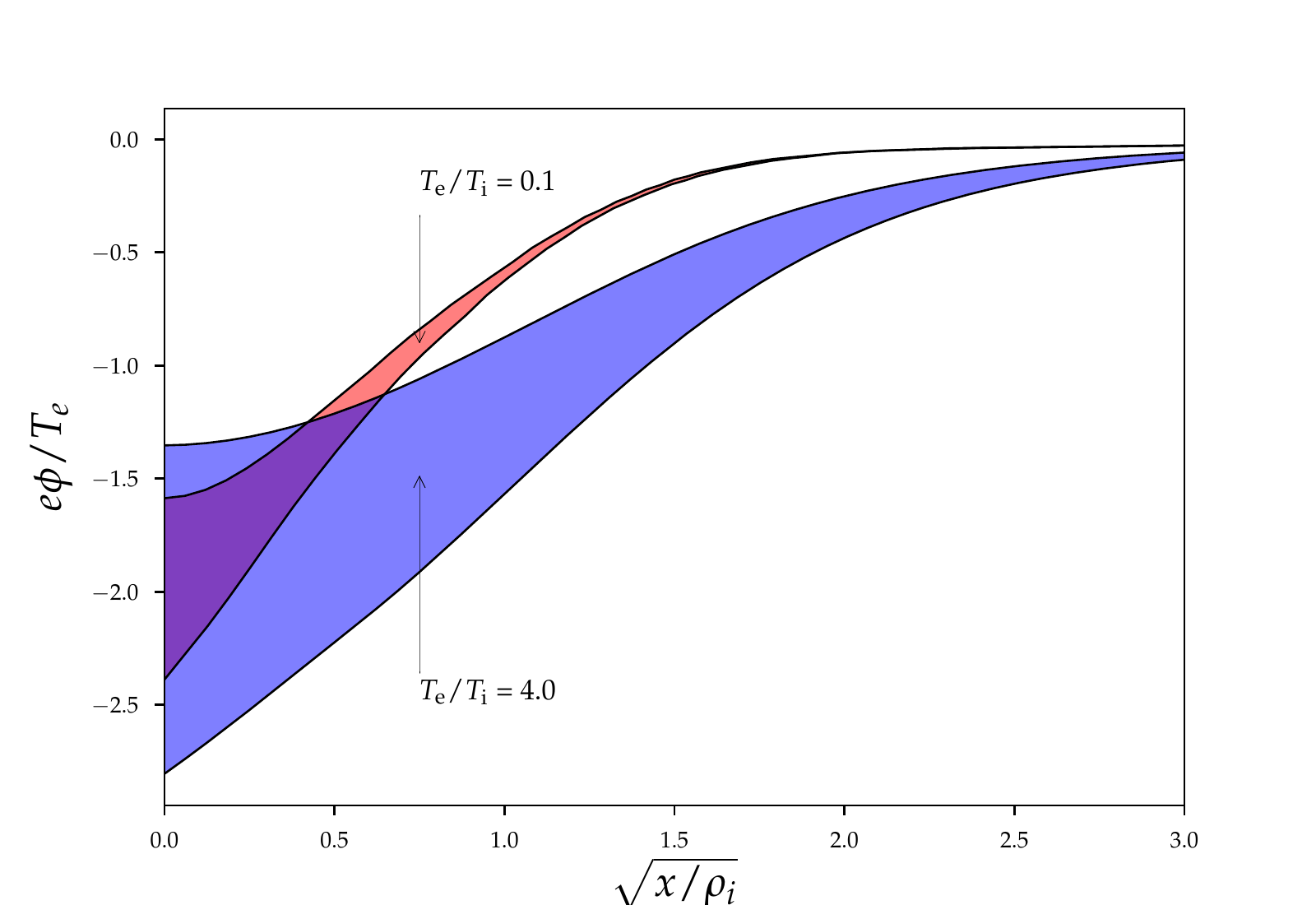}
	\caption{The difference between the critical potential solution (the upper lines) and the~$\phi_{\cc}\to -\infty$ potential solutions (the lower lines) for two different values of the temperature ratio using the boundary conditions detailed in section \ref{section:Numerical method}.}
	\label{fig:hotvscold}
\end{figure} Physically, while the large circular orbits of the hot ions meant that the reduction in density was inevitable, the cold ions are influenced entirely by the electric field. Hence, if the cutoff potential is raised towards zero, the cold ions are accelerated less and the density remains high. Thus, there is no critical cutoff for cold ions. This explains why the critical cutoff is much higher for the colder ions in figure \ref{fig:hotvscold}, since the cutoff must be raised much further before the finite orbit widths cause the electron density to exceed the ion density.

\section{Numerical method}
\label{section:Numerical method}
In the following section, we will outline how the numerical method used here differs from that used in reference \cite{Geraldini-2018}. We will first detail the entrance distribution functions for electrons and ions. After this we will discuss the alterations to the iteration scheme necessary to handle solutions with cutoffs and to find the critical cutoffs.
\subsection{Electron boundary conditions}
\label{secsec:entrance}
In place of the Maxwellian electrons that were used by \cite{Geraldini-2017,Geraldini-2018,Geraldini-2019}, our model accounts for the loss of electrons to the wall. Without a theory of the collisional presheath to provide an appropriate boundary condition on the electrons, the choice of entrance electron distribution functions is arbitrary, just as the original choice of a Maxwellian input was. We choose a distribution function that will tend to the Maxwellian in the limit of~$\phi_{\cc} \to -\infty$: a truncated Maxwellian
\begin{equation}
\label{eqn:electron boundary condition}
f_{\ee,\infty}(\textbf{v}) = \frac{2}{1+\erf(\sqrt{-\frac{\ee}{T_\ee}\phi_{\mathrm{cut}}})}\left(\frac{m_\ee}{2\pi T_\ee} \right)^{\frac{3}{2}}\exp\left(-\frac{m_\ee}{2T_\ee}\lvert\textbf{v}\rvert^{2}\right)\Theta\left(\sqrt{-2\frac{\Omega_{\ee}\phi_{\cc}}{B}} + v_{\parallel}\right),
\end{equation}
where $\erf(x)$ is the error function defined by~$\erf(x) = (2/\sqrt{\pi})\int_{0}^{x}\exp\left(-u^{2}\right)\mathrm{d}u$. To write equation~(\ref{eqn:electron boundary condition}) we have furthermore made the simplest choice of the function~$\Phi(\mu,\delphis)$ as described in section \ref{section:The electron density}, and set~$\Phi(\mu,\delphis)= 0$. In our conserved variables~$\lbrace U, \mu \rbrace$, the distribution function is
\begin{equation}
f_{\ee}(U,\mu) = \frac{2}{1+\erf(\sqrt{-\frac{e}{T_\ee}\phi_{\mathrm{cut}}})}\left(\frac{m_\ee}{2\pi T_\ee} \right)^{\frac{3}{2}}\exp\left(-\frac{m_{\ee}}{T_{\ee}}U\right).
\end{equation}
Using equation~(\ref{eqn:electron density local with v parallel bar}) we arrive at the density as a function of~$\phi(x)$ (see figure~\ref{fig:density}),
\begin{equation}
\label{eqn:nedens}
n_{\ee}(x) = n_{\ee,\infty}\exp\left(\frac{\ee\phi(x)}{T_\ee} \right)\frac{1+\erf\left(\sqrt{\frac{\ee}{T_\ee}(\phi(x)-\phi_{\mathrm{cut}})} \right)}{1+\erf\left(\sqrt{-\frac{\ee}{T_\ee}\phi_{\mathrm{cut}}} \right)}.
\end{equation}
To allow for different possible electron distribution functions at the entrance of the presheath, instead of using equation~(\ref{eqn:nedens}), the code takes as input the electron distribution function integrated over~$\mu$ on an evenly spaced velocity grid to determine the electron density using the procedure detailed in \ref{App:integrationscheme}. For the simulation results shown in this paper, we use a spacing in parallel velocity of~$0.03v_{\mathrm{t,e}}$ although resolutions as low as as~$0.1v_{\mathrm{t,e}}$ would not alter the results greatly. Since we have now included a cutoff potential, it is possible to calculate the electron current density along the magnetic field line. This is given by
\begin{equation}
\label{eqn:Electron flux}
J_{e} = \frac{v_{t,e}}{\sqrt{\pi}} \frac{n_{\ee,\infty}}{1+\erf\left(\sqrt{-\frac{e\phi_{\cc}}{T_{e}}} \right)}\exp\left(\frac{e\phi_{\cc}}{T_{e}}\right).
\end{equation}
This expression can also clearly be determined numerically for a general entrance electron distribution function.
\subsection{Ion boundary conditions}
\label{section:ion boundary conditions}
As in \cite{Geraldini-2018,Geraldini-2019}, without a scheme for determining the proper input distributions for the ions, there is an infinite number of possible distribution functions. We propose a family of possible distribution functions parametrised by~$\tilde{\tau}$,
\begin{equation}
\tilde{\tau} = \left.\frac{T_{\ii}}{ZT_{\ee}} \frac{1}{n_{\ee,\infty}}\dev{n_{\ee}}{\hat{\phi}}{}\right\vert_{\hat{\phi} = 0}.
\end{equation}
This parameter deviates from~$T_{\ii}/ZT_{\ee}$ by only a small amount, but we use this parameter as it is more convenient for the numerical handling of the kinetic Chodura condition. Equipped with this, we choose identical input distributions functions to those chosen in \cite{Geraldini-2019},
\begin{equation}
f_{\ii,\infty}(\vec{v}) = \begin{cases}
\mathcal{N}n_{\ii,\infty}\frac{4v_{z}^2}{\pi^{3/2}v_{\mathrm{t,i}}^{5}}\exp\left(-\frac{\lvert\vec{v} - uv_{\mathrm{t,i}}\hat{\vec{z}} \rvert^{2}}{v_{\mathrm{t,i}}^2} \right)\Theta(v_{z}) \text{  for } \tilde{\tau} \leq 1,  \\
\mathcal{N}n_{\ii,\infty}\frac{4v_{z}^{2}}{\pi^{3/2}v_{\mathrm{t,i}^{3}}(v_{\mathrm{t,i}}^{2} + rv_{z}^{2})}\exp\left(-\frac{\lvert \vec{v} \rvert^{2}}{v_{\mathrm{t,i}}^2} \right)\Theta(v_{z}) \text{  for } \tilde{\tau} > 1.
\end{cases}
\end{equation}
In the conserved variables for the ions, this gives
\begin{equation}
\label{eqn:ion boundary conditions}
F(U,\mugk) = \begin{cases}
\mathcal{N}n_{\ii,\infty}\frac{8(U-\Omega_{\ii}\mugk)}{\pi^{3/2}v_{\mathrm{t,i}}^{5}}\exp\left[-\frac{2\Omega_{\ii}\mugk +  \left(\sqrt{2(U-\Omega_{\ii}\mugk)} -uv_{\mathrm{t,i}}\right)^{2}}{v_{\mathrm{t,i}^{2}}} \right] \text{  for } \tilde{\tau} \leq 1,  \\
\mathcal{N}n_{\ii,\infty}\frac{8(U-\Omega_{\ii}\mugk)}{\pi^{3/2}v_{\mathrm{t,i}^{3}}(v_{\mathrm{t,i}}^{2} + 2r(U-\Omega_{\ii}\mugk))}\exp\left(-\frac{2U}{v_{\mathrm{t,i}}^2} \right) \text{  for } \tilde{\tau} > 1.
\end{cases}
\end{equation}
The parameter~$\mathcal{N}$ ensures that the distribution functions are normalised such that
\begin{equation}
\iiint f_{\ii,\infty}(\vec{v})\mathrm{d}\vec{v} = n_{\ii,\infty}.
\end{equation}
The resulting value of $\mathcal{N}$ is 
\begin{equation}
\mathcal{N} = \begin{cases}
 \left[(1+2u^{2})(1+\erf(u)) + \frac{2u}{\sqrt{\pi}}\exp(-u^{2}) \right]^{-1} \text{  for } \tilde{\tau} \leq 1, \\
 r^{3/2}\left[2\sqrt{r} - 2\sqrt{\pi}\exp\left(\frac{1}{r} \right)(1-\erf\left(\frac{1}{\sqrt{r}}\right))\right]^{-1} \text{  for } \tilde{\tau} > 1.
\end{cases}
\end{equation}

We motivated in section~\ref{section:expansion at infty} that the entrance distribution function must at least satisfy the modified kinetic Chodura condition~(\ref{eqn:Chodura condition}), and we further choose to marginally satisfy it. As such, we select the parameters~$u$ and~$r$ so that the distribution functions satisfy the modified Chodura equation~(\ref{eqn:Chodura condition}). The resulting equations determining~$u$ and~$r$ are thus
\begin{equation}
\mathcal{N}\left(1+\erf(u)\right) = \tilde{\tau}
\end{equation}
for~$\tilde{\tau} \leq 1$, and
\begin{equation}
\frac{\mathcal{N}\sqrt{\pi}}{\sqrt{r}}\exp\left(\frac{1}{r} \right)\left(1-\erf\left(\frac{1}{\sqrt{r}} \right) \right) = \tilde{\tau}
\end{equation}
for~$\tilde{\tau} > 1$. The fluid velocity in the~$z$ direction of the distribution functions is then a function of~$u$ or~$r$, respectively, and is given by
\begin{equation}
\label{eqn:Ion speed cold}
\frac{u_{z,\infty}}{v_{\mathrm{t,i}}} = \mathcal{N}\left[u(3+2u^{2})(1+\erf(u)) + \frac{2}{\sqrt{\pi}}\exp(-u^{2})(1+u^{2}) \right]  \text{ for } \tilde{\tau} \leq 1
\end{equation}
and
\begin{equation}
\label{eqn:Ion speed hot}
\frac{u_{z,\infty}}{v_{\mathrm{t,i}}} = \frac{2\mathcal{N}}{r^{2}\sqrt{\pi}}\left(r - \exp\left(\frac{1}{r}\right)E_{1}\left( \frac{1}{r}\right) \right) \text{ for } \tilde{\tau} > 1,
\end{equation}
where we have introduced the exponential integral
\begin{equation}
E_{1}(\xi) = \int_{\xi}^{\infty}\frac{\exp(-\eta)}{\eta}\,\mathrm{d}\eta.
\end{equation}
\subsection{The iteration scheme}
\label{secsec:iteration}
The iteration procedure from \cite{Geraldini-2018} remains largely unchanged with the alteration of the electron model. We continue to use the numerical method detailed in \cite{Geraldini-2018} to calculate the ion density as a function of position for a given entrance ion distribution function and potential in the presheath. We discretise the presheath into a grid~$(x_{n})$ in~$x$ and will calculate the potential and electron and ion densities at each~$x_{n}$. Let~$\phi_{m}(x_{n})$ denote the potential at position~$x_{n}$ on the~$m^{th}$ iteration. Likewise~$n_{\ee,m}(x_{n})$ and $n_{\ii,m}(x_{n})$ will denote the electron and ion densities at position~$x_{n}$ on the~$m^{th}$ iteration. We will see that there are steps in the iteration in which using the desired cutoff (denoted~$\phi_{\cc,\mathrm{desired}}$) would produce an error. As such, the cutoff potential is also iterated and~$\phi_{\cc,m}$ will denote the cutoff at the~$m^{\mathrm{th}}$ iteration. 

At each step in the iteration, we verify how close the system is to fulfilling quasineutrality condition~(\ref{eqn:Quasi}), and we deem quasineutrality to be satisfied if 
\begin{equation}
\label{eqn:numerical quasineutrality condition}
\left[\frac{1}{N+1}\sum_{\lbrace x_{n} \rbrace}\left(\frac{n_{\ii,m}(x_{n})}{n_{\ee,m}(x_{n})} - 1 \right)^2\right]^{1/2} < \epsilon.
\end{equation} 
Here~$N + 1$ is the number of points in~$x_{n}$, and~$\epsilon$ was taken to be 0.007 for our numerical solution.
The electron density is calculated from~(\ref{eqn:electron density local with v parallel bar}) using the integration scheme detailed in~\ref{App:integrationscheme}. For a given cutoff, this gives the electron density as a function of potential which we will denote~$\tilde{n}_{\ee,\phi_{\cc}}(\phi)$,
\begin{equation}
\tilde{n}_{\ee,\phi_{\cc}}: \left[\phi_{\cc},0 \right] \to \left[n_{\mathrm{e,min}}(\phi_{\cc}), n_{\ee,\infty} \right].
\end{equation} 
The electron density~(\ref{eqn:nedens}) is a monotonic increasing function of potential for the distribution functions we consider, as well as others. As a result, we can define an inverse function denoted by $\tilde{\phi}_{\phi_{\cc}}(n)$,
\begin{equation}
\tilde{\phi}_{\phi_{\cc}}: \left[n_{\mathrm{e,min}}(\phi_{\cc}), n_{\ee,\infty} \right] \to \left[\phi_{\cc},0 \right].
\end{equation}
This is achieved numerically by tabulating the electron density as a function of potential at a given cutoff and then defining a piecewise linear inverse. This tabulation used a potential grid evenly spaced between the cutoff potential and zero. Similarly, we further define~$\tilde{\phi}_{\cc}(n_{\min})$ to be the inverse function of~$n_{\ee,\min}(\phi_{\cc})$, defined in equation~(\ref{eqn:Minimum electron density in presheath}) (i.e. this returns the cutoff potential that would make~$n_{\min}$ the minimum possible electron density in the presheath). Unlike the tabulation of the electron density as a function of potential for a particular cutoff, which must be recalculated each time the cutoff is changed, this function only needs to be calculated once. As such, for our simulations, the typical spacing for the~$\tilde{\phi}_{\phi_{\cc}}$ grid was~$\sim 0.004 T_{\ee}/e$ while the grid spacing for~$\tilde{\phi}_{\cc}$ was~$0.001 T_{\ee}/\ee$. 

We may now present the iteration scheme. After specifying an initial potential profile~$\phi_{1}(x_{n})$, we begin the iteration:
\begin{enumerate}
\item{find $n_{\ii,m}(x_{n})$ from $\phi_{m}(x_{n})$;}
\item{set $\phi_{\cc,m} = \min\left(\phi_{m}(0),\phi_{\cc,\mathrm{desired}},\tilde{\phi}_{\cc}(n_{\ii,m}(0)) \right)$;}
\item{set $n_{\ee,m}(x_{n}) = \tilde{n}_{\ee,\phi_{\cc,m}}(\phi_{m}(x_{n}))$;}
\item{stop the iteration if quasineutrality satisfies (\ref{eqn:numerical quasineutrality condition});}
\item{set $\phi_{m+1}(x_{n}) = \tilde{\phi}_{\phi_{\cc,m}}\left(wn_{\ee,m}(x_{n}) + (1-w)n_{\ii,m}(x_{n})\right)$;}
\item{update the entrance ion distribution function to ensure that the modified kinetic Chodura condition~(\ref{eqn:Chodura condition}) is satisfied;}
\item{go to step (i).}
\end{enumerate}
Here~$w$ is a weight factor (which took values between 0.8 and 0) that quantifies how much the potential will change at each iteration step. After the first potential guess, each subsequent iteration is calculated by inverting the electron density relation on a weighted average of the electron and ion density. The procedure (i) used to calculate the ion density for a given potential~$\phi_{m}(x_{n})$ remains unchanged from \cite{Geraldini-2018}. The remainder of the procedure is devoted to finding the electron density and then inverting the weighted average of the densities to achieve the potential for the next iteration. This procedure is analogous to the one used in \cite{Geraldini-2018}, but is now made more complicated by the presence of a cutoff potential. This is because the cutoff potential places a lower limit on both the electron density and the potential itself. If~$\phi_{m}(x_{n}) < \phi_{\cc,\mathrm{desired}}$, then it would not be possible to find~$n_{\ee,m}(x_{n})$ using~$\tilde{n}_{e,\phi_{\cc,\mathrm{desired}}}$ since~$\phi_{m}(x_{n})$ would not be in the domain of~$\tilde{n}_{e,\phi_{\cc,\mathrm{desired}}}$. Similarly, if the ion density is lower than the electron density, then the weighted density~$wn_{\ee,m}(x_{n}) + (1-w)n_{\ii,m}(x_{n})$ can be outside the domain of~$\tilde{\phi}_{\phi_{\cc,\mathrm{desired}}}$ or~$\tilde{\phi}_{\phi(0)}$, which would make step (v) impossible. Thus, in step (ii), we choose the cutoff to be the minimum of~$\phi_{m}(0)$, $\phi_{\cc,\mathrm{desired}}$ and~$\tilde{\phi}_{\cc}(n_{\ii,m}(0))$. For the majority of solutions, the iteration will terminate with~$\phi_{\cc,m} = \phi_{\cc,\mathrm{desired}}$, indicating that a solution has been found at the required cutoff. However, it is possible for the iteration to terminate with~$\phi_{\cc,m} = \phi_{m}(0) = \tilde{\phi}_{\cc}(n_{\ii,m}(0))$. These correspond to the critical cutoff solutions discussed in section~\ref{section:criticalcutoff}. It is also evident from the iteration scheme that setting the desired cutoff any higher will result in the same critical cutoff solution. 
\section{Results}
\label{section:Results}
Figure~\ref{fig:cutoff variations potential} shows the potential throughout the presheath for a range of cutoff potentials. By plotting this against~$\sqrt{x}$, we see the linear dependence near the origin that characterises a diverging electric field at the entrance to the Debye sheath. This is what we predicted. As the cutoff potential is raised, the gradient of the potential with respect to~$\sqrt{x}$ decreases. This corresponds to the factor~$p$ in equation (\ref{eqn:psqrtfactor}) decreasing, which we predicted in section~\ref{section:expansion at zero}. Finally, when~$\phi_{\cc}= \phi(0)$, we see that the potential becomes quadratic in~$\sqrt{x}$. Thus the divergence of the electric field vanishes along with the Debye sheath for this solution. 

The shrinking of the region with diverging electric field is closely related to the weakening of the modified kinetic Bohm condition discussed in section~\ref{secsec:normalcutoff}. Figure~\ref{fig:cutoff variation pdf} shows this directly by displaying the marginalised ion distribution function in~$v_{x}$,~$f_{0}(v_{x})= \int f_{\ii}(x=0,v_{x},v_{y},v_{z})\mathrm{d}v_{y}\mathrm{d}v_{z}$,  at the entrance to the Debye sheath~(${x=0}$) for the same solutions displayed in figure~\ref{fig:cutoff variations potential}. For solutions with~$\phi_{\cc} \neq \phi(0)$, the distribution falls rapidly to zero as~$v_{x} \to 0$ which is in agreement with equation~(\ref{eqn:kinetic bohm condition alternative form}). As the cutoff potential is raised towards the critical cutoff, we find the density of particles with low~$v_{x}$ increases. This is a manifestation of the electric field being weaker and ions being accelerated less strongly into the Debye sheath. Finally, when we reach the critical solution~($\phi_{\cc} = \phi(0)$), there is a non-zero density of particles with~$v_{x} = 0$, indicating that there are particles at a turning point as they reach~$x=0$, as was expected from section~\ref{section:criticalcutoff}. In the nomenclature of section~\ref{section:ion model}, this is equivalent to the statement that type 2 and type 1 orbits are present for solutions with $\phi_{\cc}= \phi(0)$, whereas only type 2 orbits are present for solutions with~$\phi(0)>\phi_{\cc}$ (as type 1 orbits have turning points at~$x=0$).

Figure~\ref{fig:ion densities} shows the ion densities throughout the presheath for two solutions, one at the critical cutoff and one at~$\phi_{\cc}= -\infty$. The dashed and dotted lines show the closed and open orbit densities, respectively. In both cases, the ion densities drop significantly as we approach~$x=0$. However, there are qualitative differences between the two solutions. In solutions with~$\phi(0)> \phi_{\cc}$, there is a region near~$x=0$ where the closed orbit density becomes very close to zero and the open orbit density makes up almost the entire portion of the ion density. This is an identical result to the one found in \cite{Geraldini-2018} and confirms that the open orbit density is necessary to correctly understand a kinetic treatment of the ions in the presheath. The steep gradients in the potential are causing all the closed orbits to be open before reaching the wall. Again, this can be phrased in the language of type 1 and type 2 orbits discussed in section~\ref{section:ion model}: the steep gradient of the electric field makes all orbits type~2. Sufficiently close to~$x=0$, the only closed orbits that can be found are those with high energy on very large orbits (so that the restoring force of the magnetic field can compete with the large electric field) of which there are exponentially few. In contrast, when~$\phi(0) = \phi_{\cc}$, the closed orbit density is linear near the origin. Closed orbits are present arbitrarily close to the wall.

A subtle but important feature of figure~\ref{fig:ion densities} which we have not discussed yet is the increase in open orbit density at the origin when going from the non-critical to critical regimes. Of particular importance is the change of scaling of this density with~$\alpha$. In \cite{Geraldini-2018, Cohen-Ryutov-1998}, it was shown that the density would scale proportionally with~$\alpha$ in the shallow-angle limit for non-critical solutions. It was further demonstrated that, if there were a significant number of type 1 open orbits, the density would scale like~$\sqrt{\alpha}$. This feature was not realisable outside of the hot ion theory (studied in \cite{Cohen-Ryutov-1998}) as the diverging electric field did not permit type 1 orbits to exist. At the critical cutoff, however, the electric field is finite and no such limitations on type~1 orbits are present. Figure~\ref{fig:comparison_vs_alpha_a} shows the plot on logarithmic scales of density at the origin against magnetic field angle for two separate electron to ion temperature ratios. The solution with~$T_{\ee}/T_{\ii} = 0.1$ lies close to the prediction made by the hot ion theory given in equation~(\ref{eqn:Hot ion density at wall}), although the agreement worsens as $\alpha$ is reduced since the approximation of section~\ref{sec:hotions} that the ions are independent of the potential worsens as the presheath potential becomes more negative. The fact that the critical cutoff becomes more negative as $\alpha$ is decreased can be seen in figure~\ref{fig:comparison_vs_alpha_b} which shows the critical cutoffs as a function of~$\alpha$ for the same two temperature ratios. 
\begin{figure}
\centering 
\includegraphics[width=0.75\linewidth]{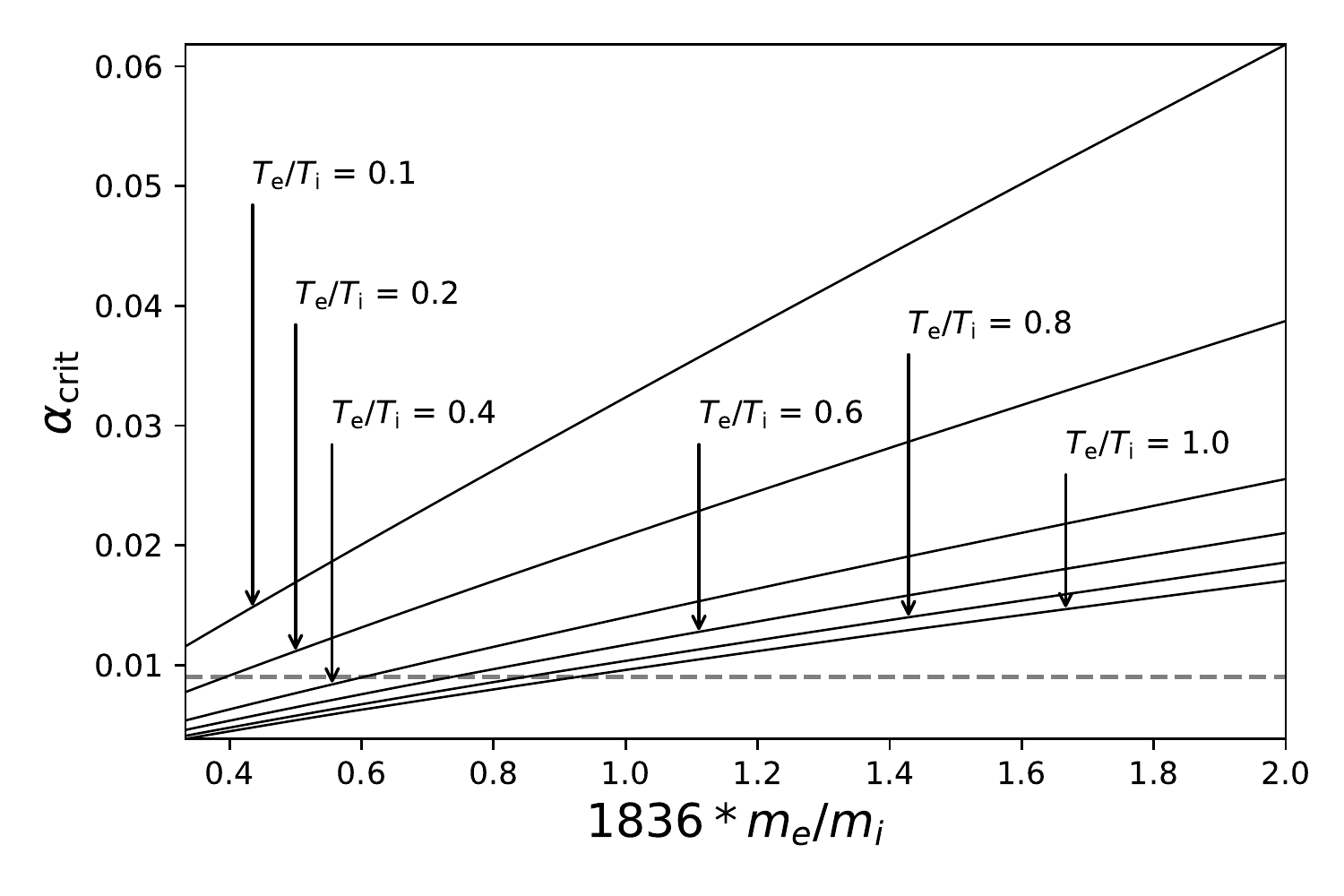}
\caption{The value of~$\alpha$ below which the ion current onto the wall will exceed the electron current as a function of the electron to ion mass ratio. The dashed line represents the lower limit in~$\alpha$ that the code can solve numerically. Below this value of~$\alpha$ the data has been extrapolated using a power law fit.}
\label{fig:critical alphas}
\end{figure}

\begin{figure}
\centering 
\includegraphics[width=0.75\linewidth]{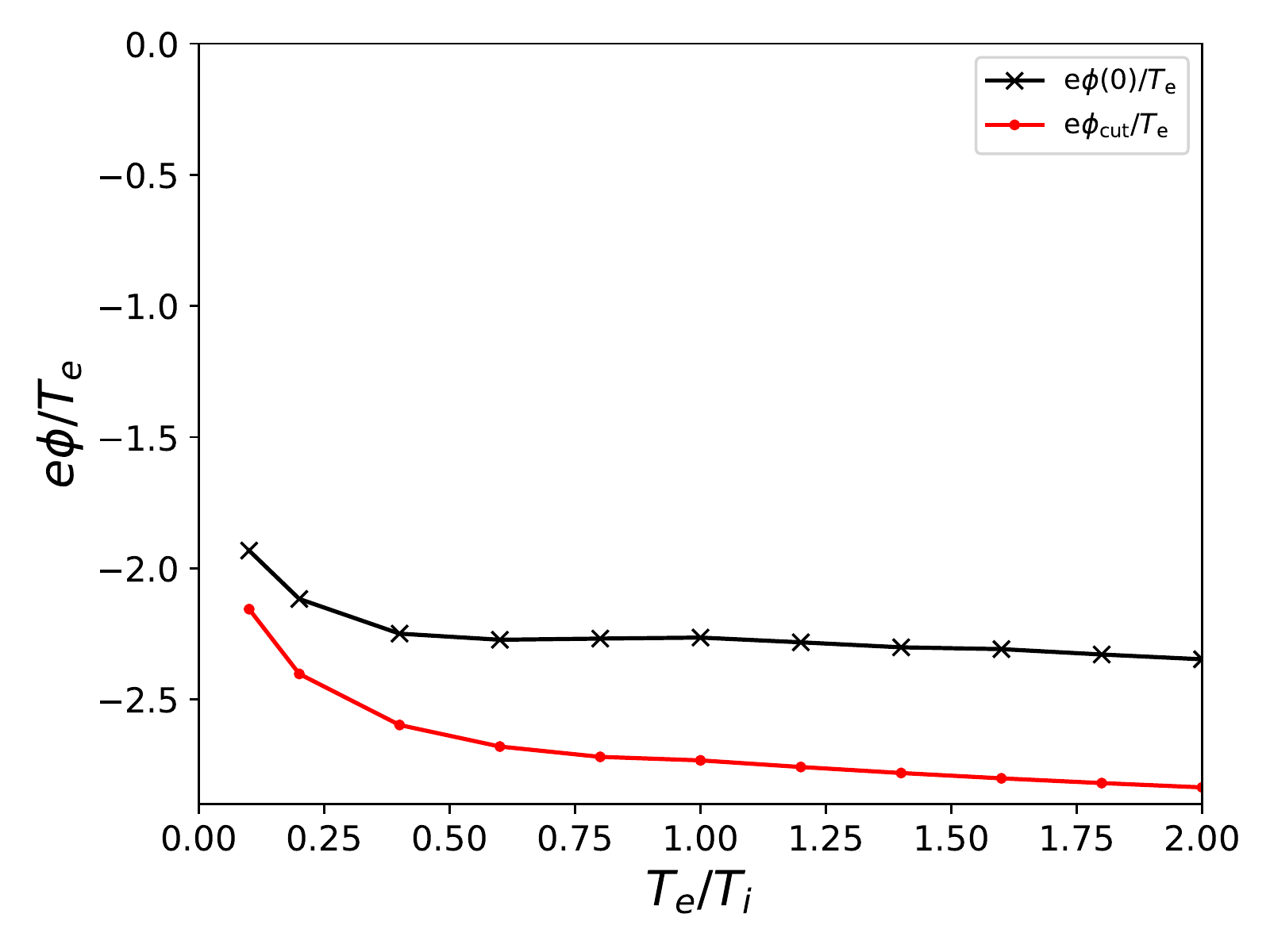}
\caption{The values of the potential drop across the magnetic presheath,~$\phi(0)$, and the total potential drop across the magnetic presheath and Debye sheath,~$-\phi_{\cc}$, at~$\alpha = 0.06$~($\sim 3.4 \degree$) for zero current solutions in the case of a Deuterium plasma. The potential drop in the Debye sheath accounts for between 10\%-20\% of the total potential drop, depending on the temperature ratio. Hence, the Debye sheath does not vanish at this angle.}
\label{fig:stangebyangle}
\end{figure}

The existence of critical cutoffs, or equivalently maximum ion densities, presents a further interesting feature relating to the wall current. Raising the cutoff raises the electron current onto the wall, as fewer electrons are repelled when the potential is less negative. However, since the cutoff cannot be raised above the critical value, for each~$\alpha$ there is a maximum electron current onto the wall. Using equation~(\ref{eqn:Electron flux}), we find that the maximum electron current density is \cite{Coulette-Manfredi-2016}
\begin{equation}
\label{eqn:Electron current}
J_{\ee,\mathrm{max}} = -e\frac{v_{\mathrm{t,e}}}{\sqrt{\pi}}n_{\ee,\mathrm{min}}(\phi_{\mathrm{crit}})\alpha.
\end{equation}
The ion current onto the wall is simply~$Ze\alpha n_{\ii,\infty}u_{z,\infty}$, with~$u_{z,\infty}$ given in equations~(\ref{eqn:Ion speed cold}-\ref{eqn:Ion speed hot}), leading to the scaling of the ion current density
\begin{equation}
J_{\ii} = ZeAc_{s}n_{\ii,\infty}\alpha,
\end{equation}
where~$A$ is some factor of order unity and~$c_{s} = \sqrt{(ZT_{\ee}+T_{\ii})/m_{\ii}}$ the characteristic ion speed. Noting that the maximum ion density at the origin falls with~$\alpha$, we see by quasineutrality that the maximum electron current is guaranteed to fall faster in~$\alpha$. There is therefore an~$\alpha$ below which the maximum electron current cannot equal the ion current. At this $\alpha$, if the system is driven to have zero total current,~$\phi(0)$ is forced to equal~$\phi_{\cc}$ (to achieve maximum electron current) and the Debye sheath will collapse. Numerically, this value of~$\alpha$ could be found by scanning through the critical cutoff solutions at different~$\alpha$ and finding the solutions for which the total current is zero. Equivalently, and for ease of calculation, we find these ~$\alpha$ values by fitting the ion density at~$x=0$ at the critical cutoff (figure~\ref{fig:comparison_vs_alpha_a}) to a power law in~$\alpha$ and solving~$J_{\ii} = J_{\ee,\mathrm{max}}$ for~$\alpha$. This critical angle is shown in figure~\ref{fig:critical alphas} as a function of the electron to ion mass ratio. The linear behaviour is exactly what would follow from a $\sqrt{\alpha}$ dependence of the density, predicted in \cite{Geraldini-2018}, when type 1 orbits are present. Note that for a Deuterium plasma with equal ion and electron temperature, the critical angle for which the Debye sheath collapses in ambipolar conditions is 0.3-0.4 degrees. This is an order of magnitude smaller than previous estimates of about 3 degrees \cite{Stangeby-2012}. We show in figure~\ref{fig:stangebyangle} that the Debye sheath does not vanish at this angle for realistic values of the temperature ratio within our model.

\section{Conclusion}
\label{section:Conclusion}
We have extended the model of Geraldini et al \cite{Geraldini-2018} to account for the effects of absorption of electrons with sufficient energy to reach the wall. Drift kinetics was used to find the kinetic equation satisfied by the electron distribution function. The density of electrons within our model was derived in equation (\ref{eqn:electron density local with v parallel bar}), which we simplified by neglecting electron orbit size. The kinetic Bohm and Chodura conditions were derived in section~\ref{section:Quasineutrality}, where we showed that the truncation of the electron distribution leads to subsonic sheaths, which agrees qualitatively with the results of \cite{Loizu-2011}. It was shown in section~\ref{sec:hotions} that the kinetic Bohm condition fully breaks down at a certain value of the wall potential, referred to as the critical cutoff. At this critical cutoff, the Debye sheath vanishes entirely, bringing the quasineutral presheath into direct contact with the wall. It was further shown in section~\ref{section:cold ion limit} that a Chodura-like fluid theory cannot capture the existence of a critical cutoff due to the lack of finite ion orbit widths (equation~(\ref{eqn:potentialdrop})). The numerical method of Geraldini et al \cite{Geraldini-2018} was altered to include the revised electron model, and the existence of critical cutoffs for finite ion temperatures was shown. Finally, it was argued that, should the wall be at a floating potential, there is a critical magnetic field angle at which the cutoff potential is equal to the critical cutoff. At this point, no Debye sheath is present. Below this angle of incidence, or equivalently for a wall potential above the critical cutoff, the behaviour of the magnetic presheath and Debye sheath is unknown. The solution to the plasma wall boundary above the critical cutoff, therefore, must break one of the assumptions on the nature of our ion or electron model. The reversal of the sheath potential,~$\phi_{\cc}> \phi(0)$, which is excluded by our assumptions, is a possible scenario.

\section*{Acknowledgements}
This work has received funding from UKRI grant number EP/T012250/1. This work has been carried out within the framework of the EUROfusion Consortium and has received funding from the Euratom research and training programme 2014–2018 and 2019–2020 under grant agreement no. 633053. The views and opinions expressed herein do not necessarily reflect those of the European Commission.

AG was supported by the US Department of Energy through grant DE-FG02-93ER-54197. This work was supported in part by the Swiss National Science Foundation

\appendix
\section{Derivation of the electron drift kinetic equation}
In section~\ref{section:Electron model}, we argued that by using variables which are approximately constant for the electron motion, the electron distribution function became independent of~$x$. As a result, finding the electron density became a one dimensional integral where one only had to take into account the deceleration of the electron along the magnetic field line. However, several approximations have been used in this argument. Firstly, the adiabatic invariant given in~(\ref{eqn:mu definition main text}) is not a constant of the motion, but varies slowly. Secondly, the distribution function at a given position need not be independent of gyrophase as changes in the electric field can distort orbits. While these are only small corrections, the electron and ion densities associated with the Debye sheath entrance are also very small and could be of the same size as these corrections. In this appendix therefore, we will provide a drift kinetic derivation of the electron model, which will allow us to track the size of the errors in our approximation.
\subsection{Drift kinetic electrons}
\label{section:Gyrokinetic electrons}
We begin with the Vlasov equation for the electrons:
\begin{equation}
\label{eqn:vlasov}
\pdev{f_{\ee}}{t} + \vec{v}\cdot\nabla f_{\ee} - \frac{e}{m_{\ee}}\left( \vec{E} + \vec{v}\times\vec{B}\right)\cdot\nabla_{v}f_{\ee} = 0.
\end{equation}
The aim of this section is to rewrite equation~(\ref{eqn:vlasov}) to achieve an expression for how the distribution function changes as we move into the presheath. Equation~(\ref{eqn:xbar dot}) implies that, despite their fast oscillations with timescale~$1/\Omega_{\ee}$, the electrons approach the wall on a timescale~$\rho_{\ii}/v_{\mathrm{t,e}}\alpha$. Furthermore, the electrons oscillate on length scales of order~$\rho_{\ee}$, much smaller than the characteristic width of the presheath. We will therefore be able to expand the electron motion in small~$\rho_{\ee}/\rho_{\ii} \sim \sqrt{m_{\ee}/m_{\ii}}$ as well as small angle~$\alpha$. Given such conditions, the classic approach to this problem is that of drift kinetics \cite{Hazeltine-1973}. Hence we will change variables to those quantities that vary slowly over a single particle oscillation. Of course, the gyrophase given in equation~(\ref{eqn:varphi definition main text}) will capture the fast motion and can be alternately written as 
\begin{equation}
\label{eqn:varphi definition}
\varphi = -\tan^{-1}\left(\frac{v_{y}-v_{E}}{v_{n}} \right).
\end{equation}
Here~$\hat{\vec{e}}_{n} = \cos\alpha\, \hat{\vec{e}}_{x} + \sin\alpha \,\hat{\vec{e}}_{z}$ denotes the unit vector perpendicular to both the magnetic field and the~$y$ direction. The quantity~$v_{n} = \vec{v} \cdot \hat{\vec{e}}_{n}$ is the component of velocity in the direction of~$\hat{\vec{e}}_{n}$ while~$\vec{v_{E}}$ denotes the~$\vec{E}\times\vec{B}$ drift velocity,
\begin{equation}
\label{eqn:E cross B drift}
\vec{v_{E}} = \frac{1}{\lvert \vec{B} \rvert}\vec{E}\times \hat{\vec{b}} = \frac{1}{B}\phi'(x)\cos\alpha\hat{\vec{e}}_{y}.
\end{equation}
We have chosen to use velocities suited to the basis~$\lbrace\hat{\vec{e}}_{n},\hat{\vec{e}}_{y},\hat{\vec{b}}\rbrace$ as it is in the plane perpendicular to the magnetic field that the electrons oscillate. Using this basis removes certain terms small in~$\alpha$ that would complicate the algebra.

Clearly the other variables we will choose are the energy~$U$ and the adiabatic invariant given in equation~(\ref{eqn:mu definition main text}) which we can rewrite as
\begin{equation}
\label{eqn:mu definition}
\mu = \frac{1}{2\Omega_{\ee}}\left( v_{n}^{2} + \left(v_{y} - v_{E}\right)^{2}\right).
\end{equation}
This variable was chosen over~$U_{\perp}$ as it is designed to vary on average zero over a single period and is also very convenient for expressions involving~$\varphi$. 

Using equations~(\ref{eqn:xbar}),~(\ref{eqn:U}),~(\ref{eqn:varphi definition}), and~(\ref{eqn:mu definition}), we change  from~$\lbrace t,\vec{r},\vec{v} \rbrace$ to the new variables~$\lbrace t, \bar{x}, y, z, U,\mu,\varphi \rbrace$. In these variables the Vlasov equation~(\ref{eqn:vlasov}) becomes 
\begin{equation}
\dot{\bar{x}}\pdev{f_{\ee}}{\bar{x}} + \dot{\mu}\pdev{f_{\ee}}{\mu} + \dot{\varphi}\pdev{f_{\ee}}{\varphi} = 0,
\end{equation}
where we have removed the~$t$,~$y$ and~$z$ related terms as the distribution is independent of these (as explained in the introduction), and we have removed the term~$\dot{U}(\partial f_{\ee}/\partial U)$ since~$\dot{U}=0$. Here, the dot operator is defined by 
\begin{equation}
\dot{Q} = \pdev{Q}{t} + \vec{v}\cdot \nabla Q - \frac{e}{m_{\ee}}\left(\vec{E} + \vec{v}\times \vec{B} \right)\cdot\nabla_{v}Q,
\end{equation}
which we will compute for our variables~$\bar{x}$,~$\mu$ and~$\varphi$. First noting that equations~(\ref{eqn:varphi definition})) and~(\ref{eqn:mu definition}) give
\begin{equation}
\label{eqn:sin varphi}
\sin\varphi = -\frac{v_{y}- v_{E}}{\sqrt{2\Omega_{\ee}\mu}},
\end{equation}
and
\begin{equation}
\label{eqn:cos varphi}
\cos\varphi =  \frac{v_{n}}{\sqrt{2\Omega_{\ee}\mu}},
\end{equation}
we find the dot operator of our variables equal to
 \begin{dmath}
\label{eqn:xbardot for electrons}
\dot{\bar{x}} = \dot{x} - \frac{\dot{v}_{y}}{\Omega_\ee} = -v_{\parallel}\sin\alpha + \sqrt{2\Omega_{\ee}\mu}(\cos\alpha - 1)\cos\varphi,
\end{dmath}
\begin{dmath}
\label{eqn:mu dot for electrons}
\dot{\mu} = \frac{-\dot{v}_{E}(v_{y}-v_{E})}{\Omega_{\ee}} = \frac{\sqrt{2\Omega_{\ee}\mu}\sin\varphi}{\Omega_{\ee}}\frac{1}{B}\phi''(x)\cos\alpha\left(v_{n}\cos\alpha - v_{\parallel}\sin\alpha \right),
\end{dmath}
and
\begin{dmath}
\label{eqn:varphi dot for electrons}
\dot{\varphi} + \Omega_{\ee} = \frac{\dot{v}_{E}v_{n}}{2\Omega_{\ee}\mu} = \frac{\cos\varphi}{\sqrt{2\Omega_{\ee}\mu}}\frac{1}{B}\phi''(x)\cos\alpha\left(v_{n}\cos\alpha - v_{\parallel}\sin\alpha  \right).
\end{dmath}
From these, we may examine the quantity~$\dot{Q}/Q$ to find the characteristic frequency of each of our variables. Equations~(\ref{eqn:xbardot for electrons}),~(\ref{eqn:mu dot for electrons}) and~(\ref{eqn:varphi dot for electrons}) give the scalings
\begin{equation}
\frac{\dot{\bar{x}}}{\rho_{\ii}} = O\left( \frac{v_{\mathrm{t,e}}}{\rho_{\ii}}\alpha \right),
\end{equation}
\begin{equation}
\frac{\Omega_{\ee}\dot{\mu}}{v_{\mathrm{t,e}}^2} = O\left(\frac{v_{\mathrm{t,e}}}{\rho_{\ii}}\sqrt{\frac{m_{\ee}}{m_{\ii}}} \right),
\end{equation}
and
\begin{equation}
\dot{\varphi}+\Omega_{\ee}  =  O\left(\frac{v_{\mathrm{t,e}}}{\rho_{\ii}}\sqrt{\frac{m_{\ee}}{m_{\ii}}} \right).
\end{equation}

For functions~$G = G(\bar{x},U,\mu,\varphi)$, we define the operator~$\mathcal{L}$ by
\begin{dmath}
\mathcal{L}(G) = \dot{\bar{x}}\pdev{G}{\bar{x}} + \dot{\mu}\pdev{G}{\mu} + (\dot{\varphi}+\Omega_{\ee})\pdev{G}{\varphi}.
\end{dmath}
Equipped with this, the evolution equation for~$f_{e}$ becomes
\begin{equation}
\mathcal{L}(f_{\ee}) - \Omega_{\ee}\pdev{f_{\ee}}{\varphi} = 0.
\end{equation}
Using our scalings for the frequency of our new variables, we may find the scaling of~$\mathcal{L}\left(G \right)$ to be
\begin{equation}
\label{eqn:scaling for G}
\mathcal{L}(G) = \underbrace{\dot{\bar{x}}\pdev{G}{\bar{x}}}_{G\frac{v_{\mathrm{t,e}}}{\rho_{\ii}}\alpha} +  \underbrace{\dot{\mu}\pdev{G}{\mu}}_{G\frac{v_{\mathrm{t,e}}}{\rho_{\ii}}\sqrt{\frac{m_{\ee}}{m_{\ii}}}} + \underbrace{(\dot{\varphi}+\Omega_{\ee})\pdev{G}{\varphi}}_{G\frac{v_{\mathrm{t,e}}}{\rho_{\ii}}\sqrt{\frac{m_{\ee}}{m_{\ii}}}}.
\end{equation}
Our choice of variables has made the characteristic frequency of the operator~$\mathcal{L}$ acting on~$G$ much smaller than the characteristic frequency of particle gyration. Note that presently the choice of ordering~(\ref{eqn:alpha ordering large}) or~(\ref{eqn:alpha ordering small}) does not change the overall scaling of~$\mathcal{L}\left(G \right)$, which is dominated by the largest terms, of order~$\sqrt{m_{\ee}/m_{\ii}}$.

Now, for functions~$G = G(\bar{x},U,\mu,\varphi)$, we further define the gyroaverage of~$G$,~$\gyg{G}$, by 
\begin{equation}
\gyg{G} = \frac{1}{2\pi}\int_{0}^{2\pi} G(\bar{x},U,\mu,\varphi)\mathrm{d}\varphi.
\end{equation}
This is an integral over the gyrophase at constant~$\bar{x}$,~$\mu$ and~$U$. The gyroaverage allows us to decompose~$G$ into its gyrophase dependent and independent pieces,
\begin{equation}
\label{eqn:Gryoaverage decomposition}
G = \gyg{G} + \tilde{G}.
\end{equation}
Decomposing~$f_{\ee}$ into its gyrophase dependent and independent pieces, we get
\begin{equation}
\mathcal{L}(\gyg{f_{\ee}}) + \mathcal{L}(\tilde{f_{\ee}}) - \Omega_{\ee}\pdev{\tilde{f_{\ee}}}{\varphi} = 0.
\end{equation}
We extract from this equation its gyrophase independent part
\begin{equation}
\label{eqn:gyrophase independent part of vlasov}
\gyg{\mathcal{L}(\gyg{f_{\ee}})} + \gyg{\mathcal{L}(\tilde{f}_{\ee})} = 0,
\end{equation}
leaving the gyrophase dependent part
\begin{equation}
\label{eqn:gyrophase dependent part of vlasov}
\Omega_{\ee}\pdev{\tilde{f}_{\ee}}{\varphi} - \widetilde{\mathcal{L}(\tilde{f}_{\ee})} = \widetilde{\mathcal{L}(\gyg{f_{\ee}})}.
\end{equation}
In order to solve equation~(\ref{eqn:gyrophase independent part of vlasov}) for the evolution of~$\gyg{f_{\ee}}$, we need to know~$\tilde{f}_{\ee}$. We can find~$\tilde{f}_{\ee}$ by solving equation~(\ref{eqn:gyrophase dependent part of vlasov}) perturbatively. We write~$\tilde{f}_{\ee} = \tilde{f}_{\ee,1} + \tilde{f}_{\ee,2} + ...$ and demand that~$\tilde{f}_{\ee,1}$ solves
\begin{equation}
\label{eqn:perturbative solution first term}
\Omega_{\ee}\pdev{\tilde{f}_{\ee,1}}{\varphi} = \widetilde{\mathcal{L}(\gyg{f_{\ee}})},
\end{equation}
and each subsequent term satisfies
\begin{equation}
\label{eqn:perturbative solution nth term}
\Omega_{\ee}\pdev{\tilde{f}_{\ee,n+1}}{\varphi} = \widetilde{\mathcal{L}(\tilde{f}_{\ee,n})}.
\end{equation}
By analysing the scalings of equations~(\ref{eqn:perturbative solution first term}) and~(\ref{eqn:perturbative solution nth term}), and using equation~(\ref{eqn:scaling for G}), we find that~$\tilde{f}_{\ee,1} \sim (m_{\ee}/m_{\ii})\gyg{f_{\ee}}$. In~\ref{section:Gyroaverages} we calculate the contribution of~$\gyg{\mathcal{L}(\gyg{f_{\ee}})}$ to equation~(\ref{eqn:gyrophase independent part of vlasov}). Using this we arrive at 
\begin{equation}
\label{eqn:characteristic equation for electrons with errors}
-\alpha\gyg{v_{\parallel}}\pdev{\gyg{f_{\ee}}}{\bar{x}} = O\left(\gyg{\mathcal{L}(\tilde{f}_{\ee,1})} ,\frac{v_{\mathrm{t,e}}}{\rho_{\ii}}\frac{m_{\ee}}{m_{\ii}}\alpha \gyg{f_{\ee}} \right).
\end{equation}
Neglecting the right hand side of equation~(\ref{eqn:characteristic equation for electrons with errors}),
\begin{equation}
\label{eqn:characteristic equation for electrons}
-\alpha\gyg{v_{\parallel}}\pdev{\gyg{f_{\ee}}}{\bar{x}} = 0.
\end{equation}
This equation tells us that the gyroaveraged distribution function, written as a function of the total energy~$U$ and adiabatic invariant~$\mu$, is independent of~$\bar{x}$. However, the left hand side of equation~(\ref{eqn:characteristic equation for electrons}) scales with~$\alpha$. Therefore, for small~$\alpha$, we must carefully justify that the neglected terms of equation (\ref{eqn:characteristic equation for electrons with errors}) are smaller than the term we retain. The second neglected term on the right hand side of equation~(\ref{eqn:characteristic equation for electrons with errors}) includes a factor of~$\alpha$, hence, is trivially smaller by a factor of mass ratio than the left hand side for all~$\alpha$. A naive glance at the other neglected term,~$\gyg{\mathcal{L}(\tilde{f}_{\ee,1})}$, would indicate that it has size~$\left(m_{\ee}/m_{\ii} \right)^{3/2}$. Since this does not scale with~$\alpha$, this would imply that, in our small~$\alpha$ ordering in equation~(\ref{eqn:alpha ordering small}), the errors in our equation for~$\gyg{f_{\ee}}$ would have size~$\sqrt{m_{\ee}/m_{\ii}}$. This is however, not the case, as we show in~\ref{section:first order correction} that the offending term which would appear to have size~$\left(m_{\ee}/m_{\ii} \right)^{3/2}$ in fact cancels, leaving terms that are smaller in mass ratio in both orderings~(\ref{eqn:alpha ordering large}) and~(\ref{eqn:alpha ordering small}).
\subsection{Gyroaverages}
\label{section:Gyroaverages}
Here we calculate the gyroaverages of the quantities~$\dot{\bar{x}},\dot{\mu}$ and $\dot{\varphi}$. To do this we must rewrite equations~(\ref{eqn:xbardot for electrons}),~(\ref{eqn:mu dot for electrons}) and,~(\ref{eqn:varphi dot for electrons}) in terms of the variables that we hold constant for the gyroaverage, and~$\varphi$. Since all our variables contain~$v_{\parallel}$, it is first helpful to show that its~$\varphi$ dependence is small by a factor of~$m_{\ee}/m_{\ii}$. We start by rewriting our equation for~$U$
\begin{dmath}
\label{eqn:expansion of U for v parallel}
U = \Omega_{\ee}\mu - \frac{\Omega_{\ee}}{B}\phi(\bar{x})+ \frac{v_{\parallel}^{2}}{2}  + v_{E}v_{y} - \frac{\Omega_{\ee}}{B}\phi'(\bar{x})(x-\bar{x}) + O\left(v_{\mathrm{t,e}}^{2}\frac{m_{\ee}}{m_{\ii}} \right).
\end{dmath}
Here we have Taylor expanded~$\phi$ around~$\bar{x}$ since we require our terms to be functions only of variables that we hold constant (this is a procedure that will be employed frequently in the gyroaveraging process). Now analysing the last three terms of equation~(\ref{eqn:expansion of U for v parallel}), we have
\begin{dmath}
\label{eqn:gyrophase dependent terms of v parallel}
v_{y}v_{E} - \frac{\Omega_{\ee}}{B}\phi'(\bar{x})(x - \bar{x}) = \frac{v_{y}}{B}\phi''(\bar{x})(x-\bar{x}) + O\left(v_{\mathrm{t,e}}^{2}\frac{m_{\ee}}{m_{\ii}} \right) = O\left(v_{\mathrm{t,e}}^{2}\frac{m_{\ee}}{m_{\ii}} \right),
\end{dmath}
where, for the first equality, we have made use of equations~(\ref{eqn:xbar}) and~(\ref{eqn:E cross B drift}), and we have Taylor expanded in the smallness of the electron gyroradius. Thus, inserting equation~(\ref{eqn:gyrophase dependent terms of v parallel}) back into equation~(\ref{eqn:expansion of U for v parallel}), we see that 
\begin{equation}
\label{eqn:v parallel in terms of good variables}
v_{\parallel} = \sqrt{2\left(U- \Omega_{\ee}\mu + \frac{\Omega_{\ee}}{B}\phi(\bar{x})\right)} + O\left(v_{\mathrm{t,e}}\frac{m_{\ee}}{m_{\ii}} \right).
\end{equation}
Thus the gyrophase dependent part of~$v_{\parallel}$ is order mass ratio smaller than the gyroaveraged piece,~$\gyg{v_{\parallel}}$. From this and equations~(\ref{eqn:xbardot for electrons}- \ref{eqn:varphi dot for electrons}), we may now calculate the gyroaverages of our relevant variables.
\begin{equation}
\gyg{\dot{\bar{x}}} = -\gyg{v_{\parallel}}\sin\alpha,
\end{equation}
\begin{equation}
\gyg{\dot{\mu}} = O\left(\frac{v_{\mathrm{t,e}}^2}{\Omega_{\ee}}\frac{v_{\mathrm{t,e}}}{\rho_{\ii}}\frac{m_{\ee}}{m_{\ii}}\alpha \right),
\end{equation}
and
\begin{equation}
\gyg{\dot{\varphi} + \Omega_{\ee}} = O\left(\frac{v_{\mathrm{t,e}}}{\rho_{\ii}}\sqrt{\frac{m_{\ee}}{m_{\ii}}} \right).
\end{equation}
Here, the gyroaverage of~$\dot{\mu}$ has not vanished because of the oscillating part of~$\phi''(x)$. We are now in a position where we can write down the lowest order gyroaveraged Vlasov equation
\begin{equation}
\label{eqn:Gyroaverage of L operator acting on Gyroaverage of f}
\gyg{\mathcal{L}(\gyg{f_{\ee}})} = -\sin\alpha\gyg{v_{\parallel}}\pdev{\gyg{f_{\ee}}}{\bar{x}} + O\left(\frac{v_{\mathrm{t,e}}}{\rho_{\ii}}\frac{m_{\ee}}{m_{\ii}}\alpha \gyg{f_{\ee}} \right).
\end{equation}
Here the error in equation~(\ref{eqn:Gyroaverage of L operator acting on Gyroaverage of f}) comes from the term~$\gyg{\dot{\mu}}\partial\gyg{f}/\partial\mu$.
\subsection{First order correction $\gyg{\mathcal{L}(\tilde{f}_{\ee})}$}
\label{section:first order correction}
Here we will show that the term~$\gyg{\mathcal{L}(\tilde{f}_{\ee})}$ can be safely neglected in equation~(\ref{eqn:characteristic equation for electrons with errors}), even in our small angle ordering~(\ref{eqn:alpha ordering small}). To do this we will show that the term scaling with~$(m_{\ee}/m_{\ii})^{3/2}$ in~$\gyg{\mathcal{L}(\tilde{f}_{\ee})}$ cancels, leaving terms that are smaller in mass ratio or that scale with~$\alpha$. First, we calculate the first order correction,~$\tilde{f}_{\ee,1}$, by noting
\begin{equation}
\tilde{\dot{\mu}} = \frac{2\mu}{B}\phi''(\bar{x})\sin\varphi\cos\varphi + O\left(\frac{v_{\mathrm{t,e}}^{2}}{\Omega_{\ee}}\frac{v_{\mathrm{t,e}}}{\rho_{\ii}}\frac{m_{\ee}}{m_{\ii}} \right),
\end{equation}
and 
\begin{equation}
\tilde{\dot{\bar{x}}} = \left(v_{\parallel} - \gyg{v_{\parallel}} \right)\sin\alpha + \sqrt{2\Omega_{\ee}\mu}\cos\varphi\left(\cos\alpha - 1 \right) = O\left(v_{\mathrm{t,e}}\frac{m_{\ee}}{m_{\ii}}\alpha,v_{\mathrm{t,e}}\alpha^{2} \right).
\end{equation}
Then solving equation~(\ref{eqn:perturbative solution first term}) for~$f_{\ee,1}$, we find
\begin{dmath}
\tilde{f}_{\ee,1} = \frac{\mu}{\Omega_{\ee}B}\left(\sin^{2}\varphi - \frac{1}{2} \right)\phi''(\bar{x})\pdev{\gyg{f_{\ee}}}{\mu} + O\left( \left( \frac{m_{\ee}}{m_{\ii}} \right)^{3/2}\gyg{f_{\ee}},\sqrt{\frac{m_{\ee}}{m_{\ii}}}\alpha^{2}\gyg{f_{\ee}}\right).
\end{dmath}
Now, calculating~$\gyg{\mathcal{L}(\tilde{f}_{\ee,1})}$ gives

\begin{dmath}
\gyg{\mathcal{L}(\tilde{f}_{\ee,1})} = -\frac{\mu}{2\Omega_{\ee}B^{2}}\left(\phi''(\bar{x})\right)^{2}\gyg{\sin 2\varphi\cos 2\varphi}\left[ \pdev{\gyg{f_{\ee}}}{\mu} + \mu\frac{\partial^{2}\gyg{f_{\ee}}}{\partial\mu^{2}} \right] + \gyg{\cos^{3}\varphi\sin\varphi}\frac{2\mu}{\Omega_{\ee}B^{2}}\left(\phi''(\bar{x})\right)^{2}\pdev{\gyg{f_{\ee}}}{\mu} + O\left( \frac{v_{\mathrm{t,e}}}{\rho_{\ii}}\gyg{f_{\ee}}\left\lbrace\left(\frac{m_{\ee}}{m_{\ii}}\right)^{2},\left(\frac{m_{\ee}}{m_{\ii}} \right)^{3/2}\alpha, \frac{m_{\ee}}{m_{\ii}}\alpha^{2} ,\sqrt{\frac{m_{\ee}}{m_{\ii}}}\alpha^{3}\right\rbrace \right) = O\left( \frac{v_{\mathrm{t,e}}}{\rho_{\ii}}\gyg{f_{\ee}}\left\lbrace\left(\frac{m_{\ee}}{m_{\ii}}\right)^{2}, \sqrt{\frac{m_{\ee}}{m_{\ii}}}\alpha^{3}\right\rbrace \right),
\end{dmath}
where the largest terms have vanished upon gyroaveraging. Thus we have shown that the corrections due to the first order gyrophase dependent piece are sufficiently small and that equation~(\ref{eqn:characteristic equation for electrons}) is still valid for our small~$\alpha$ ordering~(\ref{eqn:alpha ordering small}).

\section{Proof that $p$ is finite}
\label{section:Vanishing Divergence of Electric field}
We will first show that~$p$ defined by equation~(\ref{eqn:psqrtfactor}) is finite for the boundary conditions defined in section~\ref{section:Numerical method}. To do this we need only show that the denominator in equation~(\ref{eqn:psqrtfactor}) will always be non-zero. We rewrite the first term of the denominator as follows
\begin{dmath}
Zv_{\mathrm{B}}^4\int_{\bar{x}_{\mathrm{c}}}^{\infty}\Omega_{\ii}\mathrm{d}\bar{x}\int_{\chi_{\mathrm{M}}(\bar{x})}^{\infty}\frac{F_{\mathrm{cl}}(\mugk,U)}{\sqrt{2\left(U-\chi_{\mathrm{M}}(\bar{x})\right)}}\Delta\left[\frac{1}{v_{x_{0}}^{3}} \right]\mathrm{d}U = Zv_{\mathrm{B}}^4\int_{\bar{x}_{\mathrm{c}}}^{\infty}\Omega_{\ii}\mathrm{d}\bar{x}\int_{\chi_{\mathrm{M}}(\bar{x})}^{\infty}\frac{F_{\mathrm{cl}}(\mugk,U)}{\sqrt{2\left(U-\chi_{\mathrm{M}}(\bar{x})\right)}}\mathrm{d}U\int_{-\infty}^{0}\frac{3}{v_{x}^{4}}\hat{\Pi}\left(v_{x},V_{\mathrm{op},0} - \Delta v_{x_{0}},V_{\mathrm{op},0}\right)\mathrm{d}v_{x} = 3Zv_{\mathrm{B}}^4 \int \frac{f_{i,x,0}(v_{x})}{v_{x}^{4}}\mathrm{d}v_{x}.
\end{dmath}
Hence it will be sufficient to prove that 
\begin{equation}
\label{eqn:inequality we want to prove}
3Zv_{\mathrm{B}}^4 \int \frac{f_{i,x,0}(v_{x})}{v_{x}^{4}}\mathrm{d}v_{x} > \dev{n_{\ee}}{\hat{\phi}}{2}.
\end{equation}
To do this we use the Bohm condition (equation~(\ref{eqn:kinetic bohm condition alternative form})) and Schwarz's inequality to write
\begin{equation}
\label{eqn:lower bound on integral used in p}
\int \frac{f_{i,x,0}(v_{x})}{v_{x}^{4}}\mathrm{d}v_{x} \geq \frac{\left( \int f_{\ii,x,0}(v_{x})v_{x}^{-2}\mathrm{d}v_{x} \right)^{2}}{\int f_{i,x,0}(v_{x})\mathrm{d}v_{x}} = \frac{1}{Zv_{B}^{4}}\frac{1}{n_{\ee}(x)}\left(\dev{n_{\ee}}{\hat{\phi}}{} \right)^{2}.
\end{equation}
Since the inequality~(\ref{eqn:lower bound on integral used in p}) is always true, if the inequality~(\ref{eqn:sufficient condition for thing}) holds then we are guaranteed to satisfy the inequality~(\ref{eqn:inequality we want to prove}). Thus the inequality~(\ref{eqn:sufficient condition for thing}) is indeed a sufficient condition for~$p$ to be finite. Now specialising to our electron model with density described by~(\ref{eqn:nedens}), we may write the first derivative as 
\begin{equation}
\label{eqn:first derivative of density}
\dev{n_{\ee}}{\phi}{} = n_{\ee}(\phi) + \frac{1}{\sqrt{\pi}}\frac{n_{\ee,\infty}e^{\hat{\phi}_{\cc}}}{1+\erf\left(\sqrt{-\hat{\phi}_{\cc}} \right)}\frac{1}{\sqrt{\hat{\phi}-\hat{\phi}_{\cc}}}
= n_{\ee} + \tilde{A}(\phi),
\end{equation}
and the second derivative as
\begin{equation}
\label{eqn:second derivative of density}
\dev{n_{\ee}}{\hat{\phi}}{2} = \dev{n_{\ee}}{\hat{\phi}}{} - \frac{1}{2\sqrt{\pi}}\frac{n_{\ee,\infty}e^{\hat{\phi}_{\cc}}}{1+\erf\left(\sqrt{-\hat{\phi}_{\cc}}\right)} \frac{1}{\left(\hat{\phi} - \hat{\phi}_{\cc} \right)^{3/2}} = \dev{n_{\ee}}{\hat{\phi}}{} - \tilde{B}(\phi).
\end{equation}
Here it is only relevant to know that~$\tilde{A}$ and~$\tilde{B}$ are positive functions of~$\phi$. Equation~(\ref{eqn:second derivative of density}) then implies that the normalised second derivative of density is less than the normalised first derivative. We may then use equations~(\ref{eqn:first derivative of density}) and~(\ref{eqn:second derivative of density}) as well as the positivity of~$\tilde{A}$ and~$\tilde{B}$ to argue
\begin{equation}
\label{eqn:derivative inequality}
\frac{1}{n_{\ee}}\left(\dev{n_{\ee}}{\hat{\phi}}{} \right)^{2} = n_{\ee} + 2\tilde{A} +\frac{\tilde{A}^{2}}{n_{\ee}} > \dev{n_{\ee}}{\hat{\phi}}{} - \tilde{B} = \dev{n_{\ee}}{\hat{\phi}}{2}.
\end{equation}
We see that we manifestly satisfy our sufficient inequality~(\ref{eqn:sufficient condition for thing}). Thus,~$p$ is finite for all solutions. 
\section{Integration scheme for the electron density}
\label{App:integrationscheme}
The electron density is given by equation~(\ref{eqn:electron density local with v parallel bar}). As such, we are concerned with integrals of the following form,
\begin{equation}
I = \int_{x=a}^{x=b}\mathrm{d}\left(\sqrt{x^{2}-a^{2}} \right)g(x),
\end{equation}
where~$x$ plays the role of~$v_{\parallel}$. In the variable~$z = \sqrt{x^2 - a^2}$, this integral would now admit the trapezium rule or Simpson's rule. One simply has to `tile' the integration domain in~$z$ with tiles of width~$h$. The complication is as follows: in our context~$g(x)$ is the entrance distribution for the electrons and we anticipate situations where~$g(x)$ will not be known exactly. Instead, if the entrance distribution function is supplied by a code, we expect~$g(x)$ to be known only at certain values evenly spaced in~$x$. We adapt Simpson's rule to work with evenly spaced $x$. This can be achieved as follows. For every small interval~$[x_{1},x_{2}]$, we use
\begin{dmath}
I_{\mathrm{strip}} = \int_{x = x_{1}}^{x = x_{2}}\dd{\left(\sqrt{x^{2}-a^{2}}\right)}g(x) = \int_{x=x_{1}}^{x= x_{2}}\dd{\left(\sqrt{x^{2}-a^{2}}\right)}\left[g(x_{1}) + g'(x_{1})(x-x_{1}) + \frac{g''(x_{1})}{2}(x-x_{1})^2 + O\left((x-x_{1})^3\right)  \right].
\end{dmath}
Rewriting~$x$, we get
\begin{dmath}
\label{eqn:integral strip equation}
I_{\mathrm{strip}} = \int_{z = \sqrt{x_{1}^{2}-a^{2}}}^{z= \sqrt{x_{2}^{2}-a^{2}}}\dd{z}\left[g(x_{1}) +g'(x_{1})\left(\sqrt{z^{2}+a^{2}}-x_{1}\right) +\frac{g''(x_{1})}{2}\left(\sqrt{z^{2}+a^{2}} - x_{1} \right)^2 + O\left( (\sqrt{z^{2}+a^{2}}-x_{1})^{3} \right)\right].
\end{dmath}
The terms involving square roots can be solved by hyperbolic substitution. This gives
\begin{dmath}
I_{\mathrm{strip}} = \left(g(x_{1})-x_{1}g'(x_{1}) + \frac{x_{1}^{2}+a^{2}}{2}g''(x_{1})\right)\left[\sqrt{x_{2}^{2}-a^{2}}-\sqrt{x_{1}^{2}-a^2}\right] + \frac{(g'(x_{1})-x_{1}g''(x_{1}))}{2}\Bigg[x_{2}\sqrt{x_{2}^{2}-a^{2}}-x_{1}\sqrt{x_{1}^{2}-a^{2}}+ a^{2}\Big[\sinh^{-1}\left(\frac{\sqrt{x_{2}^{2}-a^{2}}}{a}\right)-\sinh^{-1}\left(\frac{\sqrt{x_{1}^{2}-a^{2}}}{a}\right) \Big] \Bigg] + \frac{g''(x_{1})}{6}\Big[ (x_{2}^{2}-a^{2})^{3/2} - (x_{1}^{2}-a^{2})^{3/2} \Big].
\end{dmath}
Here, the first and second derivatives of the function are calculated using finite difference methods to the required level of accuracy,
\begin{equation}
g'(x_{1}) = \frac{g(x_{1}+h)- g(x_{1}-h)}{2h} + O( h^2)
\end{equation}
and
\begin{equation}
g''(x_{1}) = \frac{g(x_{1}+h)+g(x_{1}-h) - 2g(x_{1})}{h^{2}} + O(h).
\end{equation}

We will now address the size of the errors of our approximation to the integral. A common feature of strip based methods of integration is the scaling of the error with the strip size. Due to our choice of evenly spaced strips in~$x$, however, the strips in~$z$ are not even. Therefore, we will split our discussion into one of regular strips in~$z$ with typical width~$\sim h$ and irregular strips of width~$\sim \sqrt{h}$. In both cases, the size of the neglected term is of order~$h^{3}$ from two contributions: the truncation of the Taylor expansion of~$g(x)$, and the finite difference expansion of the derivatives~$g(x)$. There are an order unity number of irregular width strips so the total error from the irregular strips scales as~$h^{7/2}$. There are of order~$1/h$ regular strips of width~$h$ and therefore the total error of these strips scales like~$h^{3}$.

\section*{References}
\bibliography{gyrokineticsbibliography}{}
\bibliographystyle{unsrt}

\end{document}